\documentclass[12pt]{article}
\usepackage{mathtools}
\usepackage{float}
\usepackage{relsize}
\usepackage{array}
\usepackage{hyperref}
\usepackage{fullpage}
\usepackage{graphicx}
\usepackage{psfrag}
\usepackage{epsf}
\usepackage{amsfonts}
\usepackage{amsmath}
\usepackage{pstricks}
\usepackage{color}
\usepackage{setspace}
\usepackage[novbox]{pdfsync}
\input epsf.sty

\allowdisplaybreaks
\usepackage{slashed}
\usepackage{amsmath}
\usepackage{amssymb}
\usepackage{graphicx}

\def\timesbox{\hbox{$\scriptscriptstyle\times$}}
\def\ant{ {{\lower 1ex  \timesbox} \atop {\raise 1.5ex  \timesbox}}}
\def\f{\frac}

\def\non{\nonumber\\}

\def\o{\omega}

\def\t{\tau}

\newcommand{\mathsym}[1]{{}}
\newcommand{\unicode}[1]{{}}

\newcommand {\expect}[1]{\left\langle #1 \right\rangle}

\newcommand\ZZZ{{\hbox{ Z\kern-1.6mm Z}}}
\newcommand{\Iop}{\relax{\rm I\kern-.18em I}}
\newcommand{\Lop}{\relax{\rm I\kern-.18em L}}
\newcommand{\dop}{\relax{\rm I\kern-.8em d}}
\newcommand{\one}{{\hbox{ 1\kern-1.2mm l}}}

\newcommand{\be}{\begin{equation}}
\newcommand{\ee}{\end{equation}}
\newcommand{\beqa}{\begin{eqnarray}}
\newcommand{\eeqa}{\end{eqnarray}}
\newcommand{\bsp}{\begin{split}}
\newcommand{\esp}{\end{split}}
\newcommand{\bgth}{\begin{gather}}
\newcommand{\egth}{\end{gather}}

\newcommand{\hf}{{1\over 2}}

\begin{document}

\title{
{\textbf{Finite Temperature Corrections to Tachyon Mass in Intersecting $D$-Branes
}}
\author{Varun Sethi$^{1}$\footnote{vsethi@physics.du.ac.in}, Sudipto Paul Chowdhury$^{2}$\footnote{paulchowdhurysudipto@gmail.com} and 
Swarnendu Sarkar$^{1}$ \footnote{ssarkar@physics.du.ac.in}\\
$^1$\small{{\em Department of Physics and Astrophysics,
University of Delhi,}} \\ 
\small{{\em Delhi 110007, India}}\\
$^2$\small{{\em Institute of Physics, Sachivalaya Marg}}\\
\small{{\em Bhubaneswar 751005, India}}\\
}}

\maketitle

\abstract{We continue with the analysis of finite temperature corrections to the Tachyon mass in intersecting branes which was initiated in \cite{1}. In this paper we extend the computation to the case of intersecting $D3$ branes by considering a setup of two intersecting branes in flat-space background. A holographic model dual to BCS superconductor consisting of intersecting $D8$ branes in $D4$ brane background was proposed in \cite{2}. The background considered here is a simplified configuration of this dual model. We compute the one-loop Tachyon amplitude in the Yang-Mills approximation and show that the result is finite. Analyzing the amplitudes further we numerically compute the transition temperature at which the Tachyon becomes massless. The analytic expressions for the one-loop amplitudes obtained here reduce to those for intersecting $D1$ branes obtained in \cite{1} as well as those for intersecting $D2$ branes.} 
\newpage

\tableofcontents

\baselineskip=18pt

\section{Introduction}\label{intro}

Applications of AdS/CFT to various condensed matter systems and QCD at strong coupling have been an active field of research for the past few years. Specific areas where these holographic techniques have been applied with varied degree of success are in the study of of quantum liquids at strong coupling \cite{LMV}-\cite{DHS}, including studies of phase transitions such as in superconductors \cite{G}-\cite{HHH}.

A holographic model for a BCS superconductor was proposed in \cite{2}. The idea behind the proposal was to apply the holographic techniques for studying superconductivity in a top-down approach. The proposal is based on the holographic QCD model constructed by Sakai and Sugimoto \cite{SS1}. This is a variant of the model proposed by Witten \cite{EW1} which includes fundamental fermions. Various interesting aspects of holographic QCD models have been explored by several authors. For a partial list of references see \cite{KK}-\cite{J1}. The Sakai-Sugimoto model consists of flavor $D8$ branes in the $D4$ brane background. The relevant $(3+1)D$ large $N$ QCD resides on the $D4$ brane which is compactified on $S^1$. The $D8$ branes intersect in the bulk at an angle. In \cite{2} two such $D8$ branes were used to model the dual holographic superconductor. Intersecting $D$-branes usually have instabilities manifesting as tachyons in the the spectrum. Such tachyonic instability that arises in the above case of intersecting $D8$ branes has been proposed to be the dual of the Cooper-pairing instability in the  $3+1$ dimensional theory in \cite{2}. 
In \cite{1}, we have demonstrated a thermal stabilization in a simplified set-up consisting of two $D1$-branes intersecting at one non-zero angle. This is done  by computing a finite temperature mass correction to the tree-level tachyon using the Yang-Mills approximation. This approximation in the case of $Dp$-branes gives a renormalizable theory only for $p \leq 3$. 

In this paper we extend the calculations initiated in \cite{1} to the case of intersecting $D3$ branes. The setup consists of two 
$D3$ branes intersecting with one non-zero angle $\theta$ in flat space. Similar to such an analogous setup for intersecting $D1$ branes the spectrum
of open strings with the two ends on each brane consists of a Tachyon as the lowest mode. The $mass^2$ for the Tachyon is given by
$-\theta/(2\pi\alpha^{'})$ \cite{BDL,SJ1,SJ2,HT}. This was rederived in the Yang-Mills approximation $(\alpha^{'}\rightarrow 0, \theta \rightarrow 0)$, with $\theta/(2\pi\alpha^{'})=q$ fixed in \cite{Hashimoto:2003xz}-\cite{Epple:2003xt}, wherein the end point of Tachyon condensation leading to smoothing out of the brane configuration was studied. Further studies on tachyon condensation in intersecting branes include \cite{Huang,Jones,J2}.

The purpose of the present computation is to analyze the finite temperature corrections to the Tachyon mass which we obtain by computing the 2-point amplitude. Since the finite temperature corrections are positive in sign, there exists a critical temperature $T_c$ at which the Tachyon mass is exactly zero. We shall compute this temperature in the limits (as mentioned above) where the Yang-Mills approximation is valid. The calculations in this paper follow along the lines of \cite{1}. The intersecting brane configuration consists of giving a non-zero expectation value to one of the scalars namely $\Phi_1^3$ which is of the form $qx$. Here $x$ denotes a coordinate along the plane of intersection. Due to the linear (in $x$) nature of the background and in the temporal gauge ($A^a_0=0$), the wave-functions that diagonalize the quadratic part of the Lagrangian consist of Hermite Polynomials instead of exponential plane-waves which we have in the usual case. The corresponding momentum modes are discrete and are denoted by $n$. The wave functions have been derived in \cite{1} for the $D1$ branes. These functions have been adapted for the present case of $D3$ branes with some modifications. The additional complication in the present case is the presence of extra momentum degrees of freedom coming from the directions transverse to the plane of intersection. There are two such extra momentum directions relative to the $D1$ brane case. We call these momentum modes ${\bf k}=(k_2,k_3)$. 

The supersymmetric Yang-Mills theory on the $D3$ brane is finite.
Although supersymmetry is completely broken for the intersecting branes, the ultraviolet property of the theory on the intersecting branes is unaffected. The theory is still finite in the ultraviolet. We shall show analytically the ultraviolet finiteness for all the one-loop two-point amplitudes computed in the paper. Specifically we show that for large $n$, independent of the value of ${\bf k}$, the contributions from the bosons and fermions in the loop cancel. Similar cancellation has also been shown for the large ${\bf k}$ for the two-point amplitude for the ``massless" fields.  

The one-loop amplitude for the Tachyon involves fields propagating in the loop that are massless at tree-level. In order to deal with the resulting infrared divergences arising out of these massless fields we first find the one-loop correction to the masses for these fields that are massless at tree-level at finite temperature. These masses are then used to modify the propagators for the ``massless" fields. The resulting one-loop Tachyon amplitude is thus infrared finite. The analytic expressions obtained here reduce to those for the intersecting $D1$ and $D2$ branes once the extra transverse momenta/momentum is set to zero. The final evaluation of the temperature corrected masses for the ``massless" as well as the Tachyon field could not however be performed analytically. We had to resort to numerics due to the complicated nature of the expressions involved.

This paper is organized as follows. In the following section (\ref{spectrum}) we analyze the quadratic part of the Yang-Mills Lagrangian with the background $\Phi_1^3=qx$. We derive the various wave-functions both for the bosons and fermions and then write out the the Lagrangian in terms of momentum modes. The propagators for the various fields can then be read out.  In section (\ref{tachyononeloop}) we compute the one-loop two-point Tachyon amplitude. In the subsection (\ref{tachyonuv}) we analytically show the finiteness of the Tachyon amplitude for large $n$ (ultraviolet) and for all values of the momenta ${\bf k}$. The one loop amplitude for the massless fields namely $\Phi_I^3 (I=1,..,6)$ and $A_{\mu}^3 (\mu=1,2,3)$ have been worked out in section (\ref{masslessoneloop}). These have been done in subsections (\ref{phi13oneloop}), (\ref{phii3oneloop}) for fields $\Phi_1^3$ and $\Phi_2^3$ respectively. The contributions from the bosons and the fermions for fields $A_{\mu}^3$ have been worked out separately in subsections (\ref{a13bosons}), (\ref{a233bosons}), (\ref{a13fermions}) and (\ref{a233fermions}). In section (\ref{numerics}) we show the numerical results for the computation of all the two point amplitudes including transition temperature at which the Tachyon becomes massless. We conclude the paper with some outlook in section (\ref{conclusion}). Various details have been worked out in the appendices. We summarize our notations in appendix \ref{notations}. We list the various parameters, variables used in the paper in Table \ref{t1} of this appendix. We review the dimensional reduction of $D=10$ super Yang-Mills to $D=4$ in appendix \ref{reduction}. The propagators and vertices for the Tachyon and various massless fields have been worked out in appendices \ref{tachyonvertices} and \ref{masslessvertices} respectively. In appendix \ref{amplitudem} we give details of the one-loop tachyon amplitude incorporating the corrected propagators for the fields with tree-level mass zero.

\section{Tree-level spectrum}\label{spectrum}

\subsection{Bosons}

In the Yang-Mills approximation the intersecting brane configuration with one angle is obtained by putting the background value of $\Phi_1^3$ equal to
$qx$. In this background we first write down the quadratic part of the bosonic action in the temporal gauge $A^a_0=0$. Various fields decouple at this quadratic level. We first write down below the various decoupled parts of the resulting from the action (\ref{actionboson}),

\beqa\label{L1}
{\cal L}(A^2_{\mu},\Phi_1^1)=-\f{1}{4}(\partial_{\mu}A^2_{\nu}-\partial_{\nu}A^2_{\mu})^2-\f{1}{2}(\partial_{\mu}\Phi_1^1)^2-\f{1}{2}q^2x^2A^2_{\mu}A^{2\mu}-qA^2_{x}\Phi^1_1+qx(\partial_{\mu}\Phi_1^1)A^{2\mu}
\eeqa

\beqa\label{L2}
{\cal L}(A^1_{\mu},\Phi_2^1)=-\f{1}{4}(\partial_{\mu}A^1_{\nu}-\partial_{\nu}A^1_{\mu})^2-\f{1}{2}(\partial_{\mu}\Phi_1^2)^2-\f{1}{2}q^2x^2A^1_{\mu}A^{1\mu}+qA^1_{x}\Phi^2_1-qx(\partial_{\mu}\Phi_1^2)A^{1\mu}
\eeqa

\beqa\label{L3}
{\cal L}(A^3_{\mu},\Phi_I,\tilde{\Phi}_J)&=&-\f{1}{4}(\partial_{\mu}A^3_{\nu}-\partial_{\nu}A^3_{\mu})^2-\f{1}{2}(\partial_{\mu}\Phi_I^a)^2-\f{1}{2}(\partial_{\mu}\Phi_1^3)^2-\f{1}{2}(\partial_{\mu}\tilde{\Phi}_J^b)^2-\f{1}{2}q^2x^2(\Phi_I^a)^2-\f{1}{2}q^2x^2(\tilde{\Phi}_J^b)^2\non
&~&(I=2,3; a=1,2)~~(J=1,2,3; b=1,2,3)
\eeqa

Now defining

\beqa
\xi=\left(\begin{array}{c}\Phi^1_1\\A^2_1\\A^2_2\\A^2_3\end{array}\right)~;~\zeta=\left(\begin{array}{c}\Phi^1_1\\A^2_1\end{array}\right)~~~;~~~\xi^{'}=\left(\begin{array}{c}\Phi^2_1\\A^1_1\\A^1_2\\A^1_3\end{array}\right)~;~\zeta^{'}=\left(\begin{array}{c}\Phi^2_1\\A^1_1\end{array}\right)
\eeqa

the quadratic bosonic part of the action can be written as,

\beqa\label{actionb}
S_b=\int d^4 z ~\left[\f{1}{2}\xi^T{\cal O}_B\xi+\f{1}{2}\xi^{'T}{\cal O}^{'}_B\xi^{'}+{\cal L}(A^3_{\mu},\Phi_I,\tilde{\Phi}_J)\right]
\eeqa

with 

\beqa
{\cal O}_B=\left(\begin{array}{cc}{\cal O}^{11}_B&{\cal O}^{12}_B\\{\cal O}^{21}_B&{\cal O}^{22}_B\end{array}\right)
\eeqa

where

\beqa
{\cal O}^{11}_B=\left(\begin{array}{cc}-\partial^2_0+\partial^2_1+\partial^2_2+\partial^2_3&-2q-qx\partial_1\\-q+qx\partial_1&-\partial_0^2+\partial_2^2+\partial_3^2-q^2x^2\end{array}\right)~~~{\cal O}^{12}_B={\cal O}^{21}_B=\left(\begin{array}{cc}qx\partial_2&-\partial_2\partial_1\\qx\partial_3&-\partial_3\partial_1\end{array}\right)
\eeqa

\beqa
{\cal O}^{22}_B=\left(\begin{array}{cc}-\partial^2_0+\partial^2_1+\partial^2_3-q^2x^2&-\partial_2\partial_3\\-\partial_2\partial_3&-\partial^2_0+\partial^2_1+\partial^2_2-q^2x^2\end{array}\right)
\eeqa

${\cal O}^{'}_B$ can be obtained from ${\cal O}_B$ by replacing $q$ in ${\cal O}_B$ by $-q$. We have identified $z^1=x$. In the following we shall identify $z^0=-i\tau$ and $(z^2,z^3)\equiv {\bf y}$.

The eigenfunctions of ${\cal O}^{11}_B$ have bee worked out in \cite{1} these are

\beqa\label{cfunctions}
\zeta_n(x)=\left(\begin{array}{c}
\phi_n(x)\\
-A_n(x)
\end{array}\right) ~~~~\tilde{\zeta}_n(x)=\left(\begin{array}{c}
\tilde{\phi}_n(x)\\
-\tilde{A}_n(x)
\end{array}\right)
\eeqa

where 

\begin{gather}
\label{bosegnfunc}
A_n(x) =   {\cal N}(n)e^{- q x^2/2} \left(H_n (\sqrt{q} x) + 2 n H_{n-2} (\sqrt{q} x) \right) \\
\phi_n(x) = {\cal N}(n)e^{- q x^2/2} \left(H_n (\sqrt{q} x) - 2 n H_{n-2} (\sqrt{q} x) \right)
\end{gather} 

and

\begin{gather}
\label{zerobosegnfunc}
\tilde{A}_n(x) =  \tilde{{\cal N}}(n) e^{- q x^2/2} \left(H_n (\sqrt{q} x) - 2 (n-1) H_{n-2} (\sqrt{q} x) \right) \\
\tilde{\phi}_n(x) = \tilde{{\cal N}}(n) e^{- q x^2/2} \left(H_n (\sqrt{q} x) + 2 (n-1) H_{n-2} (\sqrt{q} x) \right)
\end{gather}

The normalizations are given by ${\cal N}(n)=1/\sqrt{\sqrt{\pi} 2^n (4 n^2-2n) (n-2)!}$ and $\tilde{{\cal N}}(n)=1/\sqrt{\sqrt{\pi} 2^n (4 n-2)(n-1)!}$. The eigenvalues corresponding to $\zeta_n(x)$ are $(2n-1)q$ and those corresponding to $\tilde{\zeta}_n(x)$ are all zero. Thus the spectrum in the latter case is completely degenerate. In the non-zero eigenvalue sector we do not have normalizable eigenfunctions corresponding to $n=1$. However unlike this sector, in the zero eigenvalue sector we have normalizable eigenfunction for $n=1$, which is simply $H_1(\sqrt{q} x)$ but there is 
no normalizable eigenfunctions for $n=0$ in this sector.

$\zeta_n(x)$ and $\tilde{\zeta}_n(x)$ satisfies the orthogonality conditions

\beqa\label{ortho1}
\sqrt{q}\int dx \zeta^{\dagger}_n(x)\zeta_{n^{'}}(x)=\delta_{n,n^{'}} ~~~~\sqrt{q}\int dx \tilde{\zeta}^{\dagger}_n(x)\tilde{\zeta}_{n^{'}}(x)=\delta_{n,n^{'}}
\eeqa

They also satisfy,

\beqa\label{ortho2}
\sqrt{q}\int dx \zeta^{\dagger}_n(x)\tilde{\zeta}_{n^{'}}(x)=0 \mbox{~~ for all~~$n$~and~$n^{'}$~~}.
\eeqa

Similarly the eigenfunctions of the operator ${\cal O}^{'}_B$ are simply $\zeta^{'}=(\phi_n(x), A_n(x))$, 
and $\tilde{\zeta}^{'}=(\tilde{\phi}_n(x), \tilde{A}_n(x))$ with eigenvalues $(2n-1)q$ and $0$ respectively.
There is thus a two fold degeneracy for this spectrum of the theory. 
 
We now write down the mode expansion for $\zeta$ 

\beqa\label{zeta1}
\zeta(\tau, x, {\bf y})=N^{1/2}\int\f{d^2{\bf k}}{(2\pi\sqrt{q})^2}\sum_{m,n}\left[C(m,n,{\bf k})\zeta_n(x)+
\tilde{A}^2_1(m,n,{\bf k})\tilde{\zeta}_n(x)\right]e^{-i(\o_m\tau+{\bf k.y})}
\eeqa

and similarly for $\zeta^{'}$

\beqa
\zeta^{'}(\tau, x, {\bf y})=N^{1/2}\int\f{d^2{\bf k}}{(2\pi\sqrt{q})^2}\sum_{m,n}\left[C^{'}(m,n,{\bf k})\zeta^{'}_n(x)+
\tilde{A}^1_1(m,n,{\bf k})\tilde{\zeta}^{'}_n(x)\right]e^{-i(\o_m\tau+{\bf k.y})}
\eeqa

The operator ${\cal O}^{22}_B$ suggests the following mode expansions

\beqa
A_{2,3}^2=N^{1/2}\int\f{d^2{\bf k}}{(2\pi\sqrt{q})^2}\sum_{m,n}\tilde{A}_{2,3}^2(m,n,{\bf k}){\cal N}^{'}(n)e^{-qx^2/2}H_n(\sqrt{q}x)e^{-i(\o_m\tau+{\bf k.y})}
\eeqa

The normalization $\mathcal{N}^{'}(n)=1/\sqrt{\sqrt{\pi}2^n n!}$. ${\cal N}^{'}(n)e^{-qx^2/2}H_n(\sqrt{q}x)$ are eigenfunctions of the operator $(\partial_x^2-q^2x^2)$ with  corresponding eigenvalues are $\gamma_n=(2n+1)q$.

Note that 

\beqa
{\cal O}^{21}_B \zeta_n=0~~~ \mbox{and}~~~ {\cal O}^{21}_B \tilde{\zeta}_n=2(2n-1)\tilde{{\cal N}}(n)\sqrt{q}e^{-qx^2/2}H_{n-1}(\sqrt{q}x)\left(\begin{array}{c}\partial_2\\\partial_3\end{array}\right)
\eeqa

Thus,

\beqa
{\cal O}^{21}_B \zeta=N^{1/2}\int\f{d^2{\bf k}}{(2\pi\sqrt{q})^2}\sum_{m,n}\tilde{A}^2_1(m,n,{\bf k})\left[2(2n-1)\tilde{{\cal N}}(n)\sqrt{q}e^{-qx^2/2}H_{n-1}(\sqrt{q}x)\left(\begin{array}{c}\partial_2\\\partial_3\end{array}\right)e^{-i(\o_m\t+{\bf k}.{\bf y})}\right]
\eeqa

To show the effect of the above let us consider one of the cross-terms coming from the first term in the action (\ref{actionb}),

\beqa\label{cross}
\f{1}{qg^2}\int\f{d^2{\bf k}}{(2\pi\sqrt{q})^2}\sum_{m,n}\tilde{A}^2_1(m,n,{\bf k})\left[ (-ik_2)\sqrt{(2n-1)q}\right]\tilde{A}^2_2(-m,n-1,-{\bf k})
\eeqa

with appropriate change of limits of $n$ we can write (\ref{cross}) as 

\beqa\label{cross1}
\f{1}{qg^2}\int\f{d^2{\bf k}}{(2\pi\sqrt{q})^2}\sum_{m,n}(-i\tilde{A}^2_1(m,n+1,{\bf k})\left[ (k_2)(\sqrt{\gamma_n})\right]\tilde{A}^2_2(-m,n,-{\bf k})
\eeqa

Redefining $-i\tilde{A}^2_1(m,n+1,{\bf k})\equiv \tilde{A}^2_1(m,n,{\bf k})$ we can now write down the full action as

\beqa
-\f{1}{2qg^2}\int\f{d^2{\bf k}}{(2\pi\sqrt{q})^2}\sum_{m=-\infty,n=0}^{\infty,\infty}\left[\tilde{A}_i^2(m,n,{\bf k})\left(k^2\delta^{ij}-k^ik^j\right)\tilde{A}_j^2(-m,n,-{\bf k}) +|C(m,n,{\bf k})|^2\left(\o_m^2+\lambda_n+{\bf k}^2\right)\right]\non
\eeqa

where $(i,j=1,2,3)$, $k^2=(\o_m^2+\gamma_n+{\bf k}^2)$, $C(-m,n,-{\bf k})=C^{*}(m,n,{\bf k})$ and $k_x=\sqrt{\gamma_n}$

Using the same analysis as above the second term of the action (\ref{actionb}) in terms of momentum modes is,

\beqa
-\f{1}{2qg^2}\int\f{d^2{\bf k}}{(2\pi\sqrt{q})^2}\sum_{m=-\infty,n=0}^{\infty,\infty}\left[\tilde{A}_i^1(m,n,{\bf k})\left(k^2\delta^{ij}-k^ik^j\right)\tilde{A}_j^1(-m,n,-{\bf k}) +|C^{'}(m,n,{\bf k})|^2\left(\o_m^2+\lambda_n+{\bf k}^2\right)\right]\non
\eeqa

where $(i,j=1,2,3)$, $k^2=(\o_m^2+\gamma_n+{\bf k}^2)$, $C^{'}(-m,n,-{\bf k})=C^{'*}(m,n,{\bf k})$ and $k_x=\sqrt{\gamma_n}$.

${\cal L}(A^3_{\mu},\Phi_I,\tilde{\Phi}_J)$ given by (\ref{L3}) yields decoupled eigenvalue equations for the scalar fields. The scalar fields with gauge components $(a,b=1,2)$ can be expanded using the basis of harmonic oscillator wavefunctions, $\mathcal{N}^{'}(n)e^{-qx^2/2} H_n(\sqrt{q} x)$. Thus for example,

\beqa
\Phi_{I}^{1,2}=N^{1/2}\int\f{d^2{\bf k}}{(2\pi\sqrt{q})^2}\sum_{m,n}\Phi_{I}^{1,2}(m,n,{\bf k}){\cal N}^{'}(n)e^{-qx^2/2}H_n(\sqrt{q}x)e^{-i(\o_m\tau+{\bf k.y})}
\eeqa

The scalar fields with the gauge component $(a,b=3)$ and the gauge fields $A^3_{i}$ can be expanded using the basis for plane wave as

\beqa
\Phi_{J}^{3}&=&N^{1/2}\int\f{dk_xd^2{\bf k}}{(2\pi\sqrt{q})^3}\sum_{m}\Phi_{I}^{3}(m,k)e^{-i(\o_m\tau+k_x x +{\bf k.y})}~~(J=1,2,3)\\
A_{i}^{3}&=&N^{1/2}\int\f{dk_xd^2{\bf k}}{(2\pi\sqrt{q})^3}\sum_{m}A_{i}^{3}(m,k)e^{-i(\o_m\tau+k_x x +{\bf k.y})}~~(i=1,2,3)
\eeqa

We can thus write down the third term in the action (\ref{actionb}) in terms of the momentum modes as,

\beqa\label{phiAaction}
&&-\f{1}{2qg^2}\int\f{d^2{\bf k}}{(2\pi\sqrt{q})^2}\sum_{m,n}\left[|\Phi^{a}_{I}(m,n,{\bf k})|^2\left(\o_m^2+\gamma_n+{\bf k}^2\right)+|\tilde{\Phi}^{b}_{J}(m,n,{\bf k})|^2\left(\o_m^2+\gamma_n+{\bf k}^2\right)\right]\\
&&-\f{1}{2qg^2}\int\f{dk_xd^2{\bf k}}{(2\pi\sqrt{q})^3}\sum_{m}\left[|\Phi_J^{3}(m,k)|^2k^2+|\tilde{\Phi}_J^{3}(m,k)|^2k^2+
\tilde{A}_i^3(m, k)\left(k^2\delta^{ij}-k^ik^j\right)\tilde{A}_j^3(-m,-k)\right]\non
\eeqa

Here $(I=2,3; a=1,2)$, $(J=1,2,3; b=1,2,)$, $(i,j=1,2,3)$ and $k^2=(\o_m^2+k_x^2+{\bf k}^2)$.

\subsection{Fermions}

The quadratic terms involving fermions in the action (\ref{actionfermion}) are given by

\beqa\label{Lf}
{\cal L}^{'}=-\f{i}{2}\left[\bar{\lambda}^a_k\gamma^{\mu}\partial_{\mu}\lambda_k^a+\Phi_1^3\left(\bar{\lambda}^1_k\alpha^1_{kl}\lambda^2_l-\bar{\lambda}^2_k\alpha^1_{kl}\lambda^1_l\right)\right]
\eeqa

The terms with gauge indices $1$ and $2$ decouple into four different sets. Since $\alpha^1$ is off-diagonal these sets are 
$(\lambda^1_1, \lambda^2_4), (\lambda^1_2, \lambda^2_3),(\lambda^1_3, \lambda^2_2),(\lambda^1_4, \lambda^2_1)$.

Defining,

\beqa
\chi_1=\left(\begin{array}{c}\lambda^1_1\\\lambda^2_4\end{array}\right) ~~~\chi_2=\left(\begin{array}{c}\lambda^1_2\\\lambda^2_3\end{array}\right)~~~\chi_3=\left(\begin{array}{c}\lambda^1_3\\\lambda^2_2\end{array}\right)~~~\chi_4=\left(\begin{array}{c}\lambda^1_4\\\lambda^2_1\end{array}\right)
\eeqa

The Lagrangian (\ref{Lf}) can thus be written as 
\beqa
{\cal L}^{'}=-\f{i}{2}\left[\bar{\chi}_1 {\cal O}_F\chi_1+\bar{\chi}_2 {\cal O}_F\chi_2+\bar{\chi}_3 \tilde{{\cal O}}_F\chi_3+\bar{\chi}_4 \tilde{{\cal O}}_F\chi_4+\bar{\lambda}^3_k\gamma^{\mu}\partial_{\mu}\lambda_k^3\right]
\eeqa

where the operators ${\cal O}_F$ and  $\tilde{{\cal O}}_F$  are given by,

\beqa
{\cal O}_F=\left(\begin{array}{cc}\gamma^{\mu}\partial_{\mu}&qx\\qx&\gamma^{\mu}\partial_{\mu}\end{array}\right) ~~~\tilde{{\cal O}}_F=\left(\begin{array}{cc}\gamma^{\mu}\partial_{\mu}&-qx\\-qx&\gamma^{\mu}\partial_{\mu}\end{array}\right)
\eeqa

We transform $\chi_i \rightarrow U \chi_i$ where $U=\left(\begin{array}{cc}1&0\\0&\gamma^1\end{array}\right)$. The operators ${\cal O}_F$ and  $\tilde{{\cal O}}_F$ thus transform into

\beqa
{\cal O}_F\rightarrow\gamma^{i}\partial_{i}\otimes \mathbb{I}_2+\gamma^{1}\otimes{\cal O}_F^x ~~~\mbox{with}~~~{\cal O}_F^x=\left(\begin{array}{cc}\partial_{x}&qx\\-qx&-\partial_{x}\end{array}\right)
\eeqa

\beqa
\tilde{{\cal O}}_F\rightarrow\gamma^{i}\partial_{i}\otimes \mathbb{I}_2+\gamma^{1}\otimes\tilde{{\cal O}}_F^x ~~~\mbox{with}~~~\tilde{{\cal O}}_F^x=\left(\begin{array}{cc}\partial_{x}&-qx\\qx&-\partial_{x}\end{array}\right)
\eeqa

where $i=0,2,3$. The eigenfunctions of the matrix operators have been obtained in \cite{1}. Adopting the same notation as in \cite{1} the eigenfunctions of ${\cal O}_F^x$ and $\tilde{{\cal O}}_F^x$ are

\beqa
\left(\begin{array}{c}L_n(x)\\R_n(x)\end{array}\right) ~~\mbox{and}~~ \left(\begin{array}{c}L_n(x)\\-R_n(x)\end{array}\right)
\eeqa

respectively. Here,
\begin{equation}
\label{fermsolnLR}
\begin{split}
&L_n(x) =  {\cal N}_F e^{- \frac{q x^2}{2}}\left(- \frac{i}{\sqrt{2n}} H_{n}(\sqrt{q} x) 
+  H_{n-1} (\sqrt{q} x)\right)\\
&R_n(x) = {\cal N}_F e^{- \frac{q x^2}{2}}\left(- \frac{i}{\sqrt{2n}} H_{n}(\sqrt{q} x) 
-  H_{n-1} (\sqrt{q} x)\right). 
\end{split}
\end{equation}

$H_n(\sqrt{q}x)$ are the Hermite Polynomials. The normalization ${\cal N}_F=\sqrt{\sqrt{\pi}2^{n+1} (n-1)!}$. The corresponding eigenvalues are $-i\sqrt{\lambda^{'}_n}=-i\sqrt{2nq}$. 
We now list some important relations satisfied by the eigenfunctions

\begin{gather}
\label{fermorthocon}
\sqrt{q}\int dx~\psi^{\dagger}_n(x) \psi_{n^{'}}(x) = \sqrt{q}\int dx \left(L^*_n(x) 
L_{n^{'}}(x) + R^*_n(x) R_{n^{'}}(x)\right) = \delta_{n,n^{'}}.\\
\label{fermreln1}
\sqrt{q}\int dx~L^*_n(x) L_{n^{'}}(x) = \sqrt{q}\int dx~R^*_n(x) R_{n^{'}}(x) = \hf \delta_{n,n^{'}}\\
\label{fermreln2}
\sqrt{q}\int dx~\psi^{T}_n(x) \psi_{n^{'}}(x)=\sqrt{q}\int dx \left(L_n(x) L_{n^{'}}(x) 
+ R_n(x) R_{n^{'}}(x)\right) = 0\\
\label{fermreln3}
\sqrt{q}\int dx~\psi^{\dagger}_n(x) \psi^{*}_{n^{'}}(x)=\sqrt{q}\int dx \left(L^{*}_n(x) L^{*}_{n^{'}}(x) 
+ R^{*}_n(x) R^{*}_{n^{'}}(x)\right) = 0    
\end{gather}

We can now write down the mode expansions for the fermions

\beqa
\chi_i(\tau,x,{\bf y})=N^{3/4}\sum_{n,m}\int\f{d^2{\bf k}}{(2\pi\sqrt{q})^2}\left[\left(\begin{array}{c}\theta_i(\o,n,{\bf k})L_n(x)\\\gamma^{1}\theta_i(\o,n,{\bf k})R_n(x)\end{array}\right)e^{-i(\o\tau+{\bf k.y})}+\left(\begin{array}{c}\theta^*_i(\o,n,{\bf k})L^*_n(x)\\\gamma^{1}\theta^*_i(\o,n,{\bf k})R^*_n(x)\end{array}\right)e^{i(\o\tau+{\bf k.y})}\right]\non
\eeqa

where $\theta_i$ are four component fermions. Further

\beqa
\lambda^3_l(\tau,x,{\bf y})=N^{3/4}\sum_{n,m}\int\f{dk_xd^2{\bf k}}{(2\pi\sqrt{q})^3}\lambda^3_l(\o,k_x,{\bf k})e^{-i(\o_m\tau+k_x x+{\bf k.y})}
\eeqa

The quadratic action in terms of the momentum modes is then

\beqa
S_f=\f{N^{1/2}}{qg^2}\left[\sum_{i,m,n}\int\f{d^2{\bf k}}{(2\pi\sqrt{q})^2}\bar{\theta}_i(m,n,{\bf k})i\slashed{P}_+\theta_i(m,n,{\bf k})+\f{1}{2}\sum_{l,m}\int\f{dk_xd^2{\bf k}}{(2\pi\sqrt{q})^3}\bar{\lambda}^3_l(m,k_x,{\bf k})i\slashed{K}_+\lambda^3_l(m,k_x,{\bf k})\right]
\eeqa

where $\slashed{P}_+=i\o_m\gamma^0+\sqrt{\lambda^{'}_n}\gamma^1+k_2\gamma^2+k_3\gamma^3$ and $\slashed{K}_+=i\o_m\gamma^0+k_x\gamma^1+k_2\gamma^2+k_3\gamma^3$.

\section{One-loop two-point tachyon amplitude}\label{tachyononeloop}

\begin{figure}[h]
\begin{center}
\begin{psfrags}
\psfrag{c1}[][]{$C$}
\psfrag{c2}[][]{$C$}
\psfrag{c3}[][]{$C/C^{'}$}
\psfrag{c4}[][]{$\tilde{C}/\tilde{C}^{'}$}
\psfrag{c5}[][]{$(\Phi^{1}_I,\tilde{\Phi}^{1}_I,A_i^{1})/(\Phi^{2}_I,\tilde{\Phi}^{2}_I,A_i^{2})$}
\psfrag{c6}[][]{$\Phi^3_I,\tilde{\Phi}^3_I,A_i^3$}
\psfrag{c7}[][]{$A^3_1$}
\psfrag{c8}[][]{$\Phi^3_1$}
\psfrag{v1}[][]{$V_1/V_1^{'}$}
\psfrag{v2}[][]{$\tilde{V}_1/\tilde{V}_1^{'}$}
\psfrag{v3}[][]{$V^{1}_2/V^{2}_2$}
\psfrag{v4}[][]{$V^{3}_2$}
\psfrag{v5}[][]{$V_3$}
\psfrag{v6}[][]{$V^{'}_3$}
\psfrag{1}[][]{$(I)$}
\psfrag{2}[][]{$(II)$}
\psfrag{3}[][]{$(III)$}
\psfrag{4}[][]{$(IV)$}
\psfrag{5}[][]{$(V)$}
\psfrag{6}[][]{$(VI)$}
\includegraphics[ width= 11cm,angle=0]{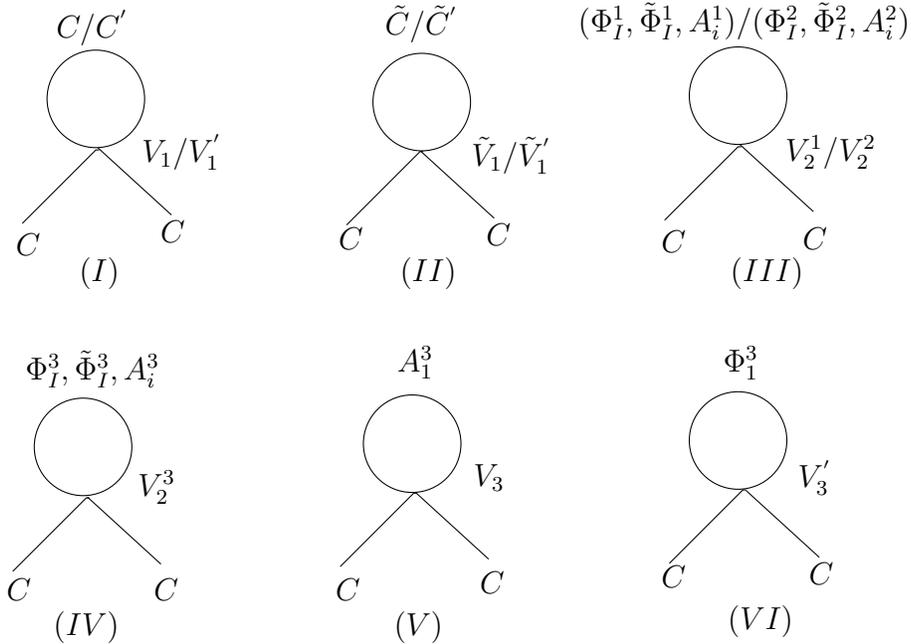}
\end{psfrags}
\caption{{Feynman diagrams contributing to $\Sigma^1_{C-C}$. The momenta on the external $C$ fields are $(\o_{m^{''}},k_x^{''},{\bf k}^{''})$ and  $({\o}_{\tilde{m}^{''}},\tilde{k}_x^{''},\tilde{{\bf k}}^{''})$. In the above diagrams $i=2,3$.}}
\label{bosfourfeyn}
\end{center}
\end{figure}

\begin{figure}[h]
\begin{center}
\begin{psfrags}
\psfrag{c1}[][]{$C$}
\psfrag{c2}[][]{$C$}
\psfrag{c3}[][]{$C^{'}$}
\psfrag{c3p}[][]{$A^3_i$}
\psfrag{c4}[][]{$\Phi_I^1/\tilde{\Phi}^1_I$}
\psfrag{c4p}[][]{$\Phi_I^3/\tilde{\Phi}_I^3$}
\psfrag{c5}[][]{$C$}
\psfrag{c5p}[][]{$\Phi^3_1$}
\psfrag{c6}[][]{$\tilde{A}_i^2$}
\psfrag{c6p}[][]{$\Phi^3_1$}
\psfrag{c7}[][]{$\tilde{A}_i^1$}
\psfrag{c7p}[][]{$A^3_j$}
\psfrag{7}[][]{$(VII)$}
\psfrag{8}[][]{$(VIII)$}
\psfrag{9}[][]{$(IX)$}
\psfrag{10}[][]{$(X)$}
\psfrag{11}[][]{$(XI)$}
\psfrag{v1}[][]{$V_4^i$}
\psfrag{v2}[][]{$V^{j*}_4$}
\psfrag{v3}[][]{$V_5$}
\psfrag{v4}[][]{$V^{*}_5$}
\psfrag{v5}[][]{$V^1_5$}
\psfrag{v6}[][]{$V^{1*}_5$}
\psfrag{v7}[][]{$\tilde{V}^{i}_5$}
\psfrag{v8}[][]{$\tilde{V}^{j*}_5$}
\psfrag{v9}[][]{$\tilde{V}^{ij}_5$}
\psfrag{v10}[][]{$\tilde{V}^{i^{'}j^{'}*}_5$}
\includegraphics[ width= 12cm,angle=0]{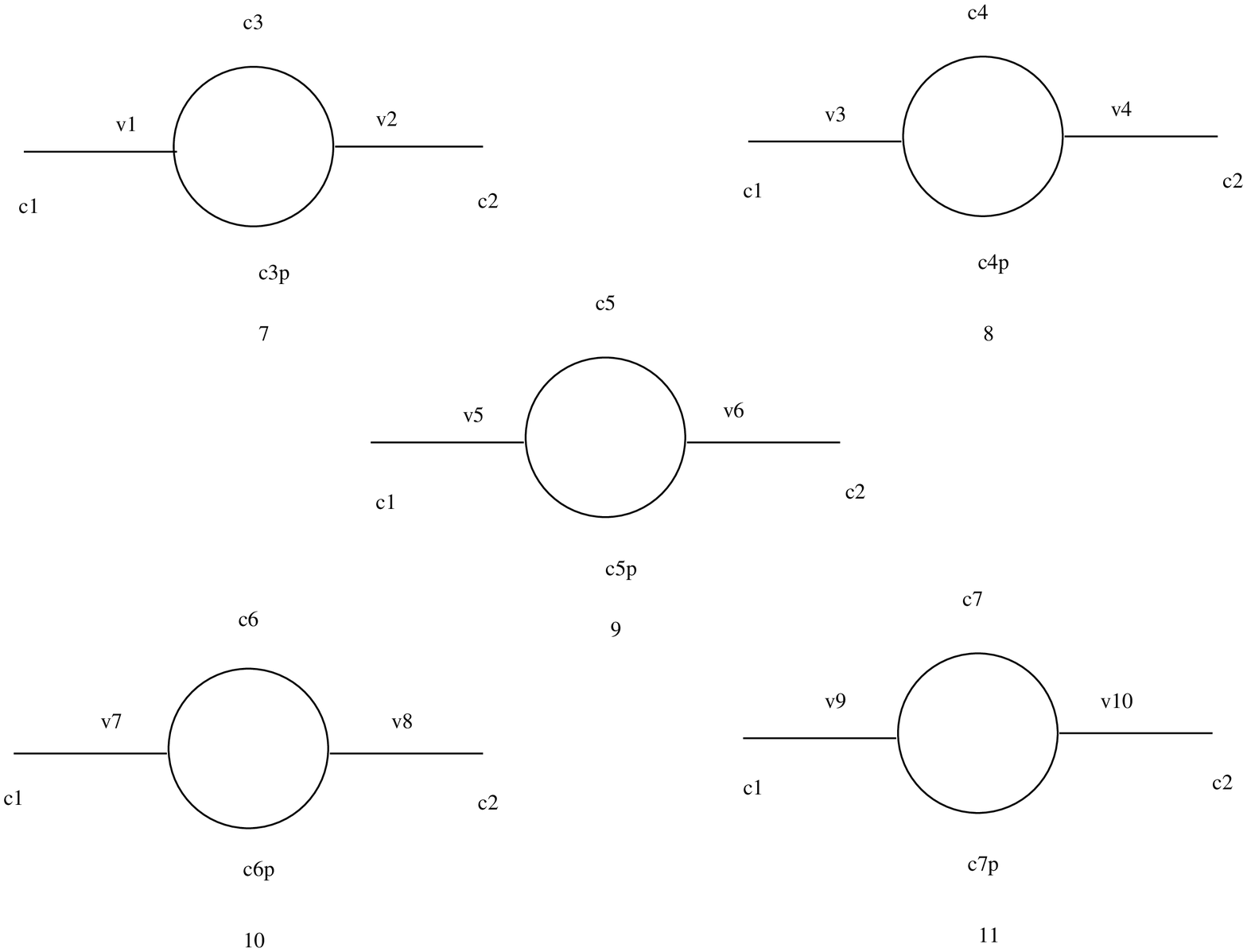}
\end{psfrags}
\caption{Feynman diagrams contributing to $\Sigma^2_{C-C}$. The momenta on the external $C$ fields are $(\o_{m^{''}},k_x^{''},{\bf k}^{''})$ and  $({\o}_{\tilde{m}^{''}},\tilde{k}_x^{''},\tilde{{\bf k}}^{''})$.}
\label{bosthreefeyn}
\end{center}
\end{figure}

In this section we compute the two-point amplitude for the $C(m^{''},n^{''},{\bf k}^{''})$ fields. The required propagators and the vertices are worked out in the appendix (\ref{tachyonvertices}). Using the four point vertices we first write down the contributions from bosons in the loop involving four-point and three-point vertices separately as shown in Figures \ref{bosfourfeyn} and \ref{bosthreefeyn}. This is given by

\beqa
\Sigma^1_{C-C}&=& \hf  N \sum_m \int \f{d^2 {\bf k}}{(2\pi\sqrt{q})^2} \left[\sum_n 
\left\{\f {(F_1+F_1^{'})(n,n,n^{''},\tilde{n}^{''})}{\o_m^2 + \lambda_n+|{\bf k}|^2} + \tilde{F}_1(n,n,n^{''},\tilde{n}^{''})P^{22}_{11}+\tilde{F}_1^{'}(n,n,n^{''},\tilde{n}^{''})P^{11}_{11}\right. \right.\nonumber\\
&+& \left.\left. F_2^{1}(n,n,n^{''},\tilde{n}^{''})\left(\f{5}{\o_m^2 + \gamma_n+|{\bf k}|^2}+P^{11}_{22}+P^{11}_{33}\right)
+F_2^{2}(n,n,n^{''},\tilde{n}^{''})\left(\f{5}{\o_m^2 + \gamma_n+|{\bf k}|^2}+P^{22}_{22}+P^{22}_{33}\right)
\right\}\right.\nonumber\\
&+& \left. \int \f{dl}{(2\pi \sqrt{q})}\left\{F^{3}_2(l,-l,n^{''},\tilde{n}^{''})\left(\f{5}{\o_m^2 + l^2+|{\bf k}|^2}+P^{33}_{22}+P^{33}_{33}\right)
+\f{F^{'}_3(l,-l,n^{''},\tilde{n}^{''})}{\o_m^2+l^2+|{\bf k}|^2}
+ F_3(l,-l,n^{''},\tilde{n}^{''})P^{33}_{11}\right\}\right] \non &\times& \delta_{m^{''}+\tilde{m}^{''}}(2\pi)^2\delta^2({\bf k}^{''}+\tilde{\bf k}^{''})
\eeqa

where the propagators $P^{ab}_{ij}$ are given by (\ref{proppab}) and (\ref{propp33}). Similarly the three point vertices with bosons give the following contribution

\beqa
\Sigma^2_{C-C}&=&-\hf qN\sum_{m,n}\int \f{dl}{(2\pi \sqrt{q})}\f{d^2 {\bf k}}{(2\pi\sqrt{q})^2}\left[\sum_{i,j=1}^3\f{F^i_4(n,l,n^{''})P^{33}_{ij}F^j_4(n,-l,\tilde{n}^{''})}{\o_m^2 + \lambda_n+|{\bf k}|^2}+\f{F_5(n,l,n^{''})F_5(n,-l,\tilde{n}^{''})}{(\o_m^2 +\gamma_n+|{\bf k}|^2)(\o_m^2 +l^2+|{\bf k}|^2)}\right.\non
&+&\left.\f{F^{1}_5(n,l,n^{''})F^{1}_5(n,-l,\tilde{n}^{''})}{(\o_m^2 +\lambda_n+|{\bf k}|^2)(\o_m^2 +l^2+|{\bf k}|^2)}+
\sum_{i,j=1}^3\f{\tilde{F}^i_5(n,l,n^{''})P^{22}_{ij}\tilde{F}^j_5(n,-l,\tilde{n}^{''})}{\o_m^2 +l^2+|{\bf k}|^2}\right.\non
&+&\left.\sum_{i,i^{'},j,j^{'}=1}^3\tilde{F}^{ij}_5(n,l,n^{''})P^{11}_{ii^{'}}P^{33}_{jj^{'}}\tilde{F}_{5}^{i^{'}j^{'}}(n,-l,\tilde{n}^{''})\right]
 \delta_{m^{''}+\tilde{m}^{''}}(2\pi)^2\delta^2({\bf k}^{''}+\tilde{\bf k}^{''})
\eeqa

We are interested in computing the two point amplitude for zero external momenta. In this case the various functions involved in the computation of the above amplitude take the following form

\begin{eqnarray}
&&F_1(0,0,n,n)+F^{'}_1(0,0,n,n)=\f{(\mathcal{N}(n))^2}{2 \sqrt{\pi}}\sqrt{q}\int^\infty_{-\infty} dx~e^{-2 qx^2}
\left(6 (H_n(\sqrt{q}x))^2- 8 n^2 (H_{n-2}(\sqrt{q}x))^2\right)\label{f1f1p}\\
&&\tilde{F}_1(0,0,n,n)+\tilde{F}^{'}_1(0,0,n,n)=\f{(\tilde{\mathcal{N}}(n+1))^2}{2 \sqrt{\pi}}\sqrt{q}\int^\infty_{-\infty} dx~e^{-2 qx^2}
\left(6 (H_{n+1}(\sqrt{q}x))^2- 8 n^2 (H_{n-1}(\sqrt{q}x))^2\right)\label{ft1ft1p}\\
&&F^1_2(0,0,n,n)+F^2_2(0,0,n,n)=\f{1}{2^{n} \sqrt{\pi} n!}\sqrt{q} \int^\infty_{-\infty} dx~e^{-2qx^2} 
(H_n(\sqrt{q} x))^2\label{f2f2p}\\ 
&&F_2^3(0,0,l,-l)=1~~~;~~~F_3(0,0,l,-l)=F_3^{'}(0,0,l,-l)=\f{1}{2}\nonumber\\
&&F_4^1(0,n,l)=-(-i)^{n-1} \f{ 4n e^{-\f{\hat{l}^2}{4}} \hat{l}^{n-1}}{\sqrt{2^{n+1} 
(4n^2-2n) (n-2)!}}~;~F_4^{i}(0,n,l)= \f{ -(-i)^{n+1}2\hat{k}_i e^{-\f{\hat{l}^2}{4}} \hat{l}^{n}}{\sqrt{2^{n+1} 
(4n^2-2n) (n-2)!}}=\hat{k}_i\tilde{{\cal F}}_4(0,n,l)\non
&&F_5(0,n,l)= -2(-i)^{(n+1)}e^{-\f{\hat{l}^2}{4}}\f{\hat{l}^{(n+1)}}{\sqrt{2^{n+1} n!}}
~~~;~~~F^{1}_5(0,n,l)= (-i)^{(n-1)}e^{-\f{\hat{l}^2}{4}}\f{2 \hat{l}^{(n+1)} + 4 n  \hat{l}^{(n-1)}}
{\sqrt{2^{n+1}  (4 n^2 - 2 n) (n-2)!}}\non
&&\tilde{F}^{1}_5(0,n,l)=-2(-i)^{(n+1)}e^{-\f{\hat{l}^2}{4}}\f{\hat{l}^{(n+2)} + \hat{l}^{n}}
{\sqrt{2^{n+2}  (4n + 2) n!}}~~~;~~~\tilde{F}^{i}_5(0,n,l)=(-i)^{n+1}\hat{k}^{'}_ie^{-\f{\hat{l}^2}{4}}\f{\hat{l}^n}{\sqrt{2^{n+1}n!}}=\hat{k}^{'}_i\tilde{{\cal F}}_5^0(0,n,l)\non
&&\tilde{F}^{11}_5(0,n,l)=-2(-i)^{(n+1)}e^{-\f{\hat{l}^2}{4}}\f{ \hat{l}^{n}}
{\sqrt{2^{n+2}  (4n + 2) n!}}~~~;~~~\tilde{F}^{i1}_5(0,n,l)=(-i)^{n+1}\hat{k}^{'}_ie^{-\f{\hat{l}^2}{4}}\f{\hat{l}^n}{\sqrt{2^{n+1}n!}}=\hat{k}^{'}_i{\cal \tilde{F}}^{1}_5(0,n,l)\non
&&\tilde{F}^{1i}_5(0,n,l)= 2 \f{ (-i)^{n+1} \hat{k}_i e^{-\f{\hat{l}^2}{4}} \hat{l}^{n+1}}{\sqrt{2^{n+2} 
(4n+2) n!}}=\hat{k}_i{\cal \tilde{F}}^{1'}_5(0,n,l)
~~~;~~~\tilde{F}^{ij}_5(0,n,l)=-2(-i)^{(n+1)}e^{-\f{\hat{l}^2}{4}}\f{\hat{l}^{(n+1)}}{\sqrt{2^{n+1} n!}}\delta^{ij}=\delta^{ij}{\cal \tilde{F}}_5(0,n,l)\non
\end{eqnarray}

In the above expressions $(i,j)=2,3$ and $\hat{k}_i=k_i/\sqrt{q}$, $\hat{l}=l/\sqrt{q}$.
\\

Using the above forms of the vertices at zero external momenta, we first write down the contributions from the four point vertices. This is given by

\beqa
\Sigma^1_{C-C}=(I)+(II)+(III)+(IV)+(V)+(VI).
\eeqa

Defining the sum over bosonic frequencies as

\beqa\label{deff}
f(\lambda_n,\beta)=N\sum_m \f{1}{\o_m^2+\lambda_n+|{\bf k}|^2}=\f{\sqrt{q}}{\sqrt{\lambda_n+|{\bf k}|^2}}\left(\hf+\f{1}{e^{\beta\sqrt{\lambda_n+|{\bf k}|^2}}-1}\right)
\eeqa

the various contributions as shown in Figure \ref{bosfourfeyn} are given as follows

\beqa
(I)&=&\hf N\sum_{m,n}\int \f{d^2 {\bf k}}{(2\pi\sqrt{q})^2} \left(F_1(0,0,n,n)+F^{'}_1(0,0,n,n)\right)\f{1}{\o_m^2+\lambda_n+|{\bf k}|^2}\non
&=&\hf \sum_{n=2}^{\infty}\int \f{d^2 {\bf k}}{(2\pi\sqrt{q})^2} \left(F_1(0,0,n,n)+F^{'}_1(0,0,n,n)\right)f(\lambda_n,\beta)
\eeqa

\beqa
(II)&=& \hf N\sum_{m,n}\int \f{d^2 {\bf k}}{(2\pi\sqrt{q})^2} \left(\tilde{F}_1(0,0,n,n)+\tilde{F}^{'}_1(0,0,n,n)\right)\f{\o_m^2+\gamma_n}{\o_m^2(\o_m^2+\gamma_n+|{\bf k}|^2)}\\
&=&\hf \sum_{n=0}^{\infty}\int \f{d^2 {\bf k}}{(2\pi\sqrt{q})^2} \left(\tilde{F}_1(0,0,n,n)+\tilde{F}^{'}_1(0,0,n,n)\right)\left[\f{|{\bf k}|^2}{\gamma_n+|{\bf k}|^2}f(\gamma_n,\beta)+\f{\gamma_n}{\gamma_n+|{\bf k}|^2}\sum_m\f{\beta\sqrt{q}}{(2\pi m)^2}\right]\nonumber
\eeqa

\beqa
(III)&=&\hf N\sum_{m,n}\int \f{d^2 {\bf k}}{(2\pi\sqrt{q})^2} \left(F_2^1(0,0,n,n)+F^2_2(0,0,n,n)\right)\left[\f{5}{\o_m^2+\gamma_n+|{\bf k}|^2)}+\f{2\o_m^2+|{\bf k}|^2}{\o_m^2(\o_m^2+\gamma_n+|{\bf k}|^2)}\right]\non
&=&\hf \sum_{n=0}^{\infty}\int \f{d^2 {\bf k}}{(2\pi\sqrt{q})^2} \left(F_2^1(0,0,n,n)+F^2_2(0,0,n,n)\right)\times\non
&&\left[(7)f(\gamma_n,\beta)
-\f{|{\bf k}|^2}{\gamma_n+|{\bf k}|^2}f(\gamma_n,\beta)+\f{|{\bf k}|^2}{\gamma_n+|{\bf k}|^2}\sum_m\f{\beta\sqrt{q}}{(2\pi m)^2}\right]
\eeqa

\beqa
(IV))&=&\hf N\sum_{m}\int \f{d l}{(2\pi\sqrt{q})}\int \f{d^2 {\bf k}}{(2\pi\sqrt{q})^2} \left(F_2^3(0,0,l,-l)\right)\left[\f{5}{(\o_m^2+l^2+|{\bf k}|^2)}+\f{2\o_m^2+|{\bf k}|^2}{\o_m^2(\o_m^2+l^2+|{\bf k}|^2)}\right]\non
&=&\hf \int \f{d l}{(2\pi\sqrt{q})} \int \f{d^2 {\bf k}}{(2\pi\sqrt{q})^2} \left(F_2^3(0,0,l,-l)\right)\times\non
&&\left[(7)f(l^2,\beta)
-\f{|{\bf k}|^2}{l^2+|{\bf k}|^2}f(l^2,\beta)+\f{|{\bf k}|^2}{l^2+|{\bf k}|^2}\sum_m\f{\beta\sqrt{q}}{(2\pi m)^2}\right]
\eeqa

\beqa
(V)&=&\hf N\sum_{m}\int \f{d l}{(2\pi\sqrt{q})}\int \f{d^2 {\bf k}}{(2\pi\sqrt{q})^2} \left(F_3^{'}(0,0,l,-l)\right)\f{\o_m^2+l^2}{\o_m^2(\o_m^2+l^2+|{\bf k}|^2)}\non
&=&\hf \int \f{d l}{(2\pi\sqrt{q})} \int \f{d^2 {\bf k}}{(2\pi\sqrt{q})^2} \left(F_3^{'}(0,0,l,-l)\right)\left[\f{|{\bf k}|^2}{l^2+|{\bf k}|^2}f(l^2,\beta)+\f{l^2}{l^2+|{\bf k}|^2}\sum_m\f{\beta\sqrt{q}}{(2\pi m)^2}\right]
\eeqa

\beqa
(VI)&=&\hf N\sum_{m}\int \f{d l}{(2\pi\sqrt{q})}\int \f{d^2 {\bf k}}{(2\pi\sqrt{q})^2} \left(F_3^{'}(0,0,l,-l)\right)\f{1}{\o_m^2+l^2+|{\bf k}|^2}\non
&=&\hf \int \f{d l}{(2\pi\sqrt{q})} \int \f{d^2 {\bf k}}{(2\pi\sqrt{q})^2} \left(F_3^{'}(0,0,l,-l)\right)f(l^2,\beta)
\eeqa

We now turn to the contribution to the two point amplitude from the three point vertices. The relevant Feynman diagrams are shown in Figure \ref{bosthreefeyn}. Thus

\beqa
\Sigma^2_{C-C}=(VII)+(VIII)+(IX)+(X)+(XI)
\eeqa

where,

\beqa
(VII)&=&-\hf N\sum_{m,n}\int \f{d l}{(2\pi\sqrt{q})} \int \f{d^2 {\bf k}}{(2\pi\sqrt{q})^2}\left[\left(q F_4^1(0,n,l)F_4^1(0,n,-l)\right)\left[\f{\o_m^2+l^2}{\o_m^2(\o_m^2+l^2+|{\bf k}|^2)(\o_m^2+\lambda_n+|{\bf k}|^2)}\right]\right.\non
&+&\left.\left(- \tilde{{\cal F}}_4(0,n,l)\tilde{{\cal F}}_4(0,n,-l)\right)\left[\f{|{\bf k}|^2(\o_m^2+|{\bf k}|^2)}{\o_m^2(\o_m^2+l^2+|{\bf k}|^2)(\o_m^2+\lambda_n+|{\bf k}|^2)}\right]\right.\non
&+&\left.\left(2 q \hat{l} F_4^1(0,n,l)\tilde{{\cal F}}_4(0,n,-l)\right)\left[\f{|{\bf k}|^2}{\o_m^2(\o_m^2+l^2+|{\bf k}|^2)(\o_m^2+\lambda_n+|{\bf k}|^2)}\right]\right]\non
&=&-\hf \sum_{n=2}^{\infty}\int \f{d l}{(2\pi\sqrt{q})} \int \f{d^2 {\bf k}}{(2\pi\sqrt{q})^2}\left[\left(q F_4^1(0,n,l)F_4^1(0,n,-l)\right)\left\{ \f{|{\bf k}|^2}{(l^2+|{\bf k}|^2)(l^2-\lambda_n)}\left(f(\lambda_n,\beta)-f(l^2,\beta)\right)\right.\right.\non
&+&\left.\left.\f{l^2}{(l^2+|{\bf k}|^2)(\lambda_n+|{\bf k}|^2)}\left(\sum_m\f{\beta\sqrt{q}}{(2\pi m)^2}-f(\lambda_n,\beta)\right) \right\}\right.\non &+&\left. \left(- \tilde{{\cal F}}_4(0,n,l)\tilde{{\cal F}}_4(0,n,-l)\right)\f{|{\bf k}|^2}{(l^2-\lambda_n)}\left\{\left(\f{\lambda_n}{\lambda_n+|{\bf k}|^2}f(\lambda_n,\beta)-\f{l^2}{l^2+|{\bf k}|^2}f(l^2,\beta)\right)\right.\right.\non
&+& \left.\left.\left(\f{|{\bf k}|^2}{\lambda_n+|{\bf k}|^2}-\f{|{\bf k}|^2}{l^2+|{\bf k}|^2}\right)
\sum_m\f{\beta\sqrt{q}}{(2\pi m)^2}\right\}\right.\non
&+&\left. \left(2 q \hat{l} F_4^1(0,n,l)\tilde{{\cal F}}_4(0,n,-l)\right)\f{|{\bf k}|^2}{(l^2-\lambda_n)}\left\{\f{1}{l^2+|{\bf k}|^2}f(l^2,\beta)-\f{1}{\lambda_n+|{\bf k}|^2}f(\lambda_n,\beta)\right.\right.\non
&+& \left.\left.\left(\f{1}{\lambda_n+|{\bf k}|^2}-\f{1}{l^2+|{\bf k}|^2}\right)
\sum_m\f{\beta\sqrt{q}}{(2\pi m)^2}\right\}\right] 
\eeqa

\beqa
(VIII)&=&-\hf N \sum_{m,n}\int \f{d l}{(2\pi\sqrt{q})} \int \f{d^2 {\bf k}}{(2\pi\sqrt{q})^2}(5)\times\left[\left(q F_5(0,n,l)F_5(0,n,-l)\right)
\f{1}{(\o_m^2+\gamma_n+|{\bf k}|^2)(\o_m^2+l^2+|{\bf k}|^2)}\right]\non
&=&-\hf \sum_{n=0}^{\infty}\int \f{d l}{(2\pi\sqrt{q})} \int \f{d^2 {\bf k}}{(2\pi\sqrt{q})^2}(5)\times\left[\left(q F_5(0,n,l)F_5(0,n,-l)\right)
\f{1}{l^2-\gamma_n}\left\{f(\gamma_n,\beta)-f(l^2,\beta)\right\}\right]
\eeqa

\beqa
(IX)&=&-\hf N \sum_{m,n}\int \f{d l}{(2\pi\sqrt{q})} \int \f{d^2 {\bf k}}{(2\pi\sqrt{q})^2}\left[\left(q F^1_5(0,n,l)F^1_5(0,n,-l)\right)
\f{1}{(\o_m^2+\lambda_n+|{\bf k}|^2)(\o_m^2+l^2+|{\bf k}|^2)}\right]\non
&=&-\hf \sum_{n=2}^{\infty}\int \f{d l}{(2\pi\sqrt{q})} \int \f{d^2 {\bf k}}{(2\pi\sqrt{q})^2}\left[\left(q F_5^1(0,n,l)F_5^1(0,n,-l)\right)
\f{1}{l^2-\lambda_n}\left\{f(\lambda_n,\beta)-f(l^2,\beta)\right\}\right]
\eeqa

\beqa
(X)&=&-\hf N\sum_{m,n}\int \f{d l}{(2\pi\sqrt{q})} \int \f{d^2 {\bf k}}{(2\pi\sqrt{q})^2}\left[\left(-q \tilde{F}_5^1(0,n,l)\tilde{F}_5^1(0,n,-l)\right)\left[\f{\o_m^2+\gamma_n}{\o_m^2(\o_m^2+l^2+|{\bf k}|^2)(\o_m^2+\gamma_n+|{\bf k}|^2)}\right]\right.\non
&+&\left.\left(- \tilde{{\cal F}}_5^0(0,n,l)\tilde{{\cal F}}_5^0(0,n,-l)\right)\left[\f{|{\bf k}|^2(\o_m^2+|{\bf k}|^2)}{\o_m^2(\o_m^2+l^2+|{\bf k}|^2)(\o_m^2+\gamma_n+|{\bf k}|^2)}\right]\right.\non
&+&\left.\left(2 q \sqrt{2n+1} \tilde{F}_5^1(0,n,l)\tilde{{\cal F}}_5^0(0,n,-l)\right)\left[\f{|{\bf k}|^2}{\o_m^2(\o_m^2+l^2+|{\bf k}|^2)(\o_m^2+\gamma_n+|{\bf k}|^2)}\right]\right]\non
&=&-\hf \sum_{n=0}^{\infty}\int \f{d l}{(2\pi\sqrt{q})} \int \f{d^2 {\bf k}}{(2\pi\sqrt{q})^2}\left[\left(-q \tilde{F}_5^1(0,n,l)\tilde{F}_5^1(0,n,-l)\right)\left\{ \f{|{\bf k}|^2}{(\gamma_n+|{\bf k}|^2)(l^2-\gamma_n)}\left(f(\gamma_n,\beta)-f(l^2,\beta)\right)\right.\right.\non
&+&\left.\left.\f{\gamma_n}{(\gamma_n+|{\bf k}|^2)(l^2+|{\bf k}|^2)}\left(\sum_m\f{\beta\sqrt{q}}{(2\pi m)^2}-f(l^2,\beta)\right) \right\}\right.\non &+&\left. \left(- \tilde{{\cal F}}_5^0(0,n,l)\tilde{{\cal F}}_5^0(0,n,-l)\right)\f{|{\bf k}|^2}{(l^2-\gamma_n)}\left\{\left(\f{\gamma_n}{\gamma_n+|{\bf k}|^2}f(\gamma_n,\beta)-\f{l^2}{l^2+|{\bf k}|^2}f(l^2,\beta)\right)\right.\right.\non
&+& \left.\left.\left(\f{|{\bf k}|^2}{\gamma_n+|{\bf k}|^2}-\f{|{\bf k}|^2}{l^2+|{\bf k}|^2}\right)
\sum_m\f{\beta\sqrt{q}}{(2\pi m)^2}\right\}\right.\non
&+&\left. \left(2 q \sqrt{2n+1} \tilde{F}_5^1(0,n,l)\tilde{{\cal F}}_5^0(0,n,-l)\right)\f{|{\bf k}|^2}{(l^2-\gamma_n)}\left\{\f{1}{l^2+|{\bf k}|^2}f(l^2,\beta)-\f{1}{\gamma_n+|{\bf k}|^2}f(\gamma_n,\beta)\right.\right.\non
&+& \left.\left.\left(\f{1}{\gamma_n+|{\bf k}|^2}-\f{1}{l^2+|{\bf k}|^2}\right)
\sum_m\f{\beta\sqrt{q}}{(2\pi m)^2}\right\}\right] 
\eeqa

\beqa
(XI)&=&-\hf Nq \sum_{m,n}\int \f{d l}{(2\pi\sqrt{q})} \int \f{d^2 {\bf k}}{(2\pi\sqrt{q})^2}\left(\f{1}{(\o_m^2)^2(\o_m^2+\gamma_n+|{\bf k}|^2)(\o_m^2+l^2+|{\bf k}|^2)}\right)\times\non
&\times&\left[\left(\tilde{F}^{11}_5(0,n,l)\tilde{F}^{11}_5(0,n,-l)\right)(\o_m^2+\gamma_n)(\o_m^2 +l^2)
+\left(-2\tilde{F}^{11}_5(0,n,l) \tilde{{\cal F}}_5(0,n,-l)\right)\left(l\sqrt{\gamma_n} |{\bf k}|^2\right)\right.\non
&+&\left.\left(-2\hat{l}\tilde{F}^{11}_5(0,n,l)\tilde{{\cal F}}^{1'}_5(0,n,-l)\right)\left( |{\bf k}|^2(\o_m^2+\gamma_n)\right)
+\left(2\sqrt{(2n+1)}\tilde{F}^{11}_5(0,n,l) \tilde{{\cal F}}^{1}_5(0,n,-l)\right)\left(|{\bf k}|^2(\o_m^2+l^2)\right)\right.\non
&+&\left.\left(-\tilde{{\cal F}}^{1}_5(0,n,l)\tilde{{\cal F}}^{1}_5(0,n,-l)\right)\left( (|{\bf k}|^2/q)(\o_m^2+l^2)(\o_m^2+|{\bf k}|^2)\right)\right.\non
&+&\left.\left(\tilde{{\cal F}}^{1'}_5(0,n,l) \tilde{{\cal F}}^{1'}_5(0,n,-l)\right)\left((|{\bf k}|^2/q)(\o_m^2+\gamma_n)(\o_m^2+|{\bf k}|^2)\right)\right.\non
&+&\left.\left(2\hat{l}\sqrt{(2n+1)}\tilde{{\cal F}}^{1}_5(0,n,l)\tilde{{\cal F}}^{1'}_5(0,n,-l)\right)\left( |{\bf k}|^2\right)^2
+\left(2\hat{l}\tilde{{\cal F}}_5(0,n,l) \tilde{{\cal F}}^{1}_5(0,n,-l)\right)\left(|{\bf k}|^2(\o_m^2+|{\bf k}|^2)\right)\right.\non
&+&\left.\left(-2\sqrt{(2n+1)}\tilde{{\cal F}}_5(0,n,l) \tilde{{\cal F}}^{1'}_5(0,n,-l)\right)\left(|{\bf k}|^2(\o_m^2+|{\bf k}|^2)\right)\right.\non
&+&\left.\left( {\cal \tilde{F}}_5(0,n,l){\cal \tilde{F}}_5(0,n,-l)\right)\left(2(\o_m^2)^2+2\o_m^2 |{\bf k}|^2+(|{\bf k}|^2)^2\right)\right]
\eeqa

Unlike the previous terms, due to the long length of the contribution $(XI)$ we have not presented here the result after doing the sum over the Matsubara frequencies.

\begin{figure}[h]
\begin{center}
\begin{psfrags}
\psfrag{c2}[][]{\scalebox{0.85}{$C(\tilde{\o}_{m^{''}},\tilde{k}_x^{''},\tilde{{\bf k}}^{''})$}}
\psfrag{c1}[][]{\scalebox{0.85}{$C(\o_{m^{''}},k_x^{''},{\bf k}^{''})$}}
\psfrag{c3}[][]{\scalebox{0.85}{$\theta_1(m,n,{\bf k})$}}
\psfrag{c4}[][]{\scalebox{0.85}{$\lambda^3_1(m^{'},n^{'},{\bf k}^{'})$}}
\psfrag{v1}[][]{\scalebox{0.85}{$F_6$}}
\psfrag{v2}[][]{\scalebox{0.85}{$F_6^{*}$}}
\psfrag{v3}[][]{\scalebox{0.85}{$V^{'\mu}_{f}$}}
\psfrag{v4}[][]{\scalebox{0.85}{$V^{'\mu*}_{f}$}}
\psfrag{a1}[][]{(a)}
\psfrag{a2}[][]{(b)}
\includegraphics[ width= 7cm,angle=0]{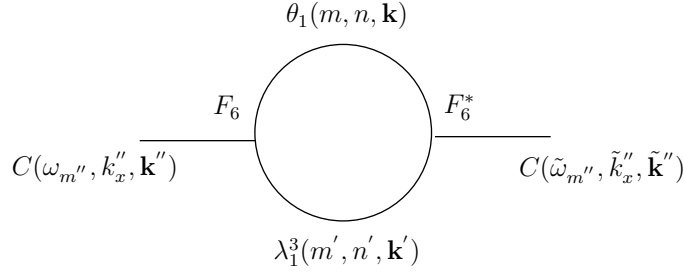}
\end{psfrags}
\caption{{Feynman diagrams involving three-point vertices $F_6$}}
\label{tachyonf}
\end{center}
\end{figure}

We now write down the contribution from fermions in the loop. The corresponding Feynman diagram is shown in Figure \ref{tachyonf}.

\beqa\label{tachfermion}
\Sigma^3_{C-C}=-4 N \sum_{m,n}\int \frac{d^2 {\bf k}}{(2\pi\sqrt{q})^2}\frac{dl}{(2\pi\sqrt{q})}\mbox{tr}\left[F_6(n^{''},n,l)\gamma^1\frac{1}{\slashed{P}_{+}}F^{*}_6(\tilde{n}^{''},n,l)\gamma^1\frac{1}{\slashed{K}^{'}_{+}}
\right](2\pi)^2\delta^{2}({\bf k}^{''}+\tilde{{\bf k}}^{''})\delta_{m^{''}+\tilde{m}^{''}}
\eeqa

where propagators $\slashed{P}_{+}=i\o_m\gamma^0+\sqrt{\lambda^{'}_n}\gamma^1+k_2\gamma^2+k_3\gamma^3 ~~;~~ \slashed{K}_{+}^{'}=i\o_{m^{'}}\gamma^0+l\gamma^1+k_2^{'}\gamma^2+k_3^{'}\gamma^3$. The momenta are related as, 
$\o_{m^{'}}=\o_m-\o_{m^{''}}$, ${\bf k}^{'}={\bf k}-{\bf k}^{''}$. The factor of $4$ in front comes from summing over the contributions from four sets of fermions (see eqn. \ref{fermionsetstachyon}).

We will now be interested in evaluating the two point function with the external momenta set to zero. ($\o_{m^{''}}=\tilde{\o}_{m^{''}}={\bf k}^{''}=\tilde{\bf k}^{''}=n^{''}=\tilde{n}^{''}=0$). Evaluating the trace in (\ref{tachfermion}) the amplitude reduces to

\beqa\label{tachfermion1}
\Sigma^3_{C-C}=&-&16N\sum_{m,n}\int \frac{d^2 {\bf k}}{(2\pi\sqrt{q})^2}\frac{d l}{(2\pi\sqrt{q})}\f{1}{2}\left[\frac{\left(l^2+\lambda^{'}_n\right)}{\left(\o_m^2+\lambda^{'}_n+|{\bf k}|^2\right)\left(\o_m^2+l^2+|{\bf k}|^2\right)}-
\frac{1}{\left(\o_m^2+\lambda^{'}_n+|{\bf k}|^2\right)}\right.\non
&-& \left.
\frac{1}{\left(\o_m^2+l^2+|{\bf k}|^2\right)}
\right]|F_6(0,n,l)|^2
\eeqa

where,

\beqa
F_6(0,n,l)=(-i)^{n+1}e^{-\f{\hat{l}^2}{4}}\f{\hat{l}^n}{\sqrt{2^{n+1}n!}}.
\eeqa

Now define the sum over the fermionic frequencies as,

\beqa\label{defg}
g(\lambda^{'}_n,\beta)=N\sum_m \f{1}{\o_m^2+\lambda^{'}_n+|{\bf k}|^2}=\f{\sqrt{q}}{\sqrt{\lambda^{'}_n+|{\bf k}|^2}}\left(\hf-\f{1}{e^{\beta\sqrt{\lambda^{'}_n+|{\bf k}|^2}}+1}\right).
\eeqa

Then after performing the sum over the Matsubara frequencies the amplitude can be written as

\beqa
\Sigma^3_{C-C}=&-&16\sum_{n}\int \frac{d^2 {\bf k}}{(2\pi\sqrt{q})^2}\frac{d l}{(2\pi\sqrt{q})}\left[\lambda^{'}_n g(\lambda^{'}_n,\beta)-l^2 g(l^2,\beta)
\right]\f{|F_6(0,n,l)|^2}{\left(l^2-\lambda^{'}_n\right)}
\eeqa

\subsection{Analysis of one loop amplitude}

\subsubsection{Ultraviolet finiteness}\label{tachyonuv}

In this section we show that the one-loop amplitude is finite in the UV (large $n$). Finiteness of the amplitudes in the UV is expected as the underlying ${\cal N}=4$ SYM theory is finite. In the present case supersymmetry is only broken by the background. Nevertheless we demonstrate this finiteness explicitly by showing the cancellation of contributions from bosons and fermions in the loop in the UV, which in this case is the large $n$ limit. This exercise is particularly useful in order to ensure that the counting of the modes in the loops has been done correctly. In the following we proceed along a slightly alternate route to demonstrate the UV finiteness than that was followed in \cite{1} for the case of intersecting $D1$ branes.

We first note that the factors coming from the three-point vertices are of the form

\beqa
\f{e^{-\f{\hat{l}^2}{2}}\hat{l}^{2n}}{2^{n+1}n!}\xrightarrow{n\rightarrow \infty}\f{1}{2\sqrt{2\pi n}}e^{\f{1}{2}(\hat{l}^2-2n)}\left(\f{\hat{l}^2}{2n}\right)^n
\eeqa

The above asymptotic expression shows that the leading large $n$ contribution from the $l$ integrals come from the region $\hat{l}^2=2n$. We will thus, in all the contributions obtained above, replace $\hat{l}^2$ by $2n$ in the propagators. This leaves us with $l$ integrals only over the vertices. We can now perform these integrals over $l$ separately noting that

\beqa\label{limit1}
\int \f{d\hat{l}}{(2\pi)}\f{e^{-\f{\hat{l}^2}{2}}\hat{l}^{2n}}{2^{n+1}n!}\xrightarrow{n\rightarrow \infty}\f{1}{(2\pi)\sqrt{2n}}.
\eeqa

To illustrate this further, consider the integral

\beqa\label{intg}
J_1=\int \f{d\hat{l}}{(2\pi)}\f{e^{-\f{\hat{l}^2}{2}}\hat{l}^{2n}}{2^{n+1}n!}{\cal I}(\hat{l}^2,\omega,|{\bf k}|^2)
\eeqa

where the function ${\cal I}(\hat{l}^2,\omega,|{\bf k}|^2)$ contains propagators and other polynomials in $\hat{l}^2$ in the numerator. According to the above in the large $n$ limit the leading value of the integral (\ref{intg}) will be given by $\f{1}{(2\pi)\sqrt{2n}}{\cal I}(2n,\omega,|{\bf k}|^2)$. As a first example let us take
${\cal I}(\hat{l}^2,\omega,|{\bf k}|^2)=\hat{l}^2$. Performing the integral (\ref{intg}) directly and then taking the large $n$ limit we get

\beqa\label{expre}
J_1=\f{2n}{2\pi \sqrt{2n}}+\f{3}{8 \pi\sqrt{2n} }+{\cal O}\left(\f{1}{n^{3/2}}\right)
\eeqa

The large $n$ leading term in the expression (\ref{expre}) can also obtained as described above.

As a second example let us consider the integral that involves both the sum over $n$ as well as the integral over $l$ (which appears in many of the expressions for the two-point function). A typical such integral is of the form,

\beqa\label{ex2}
J_2=\sum_{m,n}\int \f{d^2 {\bf k}}{(2\pi\sqrt{q})^2}\f{dl}{(2\pi\sqrt{q})}\f{e^{-\f{\hat{l}^2}{2}}\hat{l}^{2n}}{2^{n+1}n!} \frac{\hat{l}^2}{\left(\o_m^2+\lambda_n+|{\bf k}|^2\right)\left(\o_m^2+q\hat{l}^2+|{\bf k}|^2\right)}
\eeqa

We first perform the integral over $l$ in (\ref{ex2}). The result in the large $n$ limit reads (for simplicity we have set $q=1$)

\beqa\label{ex21}
 J_2=\sum_{m,n}\int \f{d^2 {\bf k}}{(2\pi)^2} \frac{1}{\left(\o_m^2+2n+|{\bf k}|^2\right)}\left[\frac{1}{2\pi\sqrt{2n}}-
\f{1+4(\o_m^2+|{\bf k}|^2)}{\sqrt{2} \pi n^{3/2}} + {\cal O}\left(\f{1}{n^{5/2}}\right)\right]
\eeqa

where we have assumed $2nq >>\o_m^2+|{\bf k}|^2$. Further terms with odd powers of $\o_m$ have been dropped. This is exactly as we had proceeded in \cite{1}. Following the other route, we have 

\beqa\label{ex22}
 J_2&=&\sum_{m,n}\int \f{d^2 {\bf k}}{(2\pi)^2} \left(\frac{1}{2\pi\sqrt{2n}}\right)\frac{2n}{\left(\o_m^2+2n+|{\bf k}|^2\right)^2}\non
&=& \sum_{m,n}\int \f{d^2 {\bf k}}{(2\pi)^2} \left(\frac{1}{2\pi\sqrt{2n}}\right)\left[\frac{1}{\left(\o_m^2+2n+|{\bf k}|^2\right)}-\frac{\o_m^2+|{\bf k}|^2}{\left(\o_m^2+2n+|{\bf k}|^2\right)^2}\right]
\eeqa

The leading term in the expressions (\ref{ex21}) and  (\ref{ex22}) match when we assume $2nq >>\o_m^2+|{\bf k}|^2$. The difference in the sub-leading pieces is expected as we have restricted ourselves to leading contribution as coming from $\hat{l}^2=2n$ in the second approach. 

Although this expansions with the assumption $2nq >>\o_m^2+|{\bf k}|^2$ is needed to show the agreement of the resulting leading behaviours in the two approaches it turns out that it will not be necessary in the ultimate demonstration of finiteness of the full two-point amplitude at large $n$. In other words in the following we work with the full expression (\ref{ex22}) and the assumption that $2nq >>\o_m^2+|{\bf k}|^2$ is not invoked. 


With these observations we proceed to write down the asymptotic contributions from each of the terms ($(I)-(XI)$).

For the terms containing four-point vertices (Feynman diagrams in Figure \ref{bosfourfeyn}) we can directly take the $n \rightarrow \infty$ limit. Using the following asymptotic forms worked out in \cite{1}

\beqa
F_1(0,0,n,n)+F_1^{'}(0,0,n,n)&=&\tilde{F}_1(0,0,n,n)+\tilde{F}_1^{'}(0,0,n,n)=\hf\left(F^1_2(0,0,n,n)+F^2_2(0,0,n,n)\right)\non &~&\xrightarrow{n\rightarrow \infty}\f{1}{(2\pi)\sqrt{2n}}\nonumber\\
\eeqa

we can now write,

\beqa
(I)\sim\hf N \sum_{m,n}\int \f{d^2 {\bf k}}{(2\pi\sqrt{q})^2}  \f{1}{(2\pi)\sqrt{2n}}\left[\f{1}{(\o_m^2+2nq+|{\bf k}|^2)}\right]
\eeqa

\beqa
(II)\sim \hf N \sum_{m,n}\int \f{d^2 {\bf k}}{(2\pi\sqrt{q})^2} \f{1}{(2\pi)\sqrt{2n}}\left[\f{\o_m^2+2nq}{\o_m^2(\o_m^2+2nq+|{\bf k}|^2)}\right]
\eeqa

\beqa
(III)=\hf N \sum_{m,n}\int \f{d^2 {\bf k}}{(2\pi\sqrt{q})^2} \f{1}{(2\pi)\sqrt{2n}}\left[\f{(5\times 2)}{\o_m^2+2nq+|{\bf k}|^2}+(2) \f{2\o_m^2+|{\bf k}|^2}{\o_m^2(\o_m^2+2nq+|{\bf k}|^2)}\right]
\eeqa

By inserting 

\beqa
\sum_{n=0}^{\infty}\f{e^{-\f{\hat{l}^2}{2}}\hat{l}^{2n}}{2^{n+1}n!}=\hf
\eeqa

in the expressions of $(IV)$, $(V)$ and $(VI)$ and doing the $l$ integral as elaborated above, we get $(IV)=(III)$ and

\beqa
(V)+(VI)\sim\hf N \sum_{m,n} \int \f{d^2 {\bf k}}{(2\pi\sqrt{q})^2} \f{1}{(2\pi)\sqrt{2n}}\left[\f{1}{\o_m^2+2nq+|{\bf k}|^2}+ \f{\o_m^2+2nq}{\o_m^2(\o_m^2+2nq+|{\bf k}|^2)}\right]
\eeqa

We now write down the asymptotic forms of the contributions $(VII)-(XI)$ corresponding to the diagrams in  Figure \ref{bosthreefeyn}. Here we implement the steps described at the beginning of this section.

\beqa
(VII)\sim-\hf N \sum_{m,n} \int \f{d^2 {\bf k}}{(2\pi\sqrt{q})^2} \f{1}{(2\pi)\sqrt{2n}}\left[\f{1}{\o_m^2}-\f{1}{\o_m^2+2nq+|{\bf k}|^2}\right]
\eeqa

\beqa
(VIII)\sim-\hf N \sum_{m,n} \int \f{d^2 {\bf k}}{(2\pi\sqrt{q})^2} \f{1}{(2\pi)\sqrt{2n}}\left[\f{(5\times 4) 2nq}{(\o_m^2+2nq+|{\bf k}|^2)^2}\right]
\eeqa

Similarly in the large $n$ limits, $(IX)=\f{1}{5}(VIII)$ and $(X)=(VII)$.

\beqa
(XI)\sim-\hf N \sum_{m,n} \int \f{d^2 {\bf k}}{(2\pi\sqrt{q})^2} \f{1}{(2\pi)\sqrt{2n}}\left[\f{8\o_m^4(2nq)+2\o_m^2|{\bf k}|^2(\o_m^2+2nq+|{\bf k}|^2)}{(\o_m^4)(\o_m^2+2nq+|{\bf k}|^2)^2}\right]
\eeqa

Adding all the contributions $(I)-(XI)$ we get

\beqa\label{bosonlargen}
\Sigma^1_{C-C}+\Sigma^2_{C-C}\sim \hf N \sum_{m,n} \int \f{d^2 {\bf k}}{(2\pi\sqrt{q})^2} \f{1}{(2\pi)\sqrt{2n}}(32)\left[\f{1}{\o_m^2+2nq+|{\bf k}|^2}-\f{2nq}{(\o_m^2+2nq+|{\bf k}|^2)^2}\right]
\eeqa

Now following the above observations it is easy to see that that the large $n$ limit of (\ref{tachfermion1}) is exactly same as (\ref{bosonlargen})  coming from fermions in the loop but with opposite sign. Thus the one loop amplitude is finite in the UV (large $n$). 

We should note that the finiteness in the large $n$ limit as shown here holds irrespective of the values of $\o_m^2$ and $|{\bf k}|^2$. That is for every fixed value of $\o_m^2$ and $|{\bf k}|^2$ the amplitudes are finite in the large $n$ limit. It would similarly be useful to give an analytic proof of the finiteness of the amplitudes for fixed values of $\o_m^2$ and $2nq$ but large $|{\bf k}|$. Unfortunately we are not able to demonstrate this analytically as the closed form expressions (in $n$) for the integrals (\ref{f1f1p},\ref{ft1ft1p},\ref{f2f2p}) that involves product of four Hermite polynomials could not be obtained. Such finiteness of the amplitudes for the massless fields however can be shown both for large $n$ and fixed $|{\bf k}|$ as well as large $|{\bf k}|$ and fixed $n$ which we shall see in the following sections.

\subsubsection{Infrared issues}\label{tachyonir}

Some of the contributions to the two point amplitude $(I-XI)$ diverges in the infrared. These infrared divergences occur from two sources: (i) There are poles of the from of $1/\o_m^2$. These poles are artifacts of the $A_0^a=0$ gauge and one should use prescriptions to remove them. See for example \cite{kapusta1}-\cite{Leib1} and references therein. In our case this means that we shall simply drop terms that are proportional to $\sum_m\f{\beta\sqrt{q}}{(2\pi m)^2}$. (ii) There are also genuine infrared divergences due to tree-level ``massless" modes propagating in the loop. These modes are the fields with the gauge index $a=3$, namely $\Phi_J^3/\tilde{\Phi}_J^3 (J=1,2,3)$ and $A_{i}^3 (i=1,2,3)$. This is because these fields do not couple to the background value of the field $\Phi_1^3$, and hence the tree-level spectrum for the fields with gauge index equal to $3$ is not gapped unlike the fields with gauge indices equal to $1,2$. To ultimately get a finite answer for the one loop amplitude we follow the procedure implemented in \cite{1}. We first find the one-loop correction to the propagators for these fields and then use the corrected propagators to evaluate the one-loop two-point tachyon amplitude. The corrected propagator will involve one-loop corrections to the zero tree-level masses for the fields $\Phi_I^3, A_{\mu}^3$. These corrections will be computed in the next section.

We now give some details on how to incorporate the one-loop masses in the propagators for the tree-level ``massless" modes.
The quadratic action for the $\Phi_I^3/\tilde{\Phi}_I^3/$ fields (see eqn \ref{phiAaction}) including the mass terms is written as,

\beqa\label{phiactionm}
&&-\f{1}{2qg^2}\int\f{dk_xd^2{\bf k}}{(2\pi\sqrt{q})^3}\sum_{m}\left[|\Phi_J^{3}(m,k)|^2\left(k^2+m^2_{\Phi_J^3}\right)+|\tilde{\Phi}_J^{3}(m,k)|^2\left(k^2+m^2_{\tilde{\Phi}_J^3}\right)\right]
\eeqa

Here  $(J=1,2,3)$ and $k^2=(\o_m^2+k_x^2+{\bf k}^2)$. Due to the remaining $SO(5)$ symmetry, except for $m^2_{\Phi_1^3}$ the masses $m^2_{\Phi_J^3}$ and $m^2_{\tilde{\Phi}_J^3}$ are all equal.  We thus write down the modified propagators as:

\beqa
\label{propphi3I}
\expect{\Phi_I^3(m,l,{\bf k})\Phi_I^3(m^{'},l^{'},{\bf k}^{'})}\rightarrow qg^2 \f{\delta_{m,-m^{'}}2\pi\delta(l+l^{'})(2\pi)^2\delta^2({\bf k}+{\bf k}^{'})}{\o_m^2+l^2+|{\bf k}|^2+m^2_{\Phi_I}}
\eeqa

For the $A^3_i$ fields writing down the modified propagator requires a bit more work. The one-loop correction breaks the  $SO(3)$ in (\ref{phiAaction}) to $SO(2)$. As a result the one-loop mass for $A^3_1$ ($m_1$) field is different from those of $A^3_2$ and $A^3_3$ fields which are equal (say $m_2$). Introducing a vector $u^i\equiv (1,0,0)$, we can write down the quadratic part of the action as

\beqa\label{Aactionm}
&&-\f{1}{2qg^2}\int\f{dk_xd^2{\bf k}}{(2\pi\sqrt{q})^3}\sum_{m}\left[\tilde{A}_i^3(m, k)\left((k^2+m_2^2)\delta^{ij}-k^ik^j+(m_1^2-m_2^2)u^iu^j\right)\tilde{A}_j^3(-m,-k)\right]
\eeqa

Here  $(i,j=1,2,3)$ and $k^2=(\o_m^2+k_x^2+{\bf k}^2)$. For the one-loop amplitude there are momentum dependent corrections as well. 
Correspondingly the term $\tilde{A}^3_i(a k^ik^j +bk_x(u^ik^j+u^jk^i)+c(k_x)^2u^iu^j)\tilde{A}^3_j$ also arises in the one-loop effective action (\ref{Aactionm}). The correction coefficients $a$, $b$ and $c$ are different due to the breaking of $SO(3)$ invariance.  
In the infrared limit however only the mass terms in (\ref{Aactionm}) provide the necessary regulation. Thus for simplicity we shall work with only the non-zero masses $m_1$ and $m_2$. Following these observations the corrected propagator takes the following form (see appendix  \ref{amplitudem} for further details)

\beqa\label{propp33m}
(qg^2) P^{33}_{ij}:\expect{\tilde{A}_i^{3}(m,l,{\bf k})\tilde{A}_j^{3}(-m,-l,-{\bf k})}\rightarrow qg^2 A \left[\delta_{ij}+B\left(k_ik_j +C(k_iu_j+u_ik_j)+D u_iu_j\right)\right]
\eeqa

where, the functions $A$, $B$, $C$ and $D$ are given by 

\beqa
&& A=\f{1}{k^2+m_2^2}~~~;~~~B=\left(\f{k^2+m_1^2}{k^2+m_1^2+m_2^2}\right)\left(\f{1}{\o_m^2+\f{k_x^2m_1^2+|{\bf k}|^2m_2^2+m_1^2m_2^2}{k^2+m_1^2+m_2^2}}\right)\non
&&C=\f{k_x\left(m_2^2-m_1^2\right)}{k^2+m_1^2}~~;~~D=\left(m_2^2-m_1^2\right)\left(\f{\o_m^2+m_2^2}{k^2+m_1^2}\right)
\eeqa 

With these modifications we can re-write the corresponding modified expressions in $(I)$-$(XI)$. For the $\Phi_I^3$ fields the mass terms in the propagators in the loop are easy to implement. However for the $A^3_i$ some more work is needed. These expressions are listed in appendix \ref{amplitudem}. The numerical results incorporating the above are analyzed in section \ref{numerics}.



\section{Two point amplitude for massless fields}\label{masslessoneloop}

In the following sections we shall compute the one-loop corrections to the masses for the fields, $\Phi_J^3/\tilde{\Phi}_J^3 (J=1,2,3)$ and $A_{i}^3 (i=1,2,3)$. For each of the two-point amplitudes we also demonstrate the cancellation of contributions from the bosons and fermions in the loop in the large $n$ limit irrespective of the values of $\o_m^2$ and $|{\bf k}|^2$ as well as for large $|{\bf k}|$ at fixed values of $\o_m^2$ and $2nq$ thereby showing that the amplitudes are UV finite.

\subsection{Two point amplitude for $\Phi_1^3$ field}\label{phi13oneloop}

In this section we compute the two point amplitude for $\Phi_1^3$ field. Using the vertices listed in appendix (\ref{verticesphi1}), we first write down the contributions to the one loop amplitude from bosons in the loop,

\begin{figure}[h]
\begin{center}
\begin{psfrags}
\psfrag{c1}[][]{\scalebox{0.95}{$\Phi^3_1$}}
\psfrag{c2}[][]{\scalebox{0.95}{$\Phi^3_1$}}
\psfrag{c3}[][]{$\Phi^{(1,2)}_I/\tilde{\Phi}^{(1,2)}_I(m,n,{\bf k})$}
\psfrag{c4}[][]{$C/C^{'}(m,n,{\bf k})$}
\psfrag{c5}[][]{$\tilde{A}_i^{(1,2)}(m,n,{\bf k})$}
\psfrag{v1}[][]{$V^{(1)}_{1}$}
\psfrag{v2}[][]{$V^{(1)}_{2}$}
\psfrag{v3}[][]{$\tilde{V}^{(1)ii}_{2}$}
\psfrag{a}[][]{$(I)$}
\psfrag{b}[][]{$(II)$}
\psfrag{c}[][]{$(III)$}
\includegraphics[ width= 10cm,angle=0]{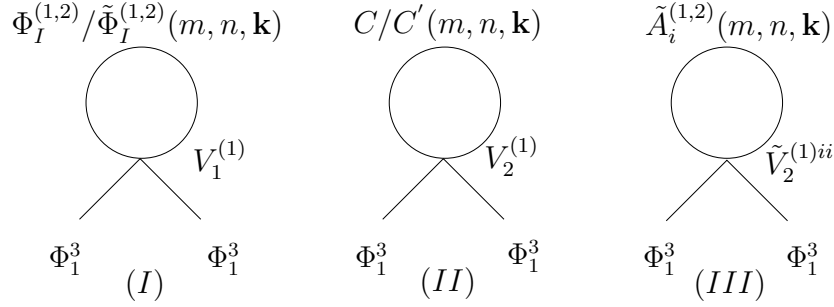}
\end{psfrags}
\caption{Feynman diagrams contributing to $\Sigma^1_{\Phi_1^3-\Phi_1^3}$. The momenta on the external $\Phi_1^3$ fields are $(\o_{m^{''}},k_x^{''},{\bf k}^{''})$ and  $({\o}_{\tilde{m}^{''}},\tilde{k}_x^{''},\tilde{{\bf k}}^{''})$}.
\label{phi14pt}
\end{center}
\end{figure}

This contribution consisting of the four-point vertices as shown in figure (\ref{phi14pt}) is given by,
\beqa\label{sigma1phi1}
\Sigma^1_{\Phi_1^3-\Phi_1^3}&=&\hf  N \int \f{d^2 {\bf k}}{(2\pi\sqrt{q})^2} \sum_{m,n} \left[
(5\times2)\f {G^{(1)}_1(n,n,k_x^{''},\tilde{k}_x^{''})}{\o_m^2 + \gamma_n+|{\bf k}|^2} + (2)\f {G^{(1)}_2(n,n,k_x^{''},\tilde{k}_x^{''})}{\o_m^2 + \lambda_n+|{\bf k}|^2}\right. \nonumber\\&+&\left.\sum_{i,j=1}^3\tilde{G}^{(1)ij}_2(n,n,k_x^{''},\tilde{k}_x^{''})(P^{11}_{ij}+P^{22}_{ij})
\right]\delta_{m^{''}+\tilde{m}^{''}}(2\pi)^2\delta^2({\bf k}^{''}+\tilde{\bf k}^{''})
\eeqa

\begin{figure}[h]
\begin{center}
\begin{psfrags}
\psfrag{c1}[][]{$\Phi^3_1$}
\psfrag{c2}[][]{$\Phi^3_1$}
\psfrag{c5}[][]{$\Phi^{(1,2)}_I$}
\psfrag{c6}[][]{$\Phi^{(1,2)}_I$}
\psfrag{c3}[][]{$C$}
\psfrag{c4}[][]{$C$}
\psfrag{c7}[][]{$C/C^{'}$}
\psfrag{c8}[][]{$\tilde{A}_i^{(1,2)}$}
\psfrag{c9}[][]{$\tilde{A}_i^{(1,2)}$}
\psfrag{c10}[][]{$\tilde{A}_j^{(1,2)}$}
\psfrag{v1}[][]{$V^{(1)}_{3}$}
\psfrag{v2}[][]{$V^{(1)*}_{3}$}
\psfrag{v3}[][]{${V}^{(1)}_{4}$}
\psfrag{v4}[][]{$V^{(1)*}_{4}$}
\psfrag{v5}[][]{$\tilde{V}^{(1)i}_{3}$}
\psfrag{v6}[][]{$\tilde{V}^{(1)j*}_{3}$}
\psfrag{v7}[][]{$\tilde{V}^{(1)ij}_{3}$}
\psfrag{v8}[][]{$\tilde{V}^{(1)ij*}_{3}$}
\psfrag{a1}[][]{$(IV)$}
\psfrag{a2}[][]{$(VI)$}
\psfrag{a3}[][]{$(V)$}
\psfrag{a4}[][]{$(VII)$}
\includegraphics[ width= 12cm,angle=0]{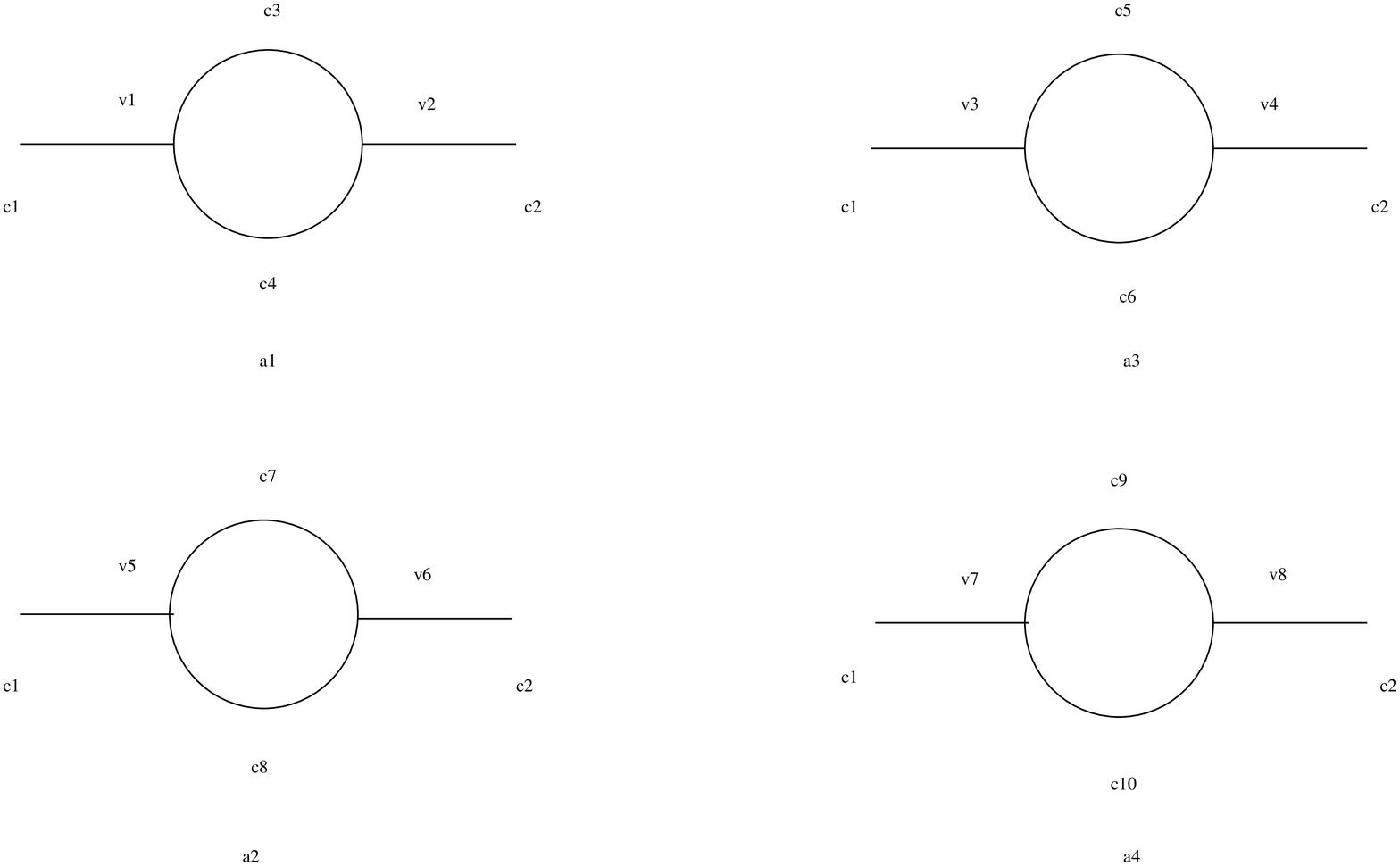}
\end{psfrags}
\caption{Feynman diagrams contributing to $\Sigma^1_{\Phi_1^3-\Phi_1^3}$. The momenta on the external $\Phi_1^3$ fields are $(\o_{m^{''}},k_x^{''},{\bf k}^{''})$ and  $({\o}_{\tilde{m}^{''}},\tilde{k}_x^{''},\tilde{{\bf k}}^{''})$}
\label{phi13pt}
\end{center}
\end{figure} 

and the contributions from the three-point vertices (see figure (\ref{phi13pt})) is 

\beqa\label{sigma2phi1}
\Sigma^2_{\Phi_1^3-\Phi_1^3}&=&-\hf qN\sum_{m,n,n^{'}}\int\f{d^2 {\bf k}}{(2\pi\sqrt{q})^2}\left[(2)\f{G^{(1)}_3(n,n^{'},k_x^{''})G^{(1)}_3(n,n^{'},\tilde{k}_x^{''})}{(\o_m^2 + \lambda_n+|{\bf k}|^2)(\o_{m^{'}}^2 +\lambda_{n^{'}}+|{\bf k}^{'}|^2)} \right.\non 
&+& \left.(5\times2)\f{G^{(1)}_4(n,n^{'},k_x^{''})G^{(1)}_4(n,n^{'},\tilde{k}_x^{''})}{(\o_m^2 + \gamma_n+|{\bf k}|^2)(\o_{m^{'}}^2 +\gamma_{n^{'}}+|{\bf k}^{'}|^2)}+ \sum_{i,j=1}^3\f{\tilde{G}^{(1)i}_3(n,n^{'},k_x^{''})\tilde{G}^{(1)j}_3(n,n^{'},\tilde{k}_x^{''})}{(\o_m^2 +\lambda_n+|{\bf k}|^2)}(P^{11}_{ij}+P^{22}_{ij})\right.\non 
&+& \left.\sum_{i,i^{'},j,j^{'}=1}^3\tilde{G}^{(1)ij}_3(n,n^{'},k_x^{''})\tilde{G}^{(1)i^{'}j^{'}}_3(n,n^{'},\tilde{k}_x^{''})(P^{11}_{ii^{'}}P^{11}_{jj^{'}}+P^{22}_{ii^{'}}P^{22}_{jj^{'}})\right]
  \delta_{m^{''}+\tilde{m}^{''}}(2\pi)^2\delta^2({\bf k}^{''}+\tilde{\bf k}^{''})
\eeqa

Where $\o_{m^{'}}=\o_m-\o_{m^{''}}$, ${\bf k}^{'}={\bf k}-{\bf k}^{''}$. For computing the above amplitudes at zero external momentum we list the various exact functions below

\beqa\label{vertexphi1n}
&&G^{(1)}_1(n,n^{'},0,0) = \delta_{n,n^{'}},~~ 
G^{(1)}_2(n,n^{'},0,0)=\hf\delta_{n,n^{'}},~~
\tilde{G}^{(1)11}_2(n,n^{'})=\hf\delta_{n,n^{'}},~~\tilde{G}^{(1)ii}_2(n,n^{'})=\delta_{n,n^{'}}\non
&&G^{(1)}_3(0,n,n^{'})
=2 \sqrt{\f{2n(n-1)(n-2)}{(2n-1)(2n-3)}} \delta_{n-1,n^{'}},~~\tilde{G}^{(1)i}_3(0,n,n^{'}) =  i\hat{k}_i\left(
 \f{\sqrt{(n-1)}}{\sqrt{2(2n-1)}} \delta_{n,n^{'}}-\f{\sqrt{n}}{\sqrt{2(2n-1)}} \delta_{n-2,n^{'}} \right)\non
&&\tilde{G}^{(1)1}_3(0,n,n^{'}) = \pm i\left(
 \f{\sqrt{2n}(n-1)}{\sqrt{(2n-1)(2n-3)}} \delta_{n-2,n^{'}}-\f{\sqrt{2(n-1)}(n+1)}{\sqrt{(2n-1)(2n+1)}} \delta_{n,n^{'}} \right),~~\tilde{G}^{(1)11}_3(0,n,n^{'})= 0\non
&&\tilde{G}^{(1)1i}_3(0,n,n^{'}) =  \mp\hat{k}_i\left(
 \f{\sqrt{(n+1)}}{\sqrt{2(2n+1)}} \delta_{n+1,n^{'}}+\f{\sqrt{n}}{\sqrt{2(2n+1)}} \delta_{n-1,n^{'}} \right)
,~~\tilde{G}^{(1)ii}_3(0,n,n^{'})=G^{(1)}_4(0,n,n^{'})=\sqrt{2n}\delta_{n-1,n^{'}}\non
\eeqa

In the above expressions $i=2,3$ and $\hat{k}_i=k_i/\sqrt{q}$.
We first analyze the contributions to the expressions (\ref{sigma1phi1}) and (\ref{sigma2phi1}) in the large $n$ limit. This will give us their ultraviolet behavior. We write down the contributions from each of the terms separately,

\beqa\label{sigma1phi1n}
\Sigma^1_{\Phi_1^3-\Phi_1^3}&\sim&\hf  N \int \f{d^2 {\bf k}}{(2\pi\sqrt{q})^2} \sum_{m,n} \left[
(5\times2)\f {1}{\o_m^2 + 2nq+|{\bf k}|^2} + (2\times\hf)\f {1}{\o_m^2 + 2nq+|{\bf k}|^2}\right. \nonumber\\&+&\left.
(2\times\hf)\f{\o_m^2+2nq}{\o_m^2(\o_m^2+2nq+|{\bf k}|^2)}+(2)\f{2\o_m^2+|{\bf k}|^2}{\o_m^2(\o_m^2+2nq+|{\bf k}|^2)}
\right]
\eeqa

\beqa\label{sigma2phi1n}
\Sigma^2_{\Phi_1^3-\Phi_1^3}&\sim&-\hf N\sum_{m,n}\int\f{d^2 {\bf k}}{(2\pi\sqrt{q})^2}\left[(2)\f{2nq}{(\o_m^2 + 2nq+|{\bf k}|)^2}  
+(5\times2)\f{2nq}{(\o_m^2 + 2nq+|{\bf k}|^2)^2}\right.\non 
&+&\left.  (2\times\hf)\f{(2nq)(\o_m^2+2nq)}{\o_m^2(\o_m^2 + 2nq+|{\bf k}|^2)^2}+(2\times\hf)\f{|{\bf k}|^2(\o_m^2+|{\bf k}|^2)}{\o_m^2(\o_m^2 + 2nq+|{\bf k}|^2)^2}+(2\times\hf)\f{2(2nq)|{\bf k}|^2}{\o_m^2(\o_m^2 + 2nq+|{\bf k}|^2)^2}\right.\non 
&+& \left.\f{2(2nq)\left[2\o_m^2(\o_m^2+|{\bf k}|^2)+(|{\bf k}|^2)^2\right]}{(\o_m^2)^2(\o_m^2 + 2nq+|{\bf k}|^2)^2}-\f{4(2nq)|{\bf k}|^2(\o_m^2+|{\bf k}|^2)}{(\o_m^2)^2(\o_m^2 + 2nq+|{\bf k}|^2)^2}\right.\non 
&+&\left.\f{|{\bf k}|^2(\o_m^2+|{\bf k}|^2)(\o_m^2+2nq)}{(\o_m^2)^2(\o_m^2 + 2nq+|{\bf k}|^2)^2}+\f{(|{\bf k}|^2)^2(2nq)}{(\o_m^2)^2(\o_m^2 + 2nq+|{\bf k}|^2)^2}
\right]
\eeqa

The full contribution from (\ref{sigma1phi1n}) and  (\ref{sigma2phi1n}) simplifies to

\beqa\label{phi1nb}
\Sigma^1_{\Phi_1^3-\Phi_1^3}+\Sigma^2_{\Phi_1^3-\Phi_1^3}&\sim&   8N  \sum_{m,n}\int \f{d^2 {\bf k}}{(2\pi\sqrt{q})^2}\left[\f {1}{\o_m^2 + 2nq+|{\bf k}|^2}-\f{2nq}{(\o_m^2 + 2nq+|{\bf k}|)^2} \right]\non
&\sim&2\sum_{n}\int \frac{d^2 {\bf k}}{(2\pi\sqrt{q})^2}\frac{1}{\sqrt{2n+|{\bf k}|^2/q}}
\eeqa

where in the last line of the above equation we have kept the zero temperature UV divergent piece.

We now compute the the exact expression for the two point amplitude for all $n$. To do this we write down term by term each of which corresponds to the Feynman diagrams in the figures (\ref{phi14pt}) and (\ref{phi13pt}). We write 

\beqa
\Sigma^1_{\Phi_1^3-\Phi_1^3}=(I)+(II)+(III)
\eeqa

where,

\beqa
\label{phi1I}
&&(I)=\hf   \sum_{n=0}^{\infty}\int \f{d^2 {\bf k}}{(2\pi\sqrt{q})^2}  (5\times 2)f(\gamma_n,\beta)~~~;~~~
(II)=\hf   \sum_{n=2}^{\infty}\int \f{d^2 {\bf k}}{(2\pi\sqrt{q})^2}  (2\times \hf)f(\lambda_n,\beta)\\
\label{phi1III}
&&(III)=\hf \sum_{n=0}^{\infty}\int \f{d^2 {\bf k}}{(2\pi\sqrt{q})^2}  \left[(2\times \hf)\left\{\f{|{\bf k}|^2}{\gamma_n+|{\bf k}|^2}f(\gamma_n,\beta)+\f{\gamma_n}{\gamma_n+|{\bf k}|^2}\sum_{m}\f{\beta\sqrt{q}}{(2\pi m)^2}\right\}\right.\\
&&+\left.(2)\left\{2f(\gamma_n,\beta)+\f{|{\bf k}|^2}{\gamma_n+|{\bf k}|^2}\sum_{m}\f{\beta\sqrt{q}}{(2\pi m)^2}-\f{|{\bf k}|^2}{\gamma_n+|{\bf k}|^2}f(\gamma_n,\beta)\right\}\right]\nonumber
\eeqa

and

\beqa
\Sigma^2_{\Phi_1^3-\Phi_1^3}=(IV)+(V)+(VI)+(VII)
\eeqa

where,

\beqa
&&(IV)=-\hf\sum_{n=2}^{\infty}\int \f{d^2 {\bf k}}{(2\pi\sqrt{q})^2}(2)\left(\f{4n(n-1)(n-2)}{(2n-1)(2n-3)}\right)\left[f(\lambda_{n-1},\beta)-f(\lambda_{n},\beta)\right]\\
&&(V)=-\hf\sum_{n=0}^{\infty}\int \f{d^2 {\bf k}}{(2\pi\sqrt{q})^2}(5\times2)n\left[f(\gamma_{n-1},\beta)-f(\gamma_{n},\beta)\right]
\eeqa

\beqa\label{phi1VI}
&&(VI)=-\hf\sum_{n=2}^{\infty}\int \f{d^2 {\bf k}}{(2\pi\sqrt{q})^2}(2)\times\\
&\times&\left[\left(\f{(n-1)(n+1)^2}{(2n-1)(2n+1)}\right)\left\{\f{-2q+|{\bf k}|^2}{\lambda_n+|{\bf k}|^2}f(\lambda_{n},\beta)-\f{|{\bf k}|^2}{\gamma_n+|{\bf k}|^2}f(\gamma_{n},\beta)+\left(\f{\gamma_n}{\lambda_n+|{\bf k}|^2}-\f{\gamma_n}{\gamma_n+|{\bf k}|^2}\right)\sum_{m}\f{\beta\sqrt{q}}{(2\pi m)^2}\right\}-\right.\non
&&\left. \left(\f{n(n-1)^2}{(2n-1)(2n-3)}\right)\left\{\f{2q+|{\bf k}|^2}{\lambda_n+|{\bf k}|^2}f(\lambda_{n},\beta)-\f{|{\bf k}|^2}{\gamma_{n-2}+|{\bf k}|^2}f(\gamma_{n-2},\beta)+\left(\f{\gamma_{n-2}}{\lambda_n+|{\bf k}|^2}-\f{\gamma_{n-2}}{\gamma_{n-2}+|{\bf k}|^2}\right)\sum_{m}\f{\beta\sqrt{q}}{(2\pi m)^2}\right\}\right.\non
&+&\left. \left(\f{(n-1)(|{\bf k}|^2/q)}{4(2n-1)}\right)\left\{\f{\lambda_n}{\lambda_n+|{\bf k}|^2}f(\lambda_{n},\beta)-\f{\gamma_n}{\gamma_n+|{\bf k}|^2}f(\gamma_{n},\beta)+\left(\f{|{\bf k}|^2}{\lambda_n+|{\bf k}|^2}-\f{|{\bf k}|^2}{\gamma_n+|{\bf k}|^2}\right)\sum_{m}\f{\beta\sqrt{q}}{(2\pi m)^2}\right\}
\right.\non
&-&\left. \left(\f{n(|{\bf k}|^2/q)}{4(2n-1)}\right)\left\{\f{\lambda_n}{\lambda_n+|{\bf k}|^2}f(\lambda_{n},\beta)-\f{\gamma_{n-2}}{\gamma_{n-2}+|{\bf k}|^2}f(\gamma_{n-2},\beta)+\left(\f{|{\bf k}|^2}{\lambda_n+|{\bf k}|^2}-\f{|{\bf k}|^2}{\gamma_{n-2}+|{\bf k}|^2}\right)\sum_{m}\f{\beta\sqrt{q}}{(2\pi m)^2}\right\}\right.\non
&+&\left.\left(\f{(n^2-1)}{(2n-1)}\right)\left\{\f{|{\bf k}|^2}{\gamma_n+|{\bf k}|^2}f(\gamma_{n},\beta)-\f{|{\bf k}|^2}{\lambda_n+|{\bf k}|^2}f(\lambda_{n},\beta)+\left(\f{|{\bf k}|^2}{\lambda_n+|{\bf k}|^2}-\f{|{\bf k}|^2}{\gamma_{n}+|{\bf k}|^2}\right)\sum_{m}\f{\beta\sqrt{q}}{(2\pi m)^2}\right\}
\right.\non
&-&\left.\left(\f{n(n-1)\sqrt{2n+1}}{(2n-1)\sqrt{2n-3}}\right)\left\{\f{|{\bf k}|^2}{\gamma_{n-2}+|{\bf k}|^2}f(\gamma_{n-2},\beta)-\f{|{\bf k}|^2}{\lambda_n+|{\bf k}|^2}f(\lambda_{n},\beta)+\left(\f{|{\bf k}|^2}{\lambda_n+|{\bf k}|^2}-\f{|{\bf k}|^2}{\gamma_{n-2}+|{\bf k}|^2}\right)\sum_{m}\f{\beta\sqrt{q}}{(2\pi m)^2}\right\}
\right]\nonumber
\eeqa

\beqa
&&(VII)=-\hf\sum_{n=0}^{\infty}\int \f{d^2 {\bf k}}{(2\pi\sqrt{q})^2}(2)\times\\
&\times&\left[\left(2n\right)\left\{\left(1-\f{|{\bf k}|^2}{\gamma_{n-1}+|{\bf k}|^2}+\hf\f{(|{\bf k}|^2)^2}{(\gamma_{n-1}+|{\bf k}|^2)^2}\right)f(\gamma_{n-1},\beta)-\left(1-\f{|{\bf k}|^2}{\gamma_{n}+|{\bf k}|^2}+\hf\f{(|{\bf k}|^2)^2}{(\gamma_{n}+|{\bf k}|^2)^2}\right)f(\gamma_{n},\beta)
\right.\right.\non  &+& \left.\left. \left(\f{|{\bf k}|^2}{\gamma_{n-1}+|{\bf k}|^2}-\f{|{\bf k}|^2}{\gamma_{n}+|{\bf k}|^2}+\f{(|{\bf k}|^2)^2}{2(\gamma_{n}+|{\bf k}|^2)^2}-\f{(|{\bf k}|^2)^2}{2(\gamma_{n-1}+|{\bf k}|^2)^2}\right)\sum_{m}\f{\beta\sqrt{q}}{(2\pi m)^2}\right.\right.\non 
&+& \left.\left.\left(\f{(|{\bf k}|^2)^2}{2(\gamma_{n-1}+|{\bf k}|^2)}-\f{(|{\bf k}|^2)^2}{2(\gamma_{n}+|{\bf k}|^2)}\right)\sum_{m}\f{\beta^3\sqrt{q}}{(2\pi m)^4}\right\}\right.\non 
&+& \left.\left(\f{(n+1)|{\bf k}|^2/q}{4(2n+1)}\right)\left\{ \f{\gamma_n|{\bf k}|^2}{(\gamma_{n}+|{\bf k}|^2)^2}f(\gamma_{n},\beta)-\left(1-\f{\gamma_n+|{\bf k}|^2}{(\gamma_{n+1}+|{\bf k}|^2)}+\f{\gamma_n|{\bf k}|^2}{(\gamma_{n+1}+|{\bf k}|^2)^2}\right)f(\gamma_{n+1},\beta)\right.\right.\non
&+& \left.\left.\left(1-\f{\gamma_n+|{\bf k}|^2}{\gamma_{n+1}+|{\bf k}|^2}+\f{\gamma_n|{\bf k}|^2}{(\gamma_{n+1}+|{\bf k}|^2)^2}-\f{\gamma_n|{\bf k}|^2}{(\gamma_{n}+|{\bf k}|^2)^2}\right)\sum_{m}\f{\beta\sqrt{q}}{(2\pi m)^2}+\left(\f{\gamma_n|{\bf k}|^2}{(\gamma_{n}+|{\bf k}|^2)}-\f{\gamma_n|{\bf k}|^2}{(\gamma_{n+1}+|{\bf k}|^2)}\right)\sum_{m}\f{\beta^3\sqrt{q}}{(2\pi m)^4}\right\}\right.\non 
&+&\left.\left(\f{n|{\bf k}|^2/q}{4(2n+1)}\right)\left\{ \left(1-\f{\gamma_n+|{\bf k}|^2}{\gamma_{n-1}+|{\bf k}|^2}+\f{\gamma_n|{\bf k}|^2}{(\gamma_{n-1}+|{\bf k}|^2)^2}\right)f(\gamma_{n-1},\beta)-\left(\f{\gamma_n|{\bf k}|^2}{(\gamma_{n}+|{\bf k}|^2)^2}\right)f(\gamma_{n},\beta)\right.\right.\non
&-& \left.\left.\left(1-\f{\gamma_n+|{\bf k}|^2}{\gamma_{n-1}+|{\bf k}|^2}+\f{\gamma_n|{\bf k}|^2}{(\gamma_{n-1}+|{\bf k}|^2)^2}-\f{\gamma_n|{\bf k}|^2}{(\gamma_{n}+|{\bf k}|^2)^2}\right)\sum_{m}\f{\beta\sqrt{q}}{(2\pi m)^2}-\left(\f{\gamma_n|{\bf k}|^2}{(\gamma_{n}+|{\bf k}|^2)}-\f{\gamma_n|{\bf k}|^2}{(\gamma_{n-1}+|{\bf k}|^2)}\right)\sum_{m}\f{\beta^3\sqrt{q}}{(2\pi m)^4}\right\}\right.\non 
&+&\left.\left(\f{(n+1)(|{\bf k}|^2)^2}{4}\right)\left\{\f{1}{(\gamma_{n}+|{\bf k}|^2)^2}f(\gamma_{n},\beta)-\f{1}{(\gamma_{n+1}+|{\bf k}|^2)^2}f(\gamma_{n+1},\beta)\right.\right.\non
&+& \left.\left. \left(\f{1}{(\gamma_{n+1}+|{\bf k}|^2)^2}-\f{1}{(\gamma_{n}+|{\bf k}|^2)^2}\right)\sum_{m}\f{\beta\sqrt{q}}{(2\pi m)^2}+\left(\f{1}{(\gamma_{n}+|{\bf k}|^2)}-\f{1}{(\gamma_{n+1}+|{\bf k}|^2)}\right)\sum_{m}\f{\beta^3\sqrt{q}}{(2\pi m)^4}\right\}\right.\non 
&+&\left.\left(\f{n(|{\bf k}|^2)^2}{4}\right)\left\{\f{1}{(\gamma_{n-1}+|{\bf k}|^2)^2}f(\gamma_{n-1},\beta)-\f{1}{(\gamma_{n}+|{\bf k}|^2)^2}f(\gamma_{n},\beta)\right.\right.\non
&+& \left.\left. \left(\f{1}{(\gamma_{n}+|{\bf k}|^2)^2}-\f{1}{(\gamma_{n-1}+|{\bf k}|^2)^2}\right)\sum_{m}\f{\beta\sqrt{q}}{(2\pi m)^2}+\left(\f{1}{(\gamma_{n-1}+|{\bf k}|^2)}-\f{1}{(\gamma_{n}+|{\bf k}|^2)}\right)\sum_{m}\f{\beta^3\sqrt{q}}{(2\pi m)^4}\right\}\right.\non 
&-&\left.\left((n+1)|{\bf k}|^2\right)\left\{\left(\f{1}{\gamma_{n+1}+|{\bf k}|^2}-\f{|{\bf k}|^2}{(\gamma_{n+1}+|{\bf k}|^2)^2}\right)f(\gamma_{n+1},\beta)-\left(\f{1}{\gamma_{n}+|{\bf k}|^2}-\f{|{\bf k}|^2}{(\gamma_{n}+|{\bf k}|^2)^2}\right)f(\gamma_{n},\beta)\right.\right.\non
&+& \left.\left. \left(\f{1}{\gamma_{n}+|{\bf k}|^2}-\f{1}{\gamma_{n+1}+|{\bf k}|^2}-\f{|{\bf k}|^2}{(\gamma_{n}+|{\bf k}|^2)^2}+\f{|{\bf k}|^2}{(\gamma_{n+1}+|{\bf k}|^2)^2}\right)\sum_{m}\f{\beta\sqrt{q}}{(2\pi m)^2}\right.\right.\non
&+& \left.\left.\left(\f{|{\bf k}|^2}{(\gamma_{n}+|{\bf k}|^2)}-\f{|{\bf k}|^2}{(\gamma_{n+1}+|{\bf k}|^2)}\right)\sum_{m}\f{\beta^3\sqrt{q}}{(2\pi m)^4}\right\}\right.\non 
&-&\left.\left(n|{\bf k}|^2\right)\left\{\left(\f{1}{\gamma_{n}+|{\bf k}|^2}-\f{|{\bf k}|^2}{(\gamma_{n}+|{\bf k}|^2)^2}\right)f(\gamma_{n},\beta)-\left(\f{1}{\gamma_{n-1}+|{\bf k}|^2}-\f{|{\bf k}|^2}{(\gamma_{n-1}+|{\bf k}|^2)^2}\right)f(\gamma_{n-1},\beta)\right.\right.\non
&+& \left.\left. \left(\f{1}{\gamma_{n-1}+|{\bf k}|^2}-\f{1}{\gamma_{n}+|{\bf k}|^2}-\f{|{\bf k}|^2}{(\gamma_{n-1}+|{\bf k}|^2)^2}+\f{|{\bf k}|^2}{(\gamma_{n}+|{\bf k}|^2)^2}\right)\sum_{m}\f{\beta\sqrt{q}}{(2\pi m)^2}\right.\right.\non
&+& \left.\left.\left(\f{|{\bf k}|^2}{(\gamma_{n-1}+|{\bf k}|^2)}-\f{|{\bf k}|^2}{(\gamma_{n}+|{\bf k}|^2)}\right)\sum_{m}\f{\beta^3\sqrt{q}}{(2\pi m)^4}\right\}\right]
\eeqa

\begin{figure}[h]
\begin{center}
\begin{psfrags}
\psfrag{c2}[][]{\scalebox{0.85}{$\Phi^3_1(\tilde{\o}_{m^{''}},\tilde{k}_x^{''},\tilde{{\bf k}}^{''})$}}
\psfrag{c1}[][]{\scalebox{0.85}{$\Phi^3_1(\o_{m^{''}},k_x^{''},{\bf k}^{''})$}}
\psfrag{c3}[][]{\scalebox{0.85}{$\theta_i(m,n,{\bf k})$}}
\psfrag{c4}[][]{\scalebox{0.85}{$\theta_i(m^{'},n^{'},{\bf k}^{'})$}}
\psfrag{v1}[][]{\scalebox{0.85}{$V^1_{f}$}}
\psfrag{v2}[][]{\scalebox{0.85}{$V^{1*}_{f}$}}
\psfrag{v3}[][]{\scalebox{0.85}{$V^{1'}_{f}$}}
\psfrag{v4}[][]{\scalebox{0.85}{$V^{1'*}_{f}$}}
\psfrag{a1}[][]{(a)}
\psfrag{a2}[][]{(b)}
\includegraphics[ width= 14cm,angle=0]{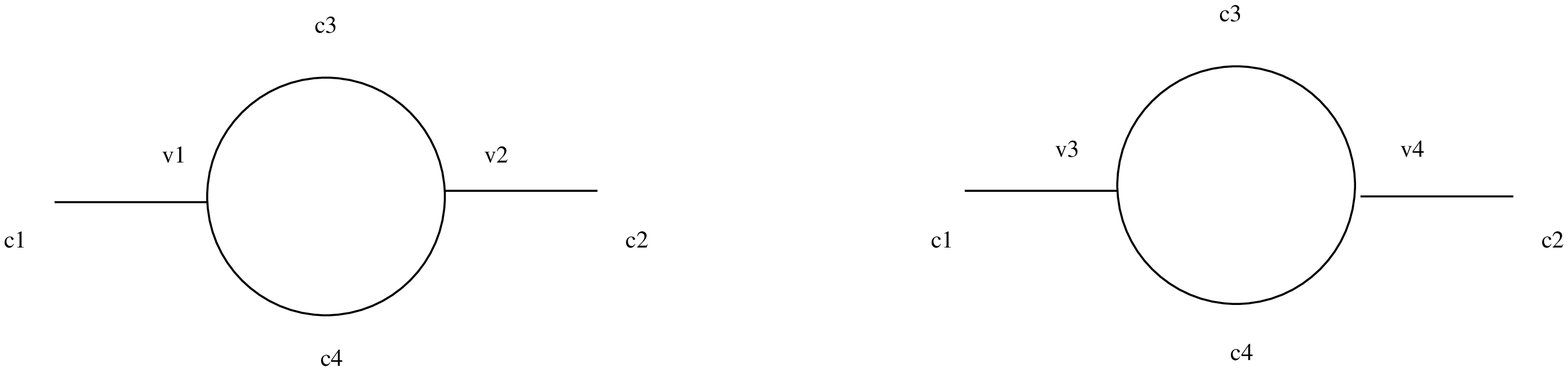}
\end{psfrags}
\caption{{Feynman diagrams involving three-point vertices $V^1_{f},~V^{1'}_f$.}}
\label{masslessp31f2}
\end{center}
\end{figure}

We now write down the contribution from fermions in the loop 

\beqa
\Sigma^3_{\Phi_1^3-\Phi_1^3}&=&4\f{N}{2}\sum_{m,n,n^{'}}\int \frac{d^2 {\bf k}}{(2\pi\sqrt{q})^2}\mbox{tr}\left[G^{1}_f(n,n^{'},k_{x}^{''})\gamma^1\frac{1}{\slashed{P}_{+}}G^{1*}_f(n,n^{'},\tilde{k}_{x}^{''})\gamma^1\frac{1}{\slashed{P}^{'}_{+}}\right.\non &-& \left.
(2) G^{1'}_f(n,n^{'},k_{x}^{''})\gamma^1\frac{1}{\slashed{P}_{+}}G^{1'*}_f(n,n^{'},\tilde{k}_{x}^{''})\gamma^1\frac{1}{\slashed{P}^{'}_{-}}
\right](2\pi)^2\delta^{2}({\bf k}^{''}+\tilde{{\bf k}}^{''})\delta_{m^{''}+\tilde{m}^{''}}
\eeqa

where propagators $\slashed{P}_{+}=i\o_m\gamma^0+\sqrt{\lambda^{'}_n}\gamma^1+k_2\gamma^2+k_3\gamma^3 ~~;~~ \slashed{P}_{+}^{'}=i\o_{m^{'}}\gamma^0+\sqrt{\lambda^{'}_{n^{'}}}\gamma^1+k_2^{'}\gamma^2+k_3^{'}\gamma^3$ 
and\\ $\slashed{P}^{'}_{-}=-i\o_m^{'}\gamma^0+\sqrt{\lambda^{'}_{n^{'}}}\gamma^1-k_2^{'}\gamma^2-k_3^{'}\gamma^3$. The momenta are related as, 
$\o_{m^{'}}=\o_m-\o_{m^{''}}$, ${\bf k}^{'}={\bf k}-{\bf k}^{''}$. The factor of $4$ in front comes from summing over the contributions from all the four sets of fermions (eqn. \ref{fermionsetsphi13}).

We will now be interested in evaluating the two point function with the external momenta set to zero. ($\o_{m^{''}}=\tilde{\o}_{m^{''}}={\bf k}^{''}=\tilde{\bf k}^{''}=0$). At zero external momenta the exact form of this contribution to the two point amplitude is

\beqa
\Sigma^3_{\Phi_1^3-\Phi_1^3}=&&2N\sum_{m,n}\int \frac{d^2 {\bf k}}{(2\pi\sqrt{q})^2}\left[\frac{\left(-\o_m^2+\sqrt{\lambda^{'}_n\lambda^{'}_{n-1}}-|{\bf k}|^2\right)}{\left(\o_m^2+\lambda^{'}_n+|{\bf k}|^2\right)\left(\o_m^2+\lambda^{'}_{n-1}+|{\bf k}|^2\right)}+
\frac{\left(-\o_m^2+\sqrt{\lambda^{'}_n\lambda^{'}_{n+1}}-|{\bf k}|^2\right)}{\left(\o_m^2+\lambda^{'}_n+|{\bf k}|^2\right)\left(\o_m^2+\lambda^{'}_{n+1}+|{\bf k}|^2\right)}\right.\non
&&- \left.
(2) \frac{\left(\o_m^2+\sqrt{\lambda^{'}_n\lambda^{'}_{n-1}}+|{\bf k}|^2\right)}{\left(\o_m^2+\lambda^{'}_n+|{\bf k}|^2\right)\left(\o_m^2+\lambda^{'}_{n-1}+|{\bf k}|^2\right)}
\right]
\eeqa

We now perform the sum over the Matsubara frequencies and
write down the the contributions as

\beqa\label{phi13fermiononeloop}
\Sigma^3_{\Phi_1^3-\Phi_1^3}&=&(2)\sum_{n=1}^{\infty}\int \frac{d^2 {\bf k}}{(2\pi\sqrt{q})^2}\left[\left\{\left(\sqrt{n(n-1)}+(n-1)\right)g(\lambda^{'}_{n-1},\beta)
-\left(\sqrt{n(n-1)}+n\right)g(\lambda^{'}_{n},\beta)\right.\right.\non &+&\left.\left.\left(\sqrt{n(n+1)}+n\right)g(\lambda^{'}_{n},\beta)
-\left(\sqrt{n(n+1)}+n+1\right)g(\lambda^{'}_{n+1},\beta)
\right\}\right.\non  &&- \left. (2) \left\{\left(\frac{\sqrt{(n-1)}}{\sqrt{n}+\sqrt{(n-1)}}\right)g(\lambda^{'}_{n-1},\beta)
+\left(\frac{\sqrt{n}}{\sqrt{n}+\sqrt{(n-1)}}\right)g(\lambda^{'}_{n},\beta)\right\}\right]\non
\eeqa

\noindent
where $g(\lambda^{'}_{n},\beta)$ is defined in equation (\ref{defg}). We can extract the zero temperature contributions from the above equation. To study the UV behaviour of the zero temperature contributions we take the large $n$ limit. In this limit the expression reduce to

\beqa\label{phi1nf}
\Sigma^3_{\Phi_1^3-\Phi_1^3}\sim -2\sum_{n}\int \frac{d^2 {\bf k}}{(2\pi\sqrt{q})^2}\frac{1}{\sqrt{2n+|{\bf k}|^2/q}}
\eeqa

Thus equations (\ref{phi1nb}) and (\ref{phi1nf}) 
show that the UV divergences cancel (for large $n$) between that contributions from the bosons and fermions in the loop.
\\
\\
\noindent
{\bf Large ${\bf k}$}

In the following we list the large ${\bf k}$ (for fixed $n$) behaviour of the integrands in $\Sigma^1_{\Phi_1^3-\Phi_1^3},\Sigma^2_{\Phi_1^3-\Phi_1^3}$ and $\Sigma^3_{\Phi_1^3-\Phi_1^3}$ at zero temperature. So the in the limit ${\bf k}>>nq$ the various expressions reduce to,

\beqa
\label{phi11largek}
&&(I) \sim  \f{5}{2}\sum_{n}\int \f{d^2 {\bf k}}{(2\pi\sqrt{q})^2} \f{\sqrt{q}}{|{\bf k}|}~~~;~~~
(II)\sim   \f{1}{4}\sum_{n}\int \f{d^2 {\bf k}}{(2\pi\sqrt{q})^2} \f{\sqrt{q}}{|{\bf k}|}~~~;~~~
(III)\sim  \f{3}{4}\sum_{n}\int \f{d^2 {\bf k}}{(2\pi\sqrt{q})^2}  \f{\sqrt{q}}{|{\bf k}|}\nonumber\\
&&(IV)=(V) \sim -\sum_{n}\int \f{d^2 {\bf k}}{(2\pi\sqrt{q})^2}\left[{\cal O}\left(\f{1}{(|{\bf k}|/\sqrt{q})^3}\right)\right]~~~;~~~
(VI)=(VII) \sim \f{1}{4}\sum_{n}\int \f{d^2 {\bf k}}{(2\pi\sqrt{q})^2} \f{\sqrt{q}}{|{\bf k}|}\nonumber\\
\eeqa

The above ${\cal O}(1/|{\bf k}|)$ contributions from the bosons in the loop in the large ${\bf k}$ limit adds up to 

\beqa
\sim 4 \sum_{n}\int \f{d^2 {\bf k}}{(2\pi\sqrt{q})^2} \f{\sqrt{q}}{|{\bf k}|}
\eeqa

This is exactly the same in magnitude and opposite in sign as the ${\cal O}(1/|{\bf k}|)$ term from (\ref{phi13fermiononeloop}). This shows that in this regime of large ${\bf k}$ and fixed $n$ the integrals converge. 

\subsection{Two point amplitude for $\Phi_I^3$ fields}\label{phii3oneloop}

There is an unbroken $SO(5)$ invariance in the theory. We thus only consider the amplitude for $\Phi_2^3$ field. Using the vertices listed in the appendix (\ref{verticesphi32}) we first compute the amplitude with bosons in the loop.

\begin{figure}[h]
\begin{center}
\begin{psfrags}
\psfrag{c1}[][]{\scalebox{0.95}{$\Phi^3_2$}}
\psfrag{c2}[][]{\scalebox{0.95}{$\Phi^3_2$}}
\psfrag{c3}[][]{$\Phi^{(1,2)}_I/\tilde{\Phi}^{(1,2)}_I(m,n,{\bf k})$}
\psfrag{c4}[][]{$C/C^{'}(m,n,{\bf k})$}
\psfrag{c5}[][]{$\tilde{A}_i^{(1,2)}(m,n,{\bf k})$}
\psfrag{v1}[][]{$V^{(2)}_{1}$}
\psfrag{v2}[][]{$V^{(2)}_{2}$}
\psfrag{v3}[][]{$\tilde{V}^{(2)ii}_{2}$}
\psfrag{a}[][]{$(I)$}
\psfrag{b}[][]{$(II)$}
\psfrag{c}[][]{$(III)$}
\includegraphics[ width= 10cm,angle=0]{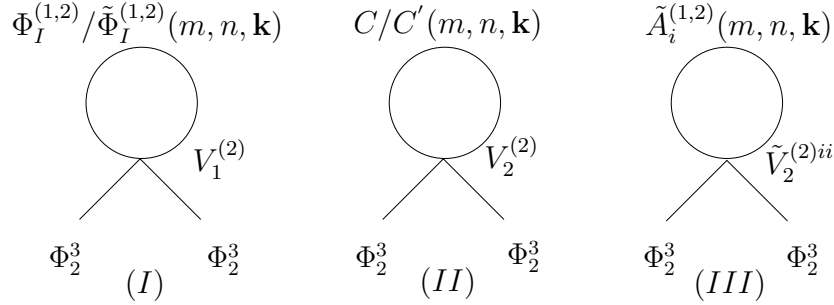}
\end{psfrags}
\caption{Feynman diagrams contributing to $\Sigma^1_{\Phi_2^3-\Phi_2^3}$. The momenta on the external $\Phi_1^3$ fields are $(\o_{m^{''}},k_x^{''},{\bf k}^{''})$ and  $({\o}_{\tilde{m}^{''}},\tilde{k}_x^{''},\tilde{{\bf k}}^{''})$}.
\label{phi24pt}
\end{center}
\end{figure}

\beqa\label{sigma1phi2}
\Sigma^1_{\Phi_2^3-\Phi_2^3}&=&\hf  N \int \f{d^2 {\bf k}}{(2\pi\sqrt{q})^2} \sum_{m,n} \left[
(4\times2)\f {G^{(2)}_1(n,n,k_x^{''},\tilde{k}_x^{''})}{\o_m^2 + \gamma_n+|{\bf k}|^2} + (2)\f {G^{(2)}_2(n,n,k_x^{''},\tilde{k}_x^{''})}{\o_m^2 + \lambda_n+|{\bf k}|^2}\right. \nonumber\\&+&\left.\sum_{i,j=1}^3\tilde{G}^{(2)ij}_2(n,n,k_x^{''},\tilde{k}_x^{''})(P^{11}_{ij}+P^{22}_{ij})
\right]\delta_{m^{''}+\tilde{m}^{''}}(2\pi)^2\delta^2({\bf k}^{''}+\tilde{\bf k}^{''})
\eeqa

\begin{figure}[h]
\begin{center}
\begin{psfrags}
\psfrag{c1}[][]{$\Phi^3_2$}
\psfrag{c2}[][]{$\Phi^3_2$}
\psfrag{c3}[][]{$\Phi^{(1,2)}_2(m,n,{\bf k})$}
\psfrag{c4}[][]{$C/C^{'}({m^{'},n^{'},{\bf k}^{'}})$}
\psfrag{c5}[][]{$\Phi^{(1,2)}_2(m,n,{\bf k})$}
\psfrag{c6}[][]{$\tilde{A}_i^{(1,2)}(m^{'},n^{'},{\bf k}^{'})$}
\psfrag{v1}[][]{$V^{(2)}_{3}$}
\psfrag{v2}[][]{$V^{(2)*}_{3}$}
\psfrag{v3}[][]{$\tilde{V}^{(2)i}_{3}$}
\psfrag{v4}[][]{$\tilde{V}^{(2)j*}_{3}$}
\psfrag{a1}[][]{$(IV)$}
\psfrag{a2}[][]{$(V)$}
\includegraphics[ width= 12cm,angle=0]{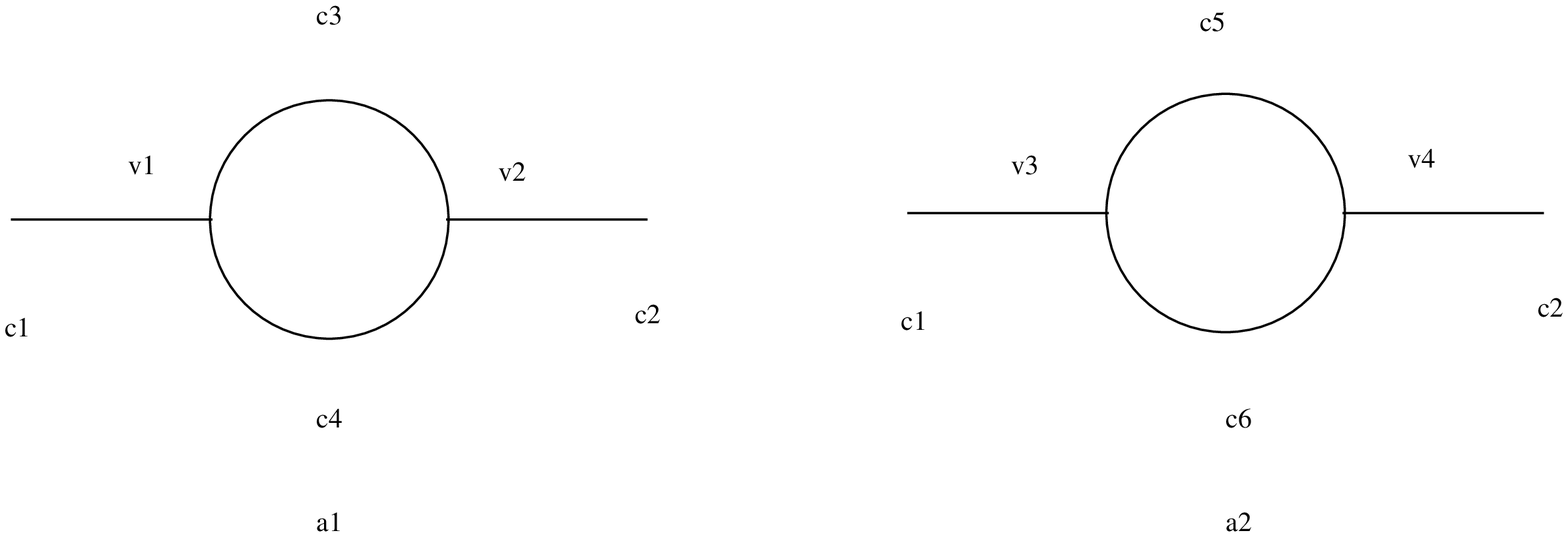}
\end{psfrags}
\caption{Feynman diagrams contributing to $\Sigma^2_{\Phi_2^3-\Phi_2^3}$. The momenta on the external $\Phi_1^3$ fields are $(\o_{m^{''}},k_x^{''},{\bf k}^{''})$ and  $({\o}_{\tilde{m}^{''}},\tilde{k}_x^{''},\tilde{{\bf k}}^{''})$}
\label{phi23pt}
\end{center}
\end{figure}

\beqa\label{sigma2phi2}
\Sigma^2_{\Phi_2^3-\Phi_2^3}&=&-\hf qN\sum_{m,n,n^{'}}\int\f{d^2 {\bf k}}{(2\pi\sqrt{q})^2}\left[(2)\f{G^{(2)}_3(n,n^{'},k_x^{''})G^{(2)}_3(n,n^{'},\tilde{k}_x^{''})}{(\o_m^2 + \gamma_n+|{\bf k}|^2)(\o_{m^{'}}^2 +\lambda_{n^{'}}+|{\bf k}^{'}|^2)} \right.\non 
&+&\left. \sum_{i,j=1}^3\f{\tilde{G}^{(2)i}_3(n,n^{'},k_x^{''})\tilde{G}^{(2)j}_3(n,n^{'},\tilde{k}_x^{''})}{(\o_m^2 +\gamma_n+|{\bf k}|^2)}(P^{11}_{ij}+P^{22}_{ij})\right]
  \delta_{m^{''}+\tilde{m}^{''}}(2\pi)^2\delta^2({\bf k}^{''}+\tilde{\bf k}^{''})
\eeqa

where $\o_{m^{'}}=\o_m-\o_{m^{''}}$, ${\bf k}^{'}={\bf k}-{\bf k}^{''}$. The vertices at zero external momentum are given by

\beqa
&&G^{(2)}_1(n,n^{'},0,0) =G^{(2)}_2(n,n^{'},0,0)=\tilde{G}^{(2)11}_2(n,n^{'},0,0)=\tilde{G}^{(2)22}_2(n,n^{'},0,0)=\tilde{G}^{(2)33}_2(n,n^{'},0,0)= \delta_{n,n^{'}}\non
&&G^{(2)}_3(n,n^{'},0,0)=0~~,~~\tilde{G}^{(2)1}_3(n,n^{'},0,0)=\mp i\sqrt{2n+1}\delta_{n,n^{'}}~~,~~\tilde{G}^{(2)i}_3(n,n^{'},0,0)=i\hat{k}_i \delta_{n,n^{'}}
\eeqa

In the above expressions $i=2,3$ and $\hat{k}_i=k_i/\sqrt{q}$.
Using these zero momentum vertices the amplitudes corresponding to the diagrams in figures (\ref{phi24pt}) and (\ref{phi23pt}) are as follows

\beqa
\Sigma^1_{\Phi_2^3-\Phi_2^3}=(I)+(II)+(III)
\eeqa

with,

\beqa
(I)=\hf N\sum_{m,n}\int\f{d^2 {\bf k}}{(2\pi\sqrt{q})^2}(4\times2)\f {1}{\o_m^2 + \gamma_n+|{\bf k}|^2}=\hf \sum_{n=0}^{\infty}\int\f{d^2 {\bf k}}{(2\pi\sqrt{q})^2}(4\times2)f(\gamma_n,\beta)
\eeqa

\beqa
(II)=\hf N\sum_{m,n}\int\f{d^2 {\bf k}}{(2\pi\sqrt{q})^2}(2)\f {1}{\o_m^2 + \lambda_n+|{\bf k}|^2}=\hf \sum_{n=2}^{\infty}\int\f{d^2 {\bf k}}{(2\pi\sqrt{q})^2}(2)f(\lambda_n,\beta)
\eeqa

\beqa
(III)&=&\hf N\sum_{m,n}\int\f{d^2 {\bf k}}{(2\pi\sqrt{q})^2}\left[(2)\f {\o_m^2+\gamma_n}{\o_m^2(\o_m^2 + \gamma_n+|{\bf k}|^2)}+(2)\f {2\o_m^2+|{\bf k}|^2}{\o_m^2(\o_m^2 + \gamma_n+|{\bf k}|^2)}\right]\\
&=&\hf \sum_{n=0}^{\infty}\int\f{d^2 {\bf k}}{(2\pi\sqrt{q})^2}(2)\left[(2)f(\gamma_n,\beta)+\sum_m\f{\beta\sqrt{q}}{(2\pi m)^2}\right]
\eeqa

and

\beqa
\Sigma^2_{\Phi_2^3-\Phi_2^3}=(IV)+(V)
\eeqa

where

\beqa
(IV)=0
\eeqa

\beqa
(V)&=&-\hf N\sum_{m,n}\int\f{d^2 {\bf k}}{(2\pi\sqrt{q})^2}\left[(2)\f {\gamma_n(\o_m^2+\gamma_n)}{\o_m^2(\o_m^2 + \gamma_n+|{\bf k}|^2)^2}+(2)\f {|{\bf k}|^2(\o_m^2+|{\bf k}|^2)}{\o_m^2(\o_m^2 + \gamma_n+|{\bf k}|^2)^2}+(4)\f {\gamma_n|{\bf k}|^2}{\o_m^2(\o_m^2 + \gamma_n+|{\bf k}|^2)^2}\right]\non
&=&-\hf N\sum_{m,n}\int\f{d^2 {\bf k}}{(2\pi\sqrt{q})^2}(2)\left[\f {1}{\o_m^2}-\f {1}{\o_m^2 + \gamma_n+|{\bf k}|^2}\right]\non
&=&-\hf \sum_{n}\int\f{d^2 {\bf k}}{(2\pi\sqrt{q})^2}(2)\left[\sum_m\f{\beta\sqrt{q}}{(2\pi m)^2}-f(\gamma_n,\beta)\right]
\eeqa

In the large $n$ limit the above amplitude reduces to

\beqa
\Sigma^1_{\Phi_2^3-\Phi_2^3}+\Sigma^2_{\Phi_2^3-\Phi_2^3}&\sim& 8N\sum_{m,n}\int\f{d^2 {\bf k}}{(2\pi\sqrt{q})^2}\f{1}{(\o_m^2 + 2nq+|{\bf k}|^2)}\\ &\sim& 4\sum_{n}\int \frac{d^2 {\bf k}}{(2\pi\sqrt{q})^2}\frac{1}{\sqrt{2n+|{\bf k}|^2/q}}~~\mbox{(zero temperature contribution)}
\eeqa

\begin{figure}[h]
\begin{center}
\begin{psfrags}
\psfrag{c2}[][]{\scalebox{0.85}{$\Phi^3_2(\tilde{\o}_{m^{''}},\tilde{k}_x^{''},\tilde{{\bf k}}^{''})$}}
\psfrag{c1}[][]{\scalebox{0.85}{$\Phi^3_2(\o_{m^{''}},k_x^{''},{\bf k}^{''})$}}
\psfrag{c3}[][]{\scalebox{0.85}{$\theta_1(m,n,{\bf k})$}}
\psfrag{c4}[][]{\scalebox{0.85}{$\theta_2(m^{'},n^{'},{\bf k}^{'})$}}
\psfrag{v1}[][]{\scalebox{0.85}{$V^2_{f}$}}
\psfrag{v2}[][]{\scalebox{0.85}{$V^{2*}_{f}$}}
\psfrag{v3}[][]{\scalebox{0.85}{$V^{2'}_{f}$}}
\psfrag{v4}[][]{\scalebox{0.85}{$V^{2'*}_{f}$}}
\psfrag{a1}[][]{(a)}
\psfrag{a2}[][]{(b)}
\includegraphics[ width= 14cm,angle=0]{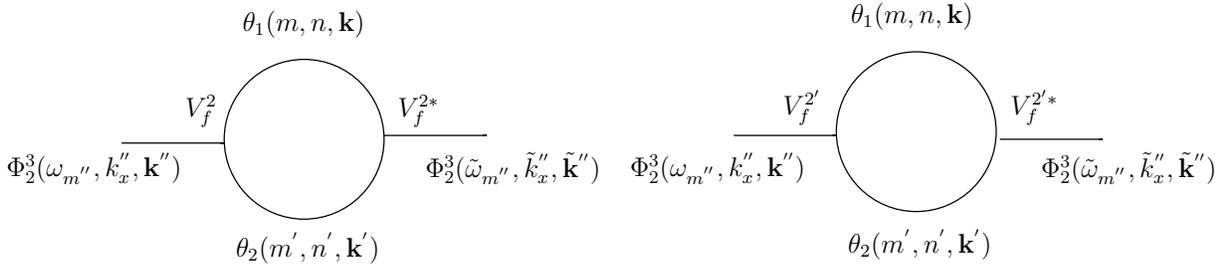}
\end{psfrags}
\caption{{Feynman diagrams involving three-point vertices $V^2_{f},~V^{2'}_f$.}}
\label{masslessp32f}
\end{center}
\end{figure} 

We now turn to the fermions. The contribution from fermions in the loop is given by 

\beqa
\Sigma^3_{\Phi_2^3-\Phi_2^3}&=&2N\sum_{m,n,n^{'}}\int \frac{d^2 {\bf k}}{(2\pi\sqrt{q})^2}\mbox{tr}\left[G^{2}_f(n,n^{'},k_{x}^{''})\gamma^1\frac{1}{\slashed{P}_{+}}G^{2*}_f(n,n^{'},\tilde{k}_{x}^{''})\gamma^1\frac{1}{\slashed{P}^{'}_{+}}\right.\non &-& \left.
 G^{2'}_f(n,n^{'},k_{x}^{''})\gamma^1\frac{1}{\slashed{P}_{+}}G^{2'*}_f(n,n^{'},\tilde{k}_{x}^{''})\gamma^1\frac{1}{\slashed{P}^{'}_{-}}
\right](2\pi)^2\delta^{2}({\bf k}^{''}+\tilde{{\bf k}}^{''})\delta_{m^{''}+\tilde{m}^{''}}
\eeqa

where propagators $\slashed{P}_{+}=i\o_m\gamma^0+\sqrt{\lambda^{'}_n}\gamma^1+k_2\gamma^2+k_3\gamma^3 ~~;~~ \slashed{P}_{+}^{'}=i\o_{m^{'}}\gamma^0+\sqrt{\lambda^{'}_{n^{'}}}\gamma^1+k_2^{'}\gamma^2+k_3^{'}\gamma^3$ 
and\\ $\slashed{P}^{'}_{-}=-i\o_m^{'}\gamma^0+\sqrt{\lambda^{'}_{n^{'}}}\gamma^1-k_2^{'}\gamma^2-k_3^{'}\gamma^3$. The momenta are related as, 
$\o_{m^{'}}=\o_m-\o_{m^{''}}$, ${\bf k}^{'}={\bf k}-{\bf k}^{''}$. The factor of $2$ in front is due to the contributions from the two sets of fermions (eqn. \ref{fermionsetsphi23}).

For zero external momenta ($\o_{m^{''}}=\tilde{\o}_{m^{''}}={\bf k}^{''}=\tilde{\bf k}^{''}=0$), $G^{2}_f(n,n^{'},0)=G^{2*}_f(n,n^{'},0)=0$. So the amplitude simplifies as

\beqa\label{phi23fermiononeloop}
\Sigma^3_{\Phi_2^3-\Phi_2^3}&=&-8N\sum_{m,n}\int \frac{d^2 {\bf k}}{(2\pi\sqrt{q})^2}\f{1}{\left(\o_m^2+\lambda^{'}_n+|{\bf k}|^2\right)}=-8\sum_{n=1}^{\infty}\int \frac{d^2 {\bf k}}{(2\pi\sqrt{q})^2}g(\lambda^{'}_n, \beta)
\eeqa

In the above equation we have summed over the Matsubara frequencies. $g(\lambda^{'}_n, \beta)$ is given by eqn. (\ref{defg}). 
To show that UV divergence cancellation we take the limit $\beta \rightarrow \infty$. The zero temperature contribution is then

\beqa
\Sigma^3_{\Phi_2^3-\Phi_2^3} \sim -4\sum_{n}\int \frac{d^2 {\bf k}}{(2\pi\sqrt{q})^2}\frac{1}{\sqrt{2n+|{\bf k}|^2/q}}
\eeqa

This is opposite in sign to the contribution from the bosons in the loop thus showing the finiteness of the one-loop amplitude in the UV.
\\
\\
\noindent
{\bf Large ${\bf k}$}

As we have done before we list behaviour of the integrands in $\Sigma^1_{\Phi_2^3-\Phi_2^3},\Sigma^2_{\Phi_2^3-\Phi_2^3}$ and $\Sigma^3_{\Phi_2^3-\Phi_2^3}$ at zero temperature in the limit ${\bf k}>>nq$.

\beqa
\label{phi23largek}
&&(I) \sim  2\sum_{n}\int \f{d^2 {\bf k}}{(2\pi\sqrt{q})^2} \f{\sqrt{q}}{|{\bf k}|}~~~;~~~
(II)\sim   \f{1}{2}\sum_{n}\int \f{d^2 {\bf k}}{(2\pi\sqrt{q})^2} \f{\sqrt{q}}{|{\bf k}|}~~~;~~~
(III)\sim  \sum_{n}\int \f{d^2 {\bf k}}{(2\pi\sqrt{q})^2}  \f{\sqrt{q}}{|{\bf k}|}\nonumber\\
&&(IV)=0~~~;~~~
(V) \sim \f{1}{2}\sum_{n}\int \f{d^2 {\bf k}}{(2\pi\sqrt{q})^2} \f{\sqrt{q}}{|{\bf k}|}\nonumber\\
\eeqa

Adding these give  

\beqa
\sim 4 \sum_{n}\int \f{d^2 {\bf k}}{(2\pi\sqrt{q})^2} \f{\sqrt{q}}{|{\bf k}|}
\eeqa

We also get the same ${\cal O}(1/|{\bf k}|)$ contribution from (\ref{phi23fermiononeloop}) having opposite sign.

\subsection{Two point amplitude for $A_{\mu}^3$ fields}\label{amu3oneloop}

In this section we compute the two point amplitude for $A_{\mu}^3$ fields. We compute the amplitudes for $\mu=1$ and $\mu \ne 1$ in the following two subsections.

\subsubsection{$\mu=1$ : Bosons}\label{a13bosons}

In this part we compute the contribution with bosons in the loop for the two-point function for the $A_{1}^3$ 

Using the vertices listed in appendix (\ref{verticesA3mu}), we first write down the contributions to the one loop amplitude from bosons in the loop,

\begin{figure}[h]
\begin{center}
\begin{psfrags}
\psfrag{c1}[][]{\scalebox{0.95}{$A^3_1$}}
\psfrag{c2}[][]{\scalebox{0.95}{$A^3_1$}}
\psfrag{c3}[][]{$\Phi^{(1,2)}_I/\tilde{\Phi}^{(1,2)}_I(m,n,{\bf k})$}
\psfrag{c4}[][]{$C/C^{'}(m,n,{\bf k})$}
\psfrag{c5}[][]{$\tilde{A}_i^{(1,2)}(m,n,{\bf k})$}
\psfrag{v1}[][]{$V^{(A_1^3)}_{1}$}
\psfrag{v2}[][]{$V^{(A_1^3)}_{2}$}
\psfrag{v3}[][]{$\tilde{V}^{(A_1^3)ii}_{2}$}
\psfrag{a}[][]{$(I)$}
\psfrag{b}[][]{$(II)$}
\psfrag{c}[][]{$(III)$}
\includegraphics[ width= 10cm,angle=0]{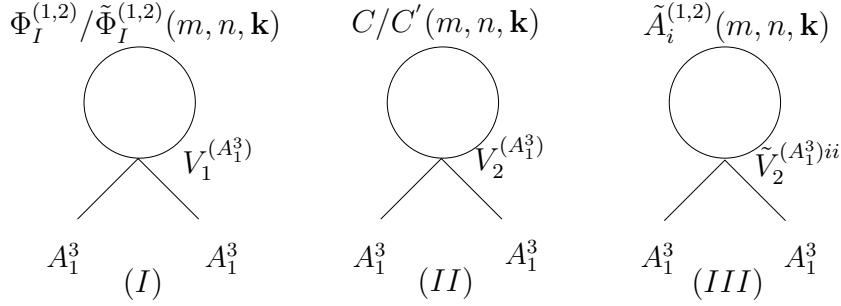}
\end{psfrags}
\caption{Feynman diagrams contributing to $\Sigma^1_{A_1^3-A_1^3}$. The momenta on the external $A_1^3$ fields are $(\o_{m^{''}},k_x^{''},{\bf k}^{''})$ and  $({\o}_{\tilde{m}^{''}},\tilde{k}_x^{''},\tilde{{\bf k}}^{''})$}.
\label{A14pt}
\end{center}
\end{figure}

The contribution consisting of the four-point vertices as shown in Figure \ref{A14pt} is given by,

\beqa\label{sigma1A1}
\Sigma^1_{A_1^3-A_1^3}&=&\hf  N \int \f{d^2 {\bf k}}{(2\pi\sqrt{q})^2} \sum_{m,n} \left[
(5\times2)\f {G^{(A_1^3)}_1(n,n,k_x^{''},\tilde{k}_x^{''})}{\o_m^2 + \gamma_n+|{\bf k}|^2} + (2)\f {G^{(A_1^3)}_2(n,n,k_x^{''},\tilde{k}_x^{''})}{\o_m^2 + \lambda_n+|{\bf k}|^2}\right. \nonumber\\&+&\left.\sum_{i,j=1}^3\tilde{G}^{(A_1^3)ij}_2(n,n,k_x^{''},\tilde{k}_x^{''})(P^{11}_{ij}+P^{22}_{ij})
\right]\delta_{m^{''}+\tilde{m}^{''}}(2\pi)^2\delta^2({\bf k}^{''}+\tilde{\bf k}^{''})
\eeqa

\begin{figure}[h]
\begin{center}
\begin{psfrags}
\psfrag{c1}[][]{$A^3_1$}
\psfrag{c2}[][]{$A^3_1$}
\psfrag{c5}[][]{$\Phi^{(1,2)}_I$}
\psfrag{c6}[][]{$\Phi^{(1,2)}_I$}
\psfrag{c3}[][]{$C$}
\psfrag{c4}[][]{$C^{'}$}
\psfrag{c7}[][]{$C/C^{'}$}
\psfrag{c8}[][]{$\tilde{A}_i^{(1,2)}$}
\psfrag{c9}[][]{$\tilde{A}_i^{(1,2)}$}
\psfrag{c10}[][]{$\tilde{A}_j^{(1,2)}$}
\psfrag{v1}[][]{$V^{(A_1^3)}_{3}$}
\psfrag{v2}[][]{$V^{(A_1^3)*}_{3}$}
\psfrag{v3}[][]{${V}^{(A_1^3)}_{4}$}
\psfrag{v4}[][]{$V^{(A_1^3)*}_{4}$}
\psfrag{v5}[][]{$\tilde{V}^{(A_1^3)i}_{3}$}
\psfrag{v6}[][]{$\tilde{V}^{(A_1^3)j*}_{3}$}
\psfrag{v7}[][]{$\tilde{V}^{(A_1^3)ij}_{3}$}
\psfrag{v8}[][]{$\tilde{V}^{(A_1^3)ij*}_{3}$}
\psfrag{a1}[][]{$(IV)$}
\psfrag{a2}[][]{$(VI)$}
\psfrag{a3}[][]{$(V)$}
\psfrag{a4}[][]{$(VII)$}
\includegraphics[ width= 12cm,angle=0]{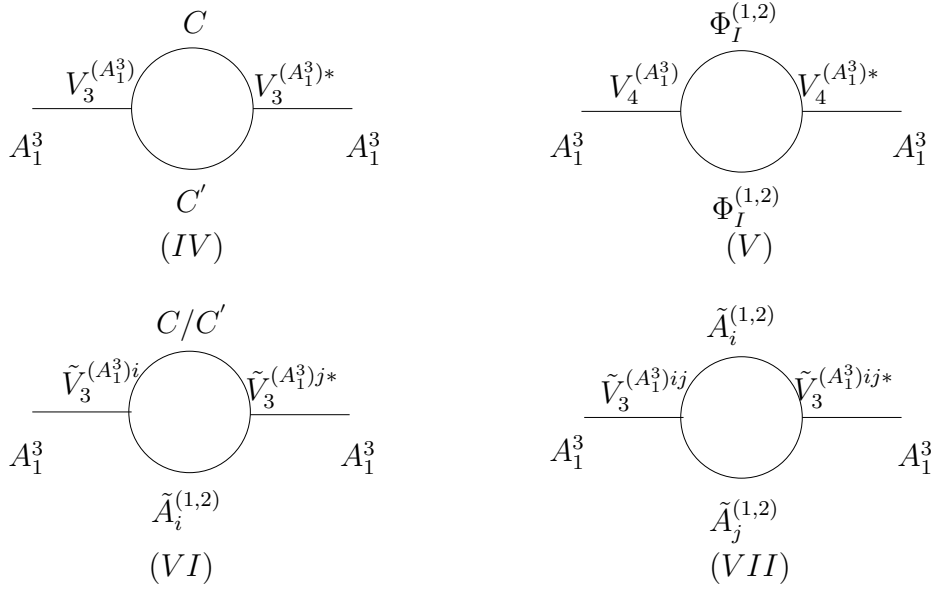}
\end{psfrags}
\caption{Feynman diagrams contributing to $\Sigma^2_{A_1^3-A_1^3}$. The momenta on the external $A_1^3$ fields are $(\o_{m^{''}},k_x^{''},{\bf k}^{''})$ and  $({\o}_{\tilde{m}^{''}},\tilde{k}_x^{''},\tilde{{\bf k}}^{''})$}
\label{A13pt}
\end{center}
\end{figure} 

and the contributions from the three-point vertices as shown in Figure \ref{A13pt}) is 

\beqa\label{sigma2A1}
\Sigma^2_{A_1^3-A_1^3}&=&-\hf qN\sum_{m,n,n^{'}}\int\f{d^2 {\bf k}}{(2\pi\sqrt{q})^2}\left[\f{G^{(1)}_3(n,n^{'},k_x^{''})G^{(1)}_3(n,n^{'},\tilde{k}_x^{''})}{(\o_m^2 + \lambda_n+|{\bf k}|^2)(\o_{m^{'}}^2 +\lambda_{n^{'}}+|{\bf k}^{'}|^2)} \right.\non 
&+& \left.(5)\f{G^{(1)}_4(n,n^{'},k_x^{''})G^{(1)}_4(n,n^{'},\tilde{k}_x^{''})}{(\o_m^2 + \gamma_n+|{\bf k}|^2)(\o_{m^{'}}^2 +\gamma_{n^{'}}+|{\bf k}^{'}|^2)}+ \sum_{i,j=1}^3\f{\tilde{G}^{(1)i}_3(n,n^{'},k_x^{''})\tilde{G}^{(1)j}_3(n,n^{'},\tilde{k}_x^{''})}{(\o_m^2 +\lambda_n+|{\bf k}|^2)}(P^{11}_{ij}+P^{22}_{ij})\right.\non 
&+& \left.\sum_{i,i^{'},j,j^{'}=1}^3\tilde{G}^{(1)ij}_3(n,n^{'},k_x^{''})\tilde{G}^{(1)i^{'}j^{'}}_3(n,n^{'},\tilde{k}_x^{''})(P^{11}_{ii^{'}}P^{22}_{jj^{'}})\right]
  \delta_{m^{''}+\tilde{m}^{''}}(2\pi)^2\delta^2({\bf k}^{''}+\tilde{\bf k}^{''})
\eeqa

where $\o_{m^{'}}=\o_m-\o_{m^{''}}$, ${\bf k}^{'}={\bf k}-{\bf k}^{''}$. Since we are interested in the amplitudes at zero external momentum  we list the various exact functions at this momentum below

\beqa\label{vertexAn}
&&G^{(A_1^3)}_1(n,n^{'},0,0) = \delta_{n,n^{'}},~~ 
G^{(A_1^3)}_2(n,n^{'},0,0)=\hf\delta_{n,n^{'}},~~
\tilde{G}^{(A_1^3)1}_2(n,n^{'},0,0)=\hf\delta_{n,n^{'}},~~\tilde{G}^{(A_1^3)i}_2(n,n^{'},0,0)=\delta_{n,n^{'}},\non
&&G^{(A_1^3)}_3(0,n,n^{'})=2 \left(\sqrt{\f{2n(n+1)(n-1)}{(2n-1)(2n+1)}} \delta_{n+1,n^{'}}-\sqrt{\f{2n(n-1)(n-2)}{(2n-1)(2n-3)}} \delta_{n-1,n^{'}}\right),\non
&&\tilde{G}^{(A_1^3)i}_3(0,n,n^{'}) =  -i\hat{k}_i\left(
 \f{\sqrt{(n-1)}}{\sqrt{2(2n-1)}} \delta_{n,n^{'}}+\f{\sqrt{n}}{\sqrt{2(2n-1)}} \delta_{n-2,n^{'}} \right),\non
&&\tilde{G}^{(A_1^3)1}_3(0,n,n^{'}) = \mp i\left(
 \f{\sqrt{2n}(n-1)}{\sqrt{(2n-1)(2n-3)}} \delta_{n-2,n^{'}}+\f{\sqrt{2(n-1)}(n+1)}{\sqrt{(2n-1)(2n+1)}} \delta_{n,n^{'}} \right),~~\tilde{G}^{(1)11}_3(0,n,n^{'})= 0,\non
&&\tilde{G}^{(A_1^3)1i}_3(0,n,n^{'}) =  \pm\hat{k}_i\left(
 \f{\sqrt{(n+1)}}{\sqrt{2(2n+1)}} \delta_{n+1,n^{'}}-\f{\sqrt{n}}{\sqrt{2(2n+1)}} \delta_{n-1,n^{'}} \right)
,\non
&&\tilde{G}^{(A_1^3)ii}_3(0,n,n^{'})=G^{(A_1^3)}_4(0,n,n^{'})=\left(\sqrt{2(n+1)}\delta_{n+1,n^{'}}-\sqrt{2n}\delta_{n-1,n^{'}}\right)
\eeqa

In the above expressions $i=2,3$ and $\hat{k}_i=k_i/\sqrt{q}$.
Using the above vertices (eqn. (\ref{vertexAn})) we can now write down the 
the contributions from the expressions (\ref{sigma1A1}) and (\ref{sigma2A1}) in the large $n$ limit. 
These are exactly equal to the expressions obtained for the $\Sigma^1_{\Phi_1^3-\Phi_1^3}$ and $\Sigma^2_{\Phi_1^3-\phi_1^3}$, eqns. (\ref{sigma1phi1n}) and  (\ref{sigma2phi1n}). Thus the full contribution $\Sigma^1_{A_1^3-A_1^3}+\Sigma^2_{A_1^3-A_1^3}$ in the large $n$ limit is given by
eqn. (\ref{phi1nb}). However the exact form (at all $n$) for some of the terms of the two point contributions differ from that of the $\Phi_1^3$ amplitude. This is because of the difference in the forms of the vertices given by eqns. (\ref{vertexAn}) and (\ref{vertexphi1n}). We thus give the contributions corresponding to the different Feynman diagrams in the Figures \ref{A14pt} and \ref{A13pt} separately below. 

\beqa
\Sigma^1_{A_1^3-A_1^3}=(I)+(II)+(III)
\eeqa

This is same as the the contribution obtained for $\Sigma^1_{\Phi_1^3-\Phi_1^3}$ where, $(I)$, $(II)$ and $(III)$ are given by eqns. (\ref{phi1I}) and (\ref{phi1III}).

We now write down the contribution towards the two point amplitude amplitude from the three point vertices as shown in Figure \ref{A13pt}. Using (\ref{deff}), this is given by

\beqa
\Sigma^2_{A_1^3-A_1^3}=(IV)+(V)+(VI)+(VII)
\eeqa

where,

\beqa
(IV)&=&-\hf\sum_{n=2}^{\infty}\int \f{d^2 {\bf k}}{(2\pi\sqrt{q})^2}\left[\left(\f{4n(n-1)(n-2)}{(2n-1)(2n-3)}\right)\left(f(\lambda_{n-1},\beta)-f(\lambda_{n},\beta)\right)\right.\\
&+&\left.\left(\f{4n(n+1)(n-1)}{(2n-1)(2n+1)}\right)\left(f(\lambda_{n},\beta)-f(\lambda_{n+1},\beta)\right)\right]\nonumber
\eeqa

\beqa
(V)=-\hf\sum_{n=0}^{\infty}\int \f{d^2 {\bf k}}{(2\pi\sqrt{q})^2}(5)\left[(n)\left(f(\gamma_{n-1},\beta)-f(\gamma_{n},\beta)\right)+(n+1)\left(f(\gamma_{n},\beta)-f(\gamma_{n+1},\beta)\right)\right]
\eeqa

$(VI)$ is same as the expression obtained for $\Phi_1^3$ case given by eqn. (\ref{phi1VI}).

\beqa
&&(VII)=-\hf\sum_{n=0}^{\infty}\int \f{d^2 {\bf k}}{(2\pi\sqrt{q})^2}\times\\
&\times&\left[\left(2n\right)\left\{\left(1-\f{|{\bf k}|^2}{\gamma_{n-1}+|{\bf k}|^2}+\hf\f{(|{\bf k}|^2)^2}{(\gamma_{n-1}+|{\bf k}|^2)^2}\right)f(\gamma_{n-1},\beta)-\left(1-\f{|{\bf k}|^2}{\gamma_{n}+|{\bf k}|^2}+\hf\f{(|{\bf k}|^2)^2}{(\gamma_{n}+|{\bf k}|^2)^2}\right)f(\gamma_{n},\beta)
\right.\right.\non  &+& \left.\left. \left(\f{|{\bf k}|^2}{\gamma_{n-1}+|{\bf k}|^2}-\f{|{\bf k}|^2}{\gamma_{n}+|{\bf k}|^2}+\f{(|{\bf k}|^2)^2}{2(\gamma_{n}+|{\bf k}|^2)^2}-\f{(|{\bf k}|^2)^2}{2(\gamma_{n-1}+|{\bf k}|^2)^2}\right)\sum_{m}\f{\beta\sqrt{q}}{(2\pi m)^2}\right.\right.\non 
&+& \left.\left.\left(\f{(|{\bf k}|^2)^2}{2(\gamma_{n-1}+|{\bf k}|^2)}-\f{(|{\bf k}|^2)^2}{2(\gamma_{n}+|{\bf k}|^2)}\right)\sum_{m}\f{\beta^3\sqrt{q}}{(2\pi m)^4}\right\}\right.\non 
&+& \left.2(n+1)\left\{\left(1-\f{|{\bf k}|^2}{\gamma_{n}+|{\bf k}|^2}+\hf\f{(|{\bf k}|^2)^2}{(\gamma_{n}+|{\bf k}|^2)^2}\right)f(\gamma_{n},\beta)-\left(1-\f{|{\bf k}|^2}{\gamma_{n+1}+|{\bf k}|^2}+\hf\f{(|{\bf k}|^2)^2}{(\gamma_{n+1}+|{\bf k}|^2)^2}\right)f(\gamma_{n+1},\beta)
\right.\right.\non  &+& \left.\left. \left(\f{|{\bf k}|^2}{\gamma_{n}+|{\bf k}|^2}-\f{|{\bf k}|^2}{\gamma_{n+1}+|{\bf k}|^2}+\f{(|{\bf k}|^2)^2}{2(\gamma_{n+1}+|{\bf k}|^2)^2}-\f{(|{\bf k}|^2)^2}{2(\gamma_{n}+|{\bf k}|^2)^2}\right)\sum_{m}\f{\beta\sqrt{q}}{(2\pi m)^2}\right.\right.\non 
&+& \left.\left.\left(\f{(|{\bf k}|^2)^2}{2(\gamma_{n}+|{\bf k}|^2)}-\f{(|{\bf k}|^2)^2}{2(\gamma_{n+1}+|{\bf k}|^2)}\right)\sum_{m}\f{\beta^3\sqrt{q}}{(2\pi m)^4}\right\}\right.\non 
&+& \left.\left(\f{(n+1)|{\bf k}|^2/q}{2(2n+1)}\right)\left\{ \f{\gamma_n|{\bf k}|^2}{(\gamma_{n}+|{\bf k}|^2)^2}f(\gamma_{n},\beta)-\left(1-\f{\gamma_n+|{\bf k}|^2}{(\gamma_{n+1}+|{\bf k}|^2)}+\f{\gamma_n|{\bf k}|^2}{(\gamma_{n+1}+|{\bf k}|^2)^2}\right)f(\gamma_{n+1},\beta)\right.\right.\non
&+& \left.\left.\left(1-\f{\gamma_n+|{\bf k}|^2}{\gamma_{n+1}+|{\bf k}|^2}+\f{\gamma_n|{\bf k}|^2}{(\gamma_{n+1}+|{\bf k}|^2)^2}-\f{\gamma_n|{\bf k}|^2}{(\gamma_{n}+|{\bf k}|^2)^2}\right)\sum_{m}\f{\beta\sqrt{q}}{(2\pi m)^2}+\left(\f{\gamma_n|{\bf k}|^2}{(\gamma_{n}+|{\bf k}|^2)}-\f{\gamma_n|{\bf k}|^2}{(\gamma_{n+1}+|{\bf k}|^2)}\right)\sum_{m}\f{\beta^3\sqrt{q}}{(2\pi m)^4}\right\}\right.\non 
&+&\left.\left(\f{n|{\bf k}|^2/q}{2(2n+1)}\right)\left\{ \left(1-\f{\gamma_n+|{\bf k}|^2}{\gamma_{n-1}+|{\bf k}|^2}+\f{\gamma_n|{\bf k}|^2}{(\gamma_{n-1}+|{\bf k}|^2)^2}\right)f(\gamma_{n-1},\beta)-\left(\f{\gamma_n|{\bf k}|^2}{(\gamma_{n}+|{\bf k}|^2)^2}\right)f(\gamma_{n},\beta)\right.\right.\non
&-& \left.\left.\left(1-\f{\gamma_n+|{\bf k}|^2}{\gamma_{n-1}+|{\bf k}|^2}+\f{\gamma_n|{\bf k}|^2}{(\gamma_{n-1}+|{\bf k}|^2)^2}-\f{\gamma_n|{\bf k}|^2}{(\gamma_{n}+|{\bf k}|^2)^2}\right)\sum_{m}\f{\beta\sqrt{q}}{(2\pi m)^2}-\left(\f{\gamma_n|{\bf k}|^2}{(\gamma_{n}+|{\bf k}|^2)}-\f{\gamma_n|{\bf k}|^2}{(\gamma_{n-1}+|{\bf k}|^2)}\right)\sum_{m}\f{\beta^3\sqrt{q}}{(2\pi m)^4}\right\}\right.\non 
&+&\left.\left(\f{(n+1)(|{\bf k}|^2)^2}{2}\right)\left\{\f{1}{(\gamma_{n}+|{\bf k}|^2)^2}f(\gamma_{n},\beta)-\f{1}{(\gamma_{n+1}+|{\bf k}|^2)^2}f(\gamma_{n+1},\beta)\right.\right.\non
&+& \left.\left. \left(\f{1}{(\gamma_{n+1}+|{\bf k}|^2)^2}-\f{1}{(\gamma_{n}+|{\bf k}|^2)^2}\right)\sum_{m}\f{\beta\sqrt{q}}{(2\pi m)^2}+\left(\f{1}{(\gamma_{n}+|{\bf k}|^2)}-\f{1}{(\gamma_{n+1}+|{\bf k}|^2)}\right)\sum_{m}\f{\beta^3\sqrt{q}}{(2\pi m)^4}\right\}\right.\non 
&+&\left.\left(\f{n(|{\bf k}|^2)^2}{2}\right)\left\{\f{1}{(\gamma_{n-1}+|{\bf k}|^2)^2}f(\gamma_{n-1},\beta)-\f{1}{(\gamma_{n}+|{\bf k}|^2)^2}f(\gamma_{n},\beta)\right.\right.\non
&+& \left.\left. \left(\f{1}{(\gamma_{n}+|{\bf k}|^2)^2}-\f{1}{(\gamma_{n-1}+|{\bf k}|^2)^2}\right)\sum_{m}\f{\beta\sqrt{q}}{(2\pi m)^2}+\left(\f{1}{(\gamma_{n-1}+|{\bf k}|^2)}-\f{1}{(\gamma_{n}+|{\bf k}|^2)}\right)\sum_{m}\f{\beta^3\sqrt{q}}{(2\pi m)^4}\right\}\right.\non 
&-&\left.\left(2(n+1)|{\bf k}|^2\right)\left\{\left(\f{1}{\gamma_{n+1}+|{\bf k}|^2}-\f{|{\bf k}|^2}{(\gamma_{n+1}+|{\bf k}|^2)^2}\right)f(\gamma_{n+1},\beta)-\left(\f{1}{\gamma_{n}+|{\bf k}|^2}-\f{|{\bf k}|^2}{(\gamma_{n}+|{\bf k}|^2)^2}\right)f(\gamma_{n},\beta)\right.\right.\non
&+& \left.\left. \left(\f{1}{\gamma_{n}+|{\bf k}|^2}-\f{1}{\gamma_{n+1}+|{\bf k}|^2}-\f{|{\bf k}|^2}{(\gamma_{n}+|{\bf k}|^2)^2}+\f{|{\bf k}|^2}{(\gamma_{n+1}+|{\bf k}|^2)^2}\right)\sum_{m}\f{\beta\sqrt{q}}{(2\pi m)^2}\right.\right.\non
&+& \left.\left.\left(\f{|{\bf k}|^2}{(\gamma_{n}+|{\bf k}|^2)}-\f{|{\bf k}|^2}{(\gamma_{n+1}+|{\bf k}|^2)}\right)\sum_{m}\f{\beta^3\sqrt{q}}{(2\pi m)^4}\right\}\right.\non 
&-&\left.\left(2n|{\bf k}|^2\right)\left\{\left(\f{1}{\gamma_{n}+|{\bf k}|^2}-\f{|{\bf k}|^2}{(\gamma_{n}+|{\bf k}|^2)^2}\right)f(\gamma_{n},\beta)-\left(\f{1}{\gamma_{n-1}+|{\bf k}|^2}-\f{|{\bf k}|^2}{(\gamma_{n-1}+|{\bf k}|^2)^2}\right)f(\gamma_{n-1},\beta)\right.\right.\non
&+& \left.\left. \left(\f{1}{\gamma_{n-1}+|{\bf k}|^2}-\f{1}{\gamma_{n}+|{\bf k}|^2}-\f{|{\bf k}|^2}{(\gamma_{n-1}+|{\bf k}|^2)^2}+\f{|{\bf k}|^2}{(\gamma_{n}+|{\bf k}|^2)^2}\right)\sum_{m}\f{\beta\sqrt{q}}{(2\pi m)^2}\right.\right.\non
&+& \left.\left.\left(\f{|{\bf k}|^2}{(\gamma_{n-1}+|{\bf k}|^2)}-\f{|{\bf k}|^2}{(\gamma_{n}+|{\bf k}|^2)}\right)\sum_{m}\f{\beta^3\sqrt{q}}{(2\pi m)^4}\right\}\right]
\eeqa

\subsubsection{$\mu=2,3$ : Bosons}\label{a233bosons}

\begin{figure}[h]
\begin{center}
\begin{psfrags}
\psfrag{c1}[][]{\scalebox{0.95}{$A^3_2$}}
\psfrag{c2}[][]{\scalebox{0.95}{$A^3_2$}}
\psfrag{c3}[][]{$\Phi^{(1,2)}_I/\tilde{\Phi}^{(1,2)}_I(m,n,{\bf k})$}
\psfrag{c4}[][]{$C/C^{'}(m,n,{\bf k})$}
\psfrag{c5}[][]{$\tilde{A}_i^{(1,2)}(m,n,{\bf k})$}
\psfrag{v1}[][]{$V^{(A_2^3)}_{1}$}
\psfrag{v2}[][]{$V^{(A_2^3)}_{2}$}
\psfrag{v3}[][]{$\tilde{V}^{(A_2^3)ii}_{2}$}
\psfrag{a}[][]{$(I)$}
\psfrag{b}[][]{$(II)$}
\psfrag{c}[][]{$(III)$}
\includegraphics[ width= 10cm,angle=0]{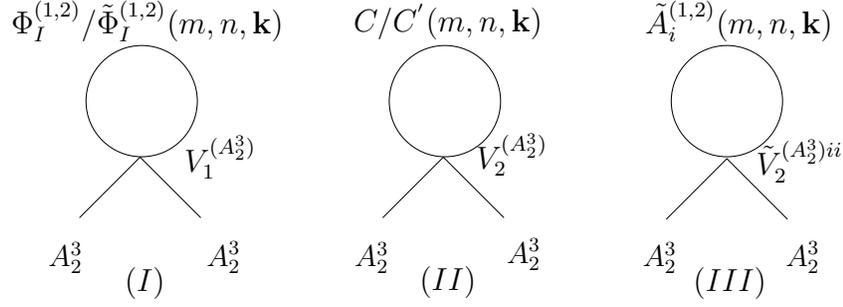}
\end{psfrags}
\caption{Feynman diagrams contributing to $\Sigma^1_{A_2^3-A_2^3}$. The momenta on the external $A_2^3$ fields are $(\o_{m^{''}},k_x^{''},{\bf k}^{''})$ and  $({\o}_{\tilde{m}^{''}},\tilde{k}_x^{''},\tilde{{\bf k}}^{''})$}.
\label{A24pt}
\end{center}
\end{figure}

We now compute the contribution with bosons in the loop for the two-point function for the $A_{2}^3$. Due to the underlying symmetry the amplitude for $\mu=3$ is same as the one for $\mu=2$. Using the vertices listed in appendix (\ref{verticesA3mu}) the contribution consisting of the four-point vertices as shown in Figure \ref{A24pt} is given by,

\beqa\label{sigma1A2}
\Sigma^1_{A_2^3-A_2^3}&=&\hf  N \int \f{d^2 {\bf k}}{(2\pi\sqrt{q})^2} \sum_{m,n} \left[
(5\times2)\f {G^{(A_2^3)}_1(n,n,k_x^{''},\tilde{k}_x^{''})}{\o_m^2 + \gamma_n+|{\bf k}|^2} + (2)\f {G^{(A_2^3)}_2(n,n,k_x^{''},\tilde{k}_x^{''})}{\o_m^2 + \lambda_n+|{\bf k}|^2}\right. \nonumber\\&+&\left.\sum_{i,j=1,3}\tilde{G}^{(A_2^3)ij}_2(n,n,k_x^{''},\tilde{k}_x^{''})(P^{11}_{ij}+P^{22}_{ij})
\right]\delta_{m^{''}+\tilde{m}^{''}}(2\pi)^2\delta^2({\bf k}^{''}+\tilde{\bf k}^{''})
\eeqa

\begin{figure}[h]
\begin{center}
\begin{psfrags}
\psfrag{c1}[][]{$A^3_2$}
\psfrag{c2}[][]{$A^3_2$}
\psfrag{c5}[][]{$\Phi^{(1,2)}_I$}
\psfrag{c6}[][]{$\Phi^{(1,2)}_I$}
\psfrag{c3}[][]{$C$}
\psfrag{c4}[][]{$C^{'}$}
\psfrag{c7}[][]{$C/C^{'}$}
\psfrag{c8}[][]{$\tilde{A}_i^{(1,2)}$}
\psfrag{c9}[][]{$\tilde{A}_i^{(1,2)}$}
\psfrag{c10}[][]{$\tilde{A}_j^{(1,2)}$}
\psfrag{v1}[][]{$V^{(A_2^3)}_{3}$}
\psfrag{v2}[][]{$V^{(A_2^3)*}_{3}$}
\psfrag{v3}[][]{${V}^{(A_2^3)}_{4}$}
\psfrag{v4}[][]{$V^{(A_2^3)*}_{4}$}
\psfrag{v5}[][]{$\tilde{V}^{(A_2^3)i}_{3}$}
\psfrag{v6}[][]{$\tilde{V}^{(A_2^3)j*}_{3}$}
\psfrag{v7}[][]{$\tilde{V}^{(A_2^3)ij}_{3}$}
\psfrag{v8}[][]{$\tilde{V}^{(A_2^3)ij*}_{3}$}
\psfrag{a1}[][]{$(IV)$}
\psfrag{a2}[][]{$(VI)$}
\psfrag{a3}[][]{$(V)$}
\psfrag{a4}[][]{$(VII)$}
\includegraphics[ width= 12cm,angle=0]{massless3pointb.eps}
\end{psfrags}
\caption{Feynman diagrams contributing to $\Sigma^2_{A_2^3-A_2^3}$. The momenta on the external $A_2^3$ fields are $(\o_{m^{''}},k_x^{''},{\bf k}^{''})$ and  $({\o}_{\tilde{m}^{''}},\tilde{k}_x^{''},\tilde{{\bf k}}^{''})$}
\label{A23pt}
\end{center}
\end{figure} 

and the contributions from the three-point vertices as shown in Figure \ref{A23pt} is 

\beqa\label{sigma2A2}
\Sigma^2_{A_2^3-A_2^3}&=&-\hf qN\sum_{m,n,n^{'}}\int\f{d^2 {\bf k}}{(2\pi\sqrt{q})^2}\left[\f{G^{(A_2^3)}_3(n,n^{'},k_x^{''})G^{(A_2^3)}_3(n,n^{'},\tilde{k}_x^{''})}{(\o_m^2 + \lambda_n+|{\bf k}|^2)(\o_{m^{'}}^2 +\lambda_{n^{'}}+|{\bf k}^{'}|^2)} \right.\non 
&+& \left.(5)\f{G^{(A_2^3)}_4(n,n^{'},k_x^{''})G^{(A_2^3)}_4(n,n^{'},\tilde{k}_x^{''})}{(\o_m^2 + \gamma_n+|{\bf k}|^2)(\o_{m^{'}}^2 +\gamma_{n^{'}}+|{\bf k}^{'}|^2)}+ \sum_{i,j=1,2}\f{\tilde{G}^{(A_2^3)i}_3(n,n^{'},k_x^{''})\tilde{G}^{(A_2^3)j}_3(n,n^{'},\tilde{k}_x^{''})}{(\o_m^2 +\lambda_n+|{\bf k}|^2)}(P^{11}_{ij}+P^{22}_{ij})\right.\non 
&+& \left.\sum_{i,i^{'},j,j^{'}=1}^3\tilde{G}^{(A_2^3)ij}_3(n,n^{'},k_x^{''})\tilde{G}^{(A_2^3)i^{'}j^{'}}_3(n,n^{'},\tilde{k}_x^{''})(P^{11}_{ii^{'}}P^{22}_{jj^{'}})\right]
  \delta_{m^{''}+\tilde{m}^{''}}(2\pi)^2\delta^2({\bf k}^{''}+\tilde{\bf k}^{''})
\eeqa

where $\o_{m^{'}}=\o_m-\o_{m^{''}}$, ${\bf k}^{'}={\bf k}-{\bf k}^{''}$. At zero external momentum the various functions have the following forms

\beqa\label{vertexA2n}
&&G^{(A_2^3)}_1(n,n^{'},0,0) = \delta_{n,n^{'}},~~ 
G^{(A_2^3)}_2(n,n^{'},0,0)=\hf\delta_{n,n^{'}},~~
\tilde{G}^{(A_1^3)1}_2(n,n^{'},0,0)=\hf\delta_{n,n^{'}},~~\tilde{G}^{(A_2^3)3}_2(n,n^{'},0,0)=\delta_{n,n^{'}},\non
&&G^{(A_2^3)}_3(0,n,n^{'})=\tilde{G}^{(A_2^3)1}_3(0,n,n^{'})=i(\hat{k}_2^{'}-\hat{k}_2)\delta_{n,n^{'}},~~\tilde{G}^{(A_2^3)2}_3(0,n,n^{'})=0,~~
\tilde{G}^{(A_2^3)12}_3(0,n,n^{'})=-i\sqrt{2n+1}\delta_{n,n^{'}}\non
&&\tilde{G}^{(A_2^3)32}_3(0,n,n^{'})=-\tilde{G}^{(A_2^3)23}_3(0,n,n^{'})=i(\hat{k}_3^{'}-\hat{k}_2^{''})\delta_{n,n^{'}},~~\tilde{G}^{(A_2^3)33}_3(0,n,n^{'})=G^{(A_2^3)}_4(0,n,n^{'})=i(\hat{k}_2^{'}-\hat{k}_2)\delta_{n,n^{'}}\non
\eeqa

In the above expressions $\hat{k}_i=k_i/\sqrt{q}$.
Using these vertices the various diagrams in Figures \ref{A24pt} and \ref{A23pt} reduce to

\beqa
(I)=\hf  N \sum_{m,n} \int \f{d^2 {\bf k}}{(2\pi\sqrt{q})^2} (5\times 2)\f{1}{\o_{m}^2 +\gamma_{n}+|{\bf k}|^2}=\hf   \sum_{n=0}^{\infty} \int \f{d^2 {\bf k}}{(2\pi\sqrt{q})^2} (5\times 2)f(\gamma_n, \beta)
\eeqa

\beqa
(II)=\hf  N \sum_{m,n} \int \f{d^2 {\bf k}}{(2\pi\sqrt{q})^2} (2)\f{1}{\o_{m}^2 +\lambda_{n}+|{\bf k}|^2}=\hf   \sum_{n=2}^{\infty} \int \f{d^2 {\bf k}}{(2\pi\sqrt{q})^2} (2)f(\lambda_n, \beta)
\eeqa

\beqa
(III)&=&\hf  N \sum_{m,n} \int \f{d^2 {\bf k}}{(2\pi\sqrt{q})^2} (2)\left[\f{\o_m^2+\gamma_n}{\o_m^2(\o_{m}^2 +\gamma_{n}+|{\bf k}|^2)}
+\f{\o_m^2+k_3^2}{\o_m^2(\o_{m}^2 +\gamma_{n}+|{\bf k}|^2)}\right]\non
&=&\hf   \sum_{n=0}^{\infty} \int \f{d^2 {\bf k}}{(2\pi\sqrt{q})^2} (2)\left[\left(1+\hf\f{|{\bf k}|^2}{\gamma_n+|{\bf k}|^2}\right)f(\gamma_n, \beta)+
\left(1-\hf\f{|{\bf k}|^2}{\gamma_n+|{\bf k}|^2}\right)\sum_{m}\f{\beta\sqrt{q}}{(2\pi m)^2}\right]
\eeqa

so that

\beqa
\Sigma^1_{A_2^3-A_2^3}=(I)+(II)+(III)
\eeqa

Similarly,

\beqa
\Sigma^2_{A_2^3-A_2^3}=(IV)+(V)+(VI)+(VII)
\eeqa

where

\beqa
(IV)=-\hf N\sum_{m,n}\int\f{d^2 {\bf k}}{(2\pi\sqrt{q})^2}\f{4k_2^2}{(\o_{m}^2 +\lambda_{n}+|{\bf k}|^2)^2}=-\hf \sum_{n=2}^{\infty}\int\f{d^2 {\bf k}}{(2\pi\sqrt{q})^2}\left(-2|{\bf k}|^2f^{'}(\lambda_n, \beta)\right)
\eeqa

\beqa
(V)=-\hf N\sum_{m,n}\int\f{d^2 {\bf k}}{(2\pi\sqrt{q})^2}(5)\f{4k_2^2}{(\o_{m}^2 +\gamma_{n}+|{\bf k}|^2)^2}=-\hf \sum_{n=0}^{\infty}\int\f{d^2 {\bf k}}{(2\pi\sqrt{q})^2}(5)\left(-2|{\bf k}|^2f^{'}(\gamma_n, \beta)\right)
\eeqa

\beqa
(VI)=0
\eeqa

\beqa
(VII)&=&-\hf N\sum_{m,n}\int\f{d^2 {\bf k}}{(2\pi\sqrt{q})^2}\f{A+B\o_m^2}{(\o_m^2)(\o_{m}^2 +\gamma_{n}+|{\bf k}|^2)^2}\\
&=&-\hf \sum_{n=0}^{\infty}\int\f{d^2 {\bf k}}{(2\pi\sqrt{q})^2}\left[\left(\f{A}{\gamma_{n}+|{\bf k}|^2}-B\right)f^{'}(\gamma_n, \beta)-\left(\f{A}{(\gamma_{n}+|{\bf k}|^2)^2}\right)\left(f(\gamma_n, \beta)-\sum_{m}\f{\beta\sqrt{q}}{(2\pi m)^2}\right)\right]\nonumber
\eeqa

where

\beqa
A&=& 2\gamma_n^2+3\gamma_n|{\bf k}|^2+|{\bf k}|^4\non
B&=& 2\gamma_n+5|{\bf k}|^2\nonumber
\eeqa

It is now not difficult to show that in the large $n$ limit the full contribution to the two point amplitude reduces to

\beqa
\Sigma^1_{A_2^3-A_2^3}+\Sigma^2_{A_2^3-A_2^3}\sim 8 N\sum_{m,n}\int\f{d^2 {\bf k}}{(2\pi\sqrt{q})^2}\left[\f{1}{\o_{m}^2 +2nq+|{\bf k}|^2}-
\f{|{\bf k}|^2}{(\o_{m}^2 +2nq+|{\bf k}|^2)^2}\right]
\eeqa
\\
\\
At zero temperature in the limit ${\bf k}>>nq$ the various ${\cal O}(1/|{\bf k}|)$ contributions reduce to,

\beqa
\label{A23largek}
&&(I) \sim  \f{5}{2}\sum_{n}\int \f{d^2 {\bf k}}{(2\pi\sqrt{q})^2} \f{\sqrt{q}}{|{\bf k}|}~~~;~~~
(II)\sim   \f{1}{2}\sum_{n}\int \f{d^2 {\bf k}}{(2\pi\sqrt{q})^2} \f{\sqrt{q}}{|{\bf k}|}~~~;~~~
(III)\sim  \f{3}{4}\sum_{n}\int \f{d^2 {\bf k}}{(2\pi\sqrt{q})^2}  \f{\sqrt{q}}{|{\bf k}|}\nonumber\\
&&(IV) \sim -\f{1}{4}\sum_{n}\int \f{d^2 {\bf k}}{(2\pi\sqrt{q})^2}  \f{\sqrt{q}}{|{\bf k}|}~~~;~~~
(V) \sim -\f{5}{4}\sum_{n}\int \f{d^2 {\bf k}}{(2\pi\sqrt{q})^2}  \f{\sqrt{q}}{|{\bf k}|}\nonumber\\
&&(VI)=0~~~;~~~(VII)\sim -\f{1}{4}\sum_{n}\int \f{d^2 {\bf k}}{(2\pi\sqrt{q})^2}  \f{\sqrt{q}}{|{\bf k}|}
\eeqa

which adds up to

\beqa\label{A23largek1}
\sim 2 \sum_{n}\int \f{d^2 {\bf k}}{(2\pi\sqrt{q})^2} \f{\sqrt{q}}{|{\bf k}|}.
\eeqa

\subsubsection{$\mu=1$ : Fermions}\label{a13fermions}

\begin{figure}[h]
\begin{center}
\begin{psfrags}
\psfrag{c2}[][]{\scalebox{0.85}{$A^3_{\mu}(\tilde{\o}_{m^{''}},\tilde{k}_x^{''},\tilde{{\bf k}}^{''})$}}
\psfrag{c1}[][]{\scalebox{0.85}{$A^3_{\mu}(\o_{m^{''}},k_x^{''},{\bf k}^{''})$}}
\psfrag{c3}[][]{\scalebox{0.85}{$\theta_1(m,n,{\bf k})$}}
\psfrag{c4}[][]{\scalebox{0.85}{$\theta_4(m^{'},n^{'},{\bf k}^{'})$}}
\psfrag{v1}[][]{\scalebox{0.85}{$V^{\mu}_{f}$}}
\psfrag{v2}[][]{\scalebox{0.85}{$V^{\mu*}_{f}$}}
\psfrag{v3}[][]{\scalebox{0.85}{$V^{'\mu}_{f}$}}
\psfrag{v4}[][]{\scalebox{0.85}{$V^{'\mu*}_{f}$}}
\psfrag{a1}[][]{(a)}
\psfrag{a2}[][]{(b)}
\includegraphics[ width= 14cm,angle=0]{massless3pointf.eps}
\end{psfrags}
\caption{{Feynman diagrams involving three-point vertices $V^{\mu}_{f},~V^{'\mu}_f$.}}
\label{masslessp32f}
\end{center}
\end{figure} 

The contribution from fermions in the loop is given by 

\beqa\label{amuf}
\Sigma^3_{A_{\mu}^3-A_{\mu}^3}&=&(-1)^{\mu}2N\sum_{m,n,n^{'}}\int \frac{d^2 {\bf k}}{(2\pi\sqrt{q})^2}\mbox{tr}\left[G_f(n,n^{'},k_{x}^{''};\mu)\gamma^{\mu}\gamma^1\frac{1}{\slashed{P}_{+}}G_f(n,n^{'},\tilde{k}_{x}^{''};\mu)\gamma^{\mu}\gamma^1\frac{1}{\slashed{P}^{'}_{+}}\right.\non &-& \left.
 G^{'}_f(n,n^{'},k_{x}^{''};\mu)\gamma^{\mu}\gamma^1\frac{1}{\slashed{P}_{+}}G^{'*}_f(n,n^{'},\tilde{k}_{x}^{''};\mu)\gamma^{\mu}\gamma^1\frac{1}{\slashed{P}^{'}_{-}}
\right](2\pi)^2\delta^{2}({\bf k}^{''}+\tilde{{\bf k}}^{''})\delta_{m^{''}+\tilde{m}^{''}}
\eeqa

where $(-1)^{\mu}=+1$ for $\mu=2,3$ and $(-1)^{\mu}=-1$ for $\mu=1$. The propagators $\slashed{P}_{+}=i\o_m\gamma^0+\sqrt{\lambda^{'}_n}\gamma^1+k_2\gamma^2+k_3\gamma^3 ~~;~~ \slashed{P}_{+}^{'}=i\o_{m^{'}}\gamma^0+\sqrt{\lambda^{'}_{n^{'}}}\gamma^1+k_2^{'}\gamma^2+k_3^{'}\gamma^3$ 
and $\slashed{P}^{'}_{-}=-i\o_m^{'}\gamma^0+\sqrt{\lambda^{'}_{n^{'}}}\gamma^1-k_2^{'}\gamma^2-k_3^{'}\gamma^3$. The momenta are related as, 
$\o_{m^{'}}=\o_m-\o_{m^{''}}$, ${\bf k}^{'}={\bf k}-{\bf k}^{''}$. The factor of $2$ in front is due to the contributions from the two sets of fermions (see eqn. \ref{fermionsetsA3}).

For zero external momenta ($\o_{m^{''}}=\tilde{\o}_{m^{''}}={\bf k}^{''}=\tilde{\bf k}^{''}=0$), we write down the amplitudes separately for $\mu=1$ and $\mu=2,3$ below.

\beqa
\Sigma^3_{A_1^3-A_1^3}&=&2N\sum_{m,n}\int \frac{d^2 {\bf k}}{(2\pi\sqrt{q})^2}\left[\frac{\left(-\o_m^2+\sqrt{\lambda^{'}_n\lambda^{'}_{n-1}}-|{\bf k}|^2\right)}{\left(\o_m^2+\lambda^{'}_n+|{\bf k}|^2\right)\left(\o_m^2+\lambda^{'}_{n-1}+|{\bf k}|^2\right)}-\frac{\left(\o_m^2+\sqrt{\lambda^{'}_n\lambda^{'}_{n-1}}+|{\bf k}|^2\right)}{\left(\o_m^2+\lambda^{'}_n+|{\bf k}|^2\right)\left(\o_m^2+\lambda^{'}_{n-1}+|{\bf k}|^2\right)}
\right.\non
&+&  \left.\frac{\left(-\o_m^2+\sqrt{\lambda^{'}_n\lambda^{'}_{n+1}}-|{\bf k}|^2\right)}{\left(\o_m^2+\lambda^{'}_n+|{\bf k}|^2\right)\left(\o_m^2+\lambda^{'}_{n+1}+|{\bf k}|^2\right)}-\frac{\left(\o_m^2+\sqrt{\lambda^{'}_n\lambda^{'}_{n+1}}+|{\bf k}|^2\right)}{\left(\o_m^2+\lambda^{'}_n+|{\bf k}|^2\right)\left(\o_m^2+\lambda^{'}_{n+1}+|{\bf k}|^2\right)}\right]
\eeqa

Summing over the Matsubara frequencies we write down various contributions and using the form for $g(\lambda^{'}_{n},\beta)$ given by (\ref{defg}), we get

\beqa
&~&\Sigma^3_{A_{1}^3-A_{1}^3}=2\sum_{n=1}^{\infty}\int \frac{d^2 {\bf k}}{(2\pi\sqrt{q})^2}\left[\left\{\left(\sqrt{n(n-1)}+(n-1)\right)g(\lambda^{'}_{n-1},\beta)-\left(\sqrt{n(n-1)}+n\right)g(\lambda^{'}_{n},\beta)\right.\right.\non
&-&\left.\left.\left(\sqrt{n(n+1)}+(n+1)\right)g(\lambda^{'}_{n+1},\beta)+\left(\sqrt{n(n+1)}+n\right)g(\lambda^{'}_{n},\beta)\right\}\right.\non
&-& \left. \left\{\left(\f{\sqrt{(n-1)}}{\sqrt{n}+\sqrt{(n-1)}}\right)g(\lambda^{'}_{n-1},\beta)+\left(\frac{\sqrt{n}}{\sqrt{n}+\sqrt{(n-1)}}\right)g(\lambda^{'}_{n},\beta)\right.\right.\non
&+& \left.\left. \left(\f{\sqrt{(n+1)}}{\sqrt{n}+\sqrt{(n+1)}}\right)g(\lambda^{'}_{n+1},\beta)+\left(\frac{\sqrt{n}}{\sqrt{n}+\sqrt{(n+1)}}\right)g(\lambda^{'}_{n},\beta)\right\}\right]
\eeqa

\noindent
In the large $n$ limit and at zero temperature ($\beta \rightarrow \infty$) the above contribution has the form

\beqa
\Sigma^3_{A_{1}^3-A_{1}^3}\sim -2\sum_{n}\int \frac{d^2 {\bf k}}{(2\pi\sqrt{q})^2}\frac{1}{\sqrt{2n+|{\bf k}|^2/q}}
\eeqa

This is equal in magnitude to the contribution from bosons in the loop but with opposite sign thus showing that the amplitude is ultraviolet finite.
A similar analysis such as the one that was done for the two-point amplitude for the $\Phi_1^3$ field at zero temperature and in the limit ${\bf k}>>nq$ can again be done here. The ${\cal O}(1/|{\bf k}|)$ contributions from the bosons and the fermions in the loop can be shown to cancel as done exactly at the end of section (\ref{phi13oneloop}).

\subsubsection{$\mu=2,3$ : Fermions}\label{a233fermions}

From eqn. (\ref{amuf}) we can now write down the contribution from the fermions in the loop for $\mu=2,3$

\beqa
\Sigma^3_{A_{\mu}^3-A_{\mu}^3}=-8N\sum_{m,n}\int \frac{d^2 {\bf k}}{(2\pi\sqrt{q})^2}\f{\o_m^2+\lambda^{'}_n}{\left(\o_m^2+\lambda^{'}_n+|{\bf k}|^2\right)^2}
\eeqa

Now summing over the Matsubara frequencies we arrive at the following expressions

\beqa
\Sigma^3_{A_{\mu}^3-A_{\mu}^3}=-8\sum_{n=1}^{\infty}\int \frac{d^2 {\bf k}}{(2\pi\sqrt{q})^2}\left[g(\lambda^{'}_n,\beta)+|{\bf k}|^2g^{'}(\lambda^{'}_n,\beta)\right]
\eeqa

\noindent
where $g(\lambda^{'}_n,\beta)$ is defined in (\ref{defg}). From this the zero temperature contribution in the large $n$ limit is

\beqa
\Sigma^3_{A_{\mu}^3-A_{\mu}^3}\sim -4\sum_{n}\int \frac{d^2 {\bf k}}{(2\pi\sqrt{q})^2}\f{2n }{\left(2n+|{\bf k}|^2/q\right)^{3/2}}
\eeqa

Thus confirming the cancellation with the contribution from bosons in the loop. Similarly we can expand the integrand at zero temperature and in the limit ${\bf k}>>nq$, which gives

\beqa
\Sigma^3_{A_{\mu}^3-A_{\mu}^3} \sim -2 \sum_{n}\int \f{d^2 {\bf k}}{(2\pi\sqrt{q})^2} \f{\sqrt{q}}{|{\bf k}|}.
\eeqa

Comparing with (\ref{A23largek1}) we see similar convergence of the full amplitude for large ${\bf k}$ with $nq$ fixed. 

\section{Numerical Results}\label{numerics}

This section is devoted to study the two-point functions for the massless as well as the Tachyon fields numerically. The analytical expressions for the two-point amplitudes derived in the previous sections can be reduced to the case of intersecting $D2$ branes by setting one of the momenta transverse to the $x$ direction as zero. However here we restrict ourselves only to the case of intersecting $D3$ branes. We give the numerical results for the tree-level massless modes and the Tachyon in separate subsections. As explained before the corrections to the masses for the tree-level massless modes are needed to cure the the infrared divergences in the Tachyon two-point amplitude. The computations are done for the Yang-Mills coupling, $g^2=1/100$. 

\subsection{Numerical Results: Tree-level massless modes}

In this section we discuss the numerical computation of the two-point functions for the modes that are massless at tree-level, namely the $\Psi \equiv (\Phi_1^3, \Phi_I^3~(I=2,\cdots,6), A_1^3~ \mbox{and}~ A_i^3~(i=2,3))$ fields. The corresponding analytical results are given in section \ref{masslessoneloop}. Explicit analytical computations are done for $I=2$ and $i=2$ in the earlier sections. Because of the remaining $SO(5)$ and $SO(2)$ symmetries the two point functions for other values of $I$ and $i$ are the same. 

The small $|{\bf k}|$ expansions of the integrands in the two point functions, $g^2(\Sigma^1_{\Psi-\Psi}+\Sigma^2_{\Psi-\Psi}+\Sigma^3_{\Psi-\Psi})$ (as functions of $|{\bf k}|$) for the various fields at zero temperature 
are as follows
 
\beqa
&&\Phi_1^3: 0.0196642 - 0.00164676~|{\bf k}|^2/q + {\cal O}((|{\bf k}|^2/q)^2)\\
&&\Phi_I^3: 0.0317415 - 0.0254426~|{\bf k}|^2/q + {\cal O}((|{\bf k}|^2/q)^2)~~~(I=2,\cdots,6) \\
&&A_1^3:  0.0249911 - 0.00457901~|{\bf k}|^2/q + {\cal O}((|{\bf k}|^2/q)^2)\\
&&A_i^3 : 0.0317415 - 0.0508852 ~|{\bf k}|^2/q + {\cal O}((|{\bf k}|^2/q)^2) ~~~(i=2,3)
\eeqa 

The leading terms in the above expansions gives the zero temperature corrections for the intersecting $D1$ branes. Also note that the leading
terms for $\Phi_I^3$ and $A_i^3$ are the same as in 1+1 dimensions $A_i^3$ combine with $\Phi_I^3$ into a $SO(7)$ multiplet. 

It has been shown that in the large $|{\bf k}|$ limit, the contribution to the two-point functions from the bosons and fermions in the loop cancel each other. Thus the amplitude is finite in large $|{\bf k}|$. For each of the fields this has been shown towards the end of sections \ref{phi13oneloop}, \ref{phii3oneloop} and \ref{amu3oneloop}. The zero temperature quantum corrections computed numerically gives the following values for the masses, $m^2/q$

\beqa
\Phi_1^3: 0.0391796~~\Phi^3_I: 0.0182371~~A^3_1: 0.0373535~~A^3_2:0.0000239
\eeqa

\begin{figure}[htbp]
  \centering
\begin{psfrags}
\psfrag{m}[][]{\scalebox{0.8}{$m^2_{\Phi_1^3}/q$}}
\psfrag{b}[][]{\scalebox{0.8}{$\sqrt{q}\beta$}}
  \begin{minipage}[b]{0.3\textwidth}
    \includegraphics[width=\textwidth]{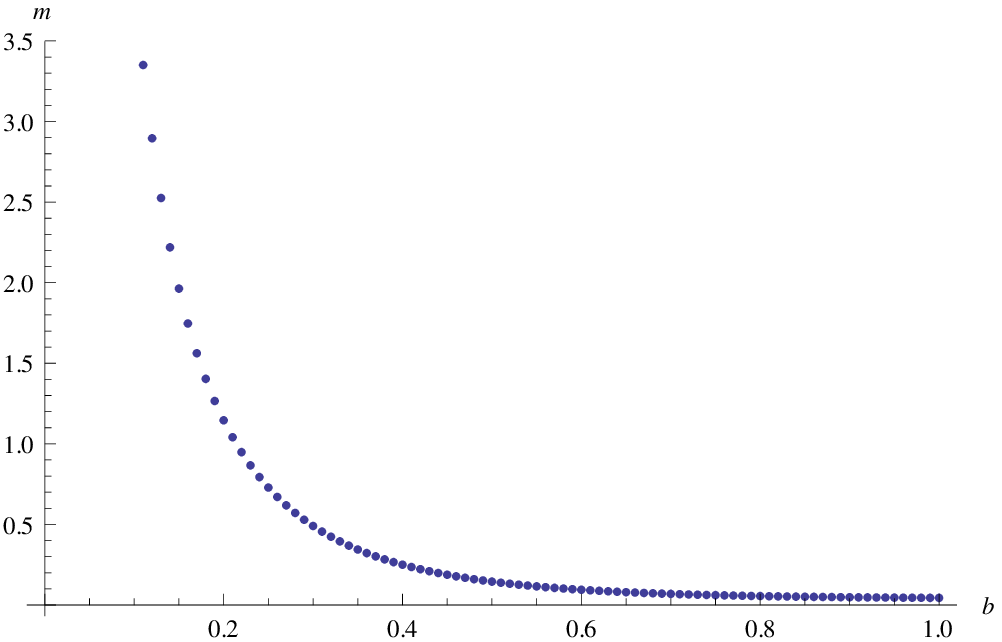}
    \caption{Plot of $m^2_{\Phi_1^3}/q$ vs $\sqrt{q}\beta$}
\label{m13b}  
\end{minipage}
\end{psfrags}  
\hspace{0.5in}
\begin{psfrags}
\psfrag{m}[][]{\scalebox{0.8}{$m^2_{\Phi_1^3}/q$}}  
\psfrag{t}[][]{\scalebox{0.8}{$\frac{T}{\sqrt{q}}$}}
\begin{minipage}[b]{0.3\textwidth}
    \includegraphics[width=\textwidth]{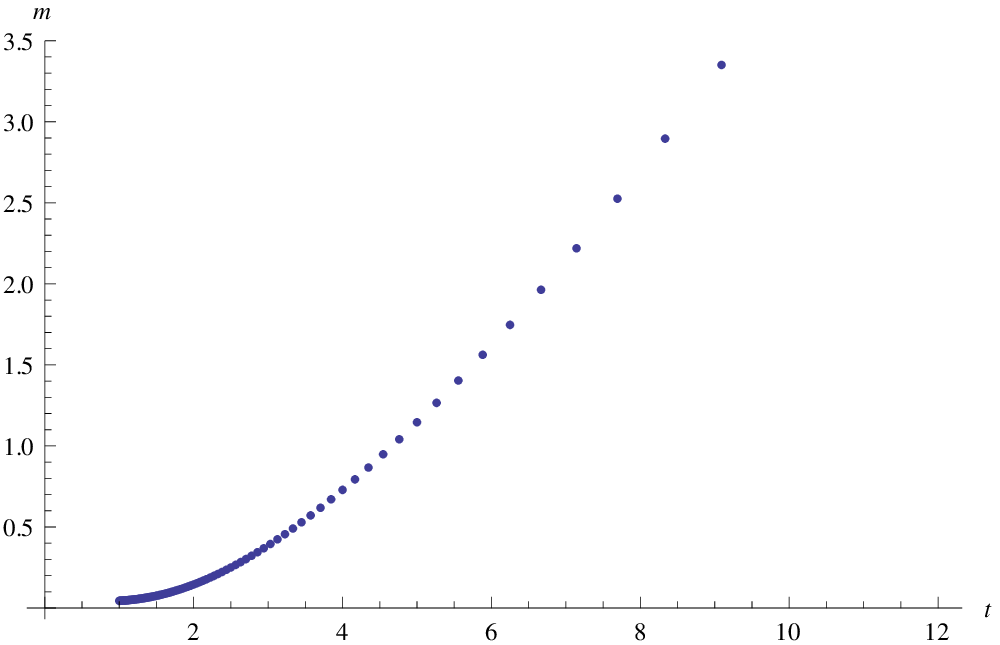}
    \caption{Plot of $m^2_{\Phi_1^3}/q$ vs $\frac{T}{\sqrt{q}}$}
\label{m13t}   
\end{minipage}
\end{psfrags}
\end{figure}

\begin{figure}[htbp]
  \centering
\begin{psfrags}
\psfrag{m}[][]{\scalebox{0.8}{$m^2_{\Phi_I^3}/q$}}
\psfrag{b}[][]{\scalebox{0.8}{$\sqrt{q}\beta$}}
  \begin{minipage}[b]{0.3\textwidth}
    \includegraphics[width=\textwidth]{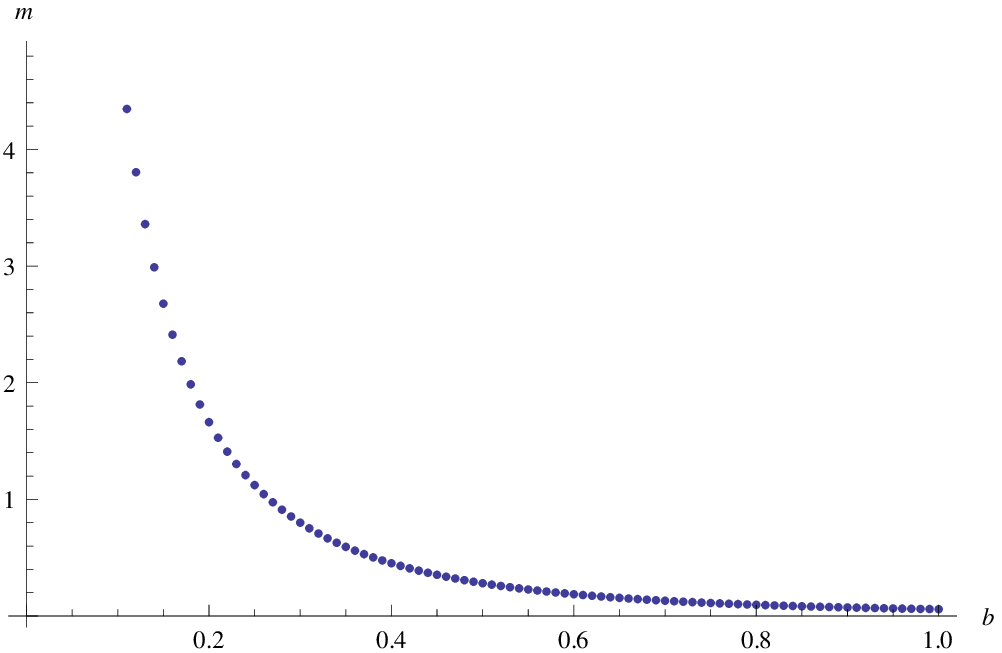}
    \caption{Plot of $m^2_{\Phi_I^3}/{q}$ vs $\sqrt{q}\beta$}
\label{mi3b}    
\end{minipage}
\end{psfrags}  
\hspace{0.5in}
\begin{psfrags}
\psfrag{m}[][]{\scalebox{0.8}{$m^2_{\Phi_I^3}/q$}}  
\psfrag{t}[][]{\scalebox{0.8}{$\frac{T}{\sqrt{q}}$}}
\begin{minipage}[b]{0.3\textwidth}
    \includegraphics[width=\textwidth]{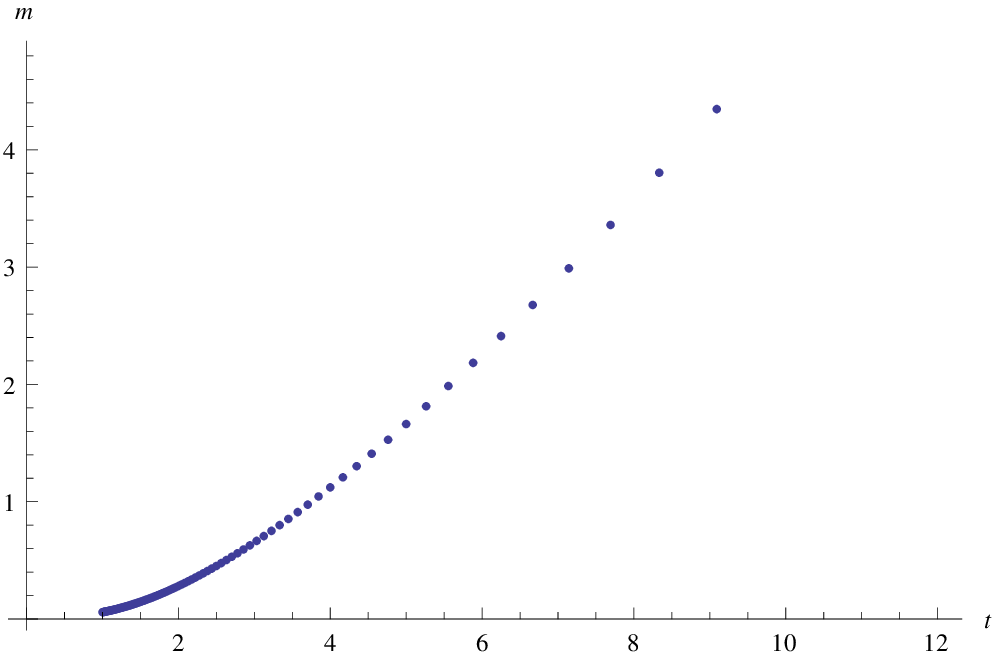}
    \caption{Plot of ${m^2_{\Phi_I^3}}/{q}$ vs $\frac{T}{\sqrt{q}}$}
\label{mi3t}    
\end{minipage}
\end{psfrags}
\end{figure}

\begin{figure}[htbp]
  \centering
\begin{psfrags}
\psfrag{m}[][]{\scalebox{0.8}{$m^2_{A_1^3}/q$}}
\psfrag{b}[][]{\scalebox{0.8}{$\sqrt{q}\beta$}}
  \begin{minipage}[b]{0.3\textwidth}
    \includegraphics[width=\textwidth]{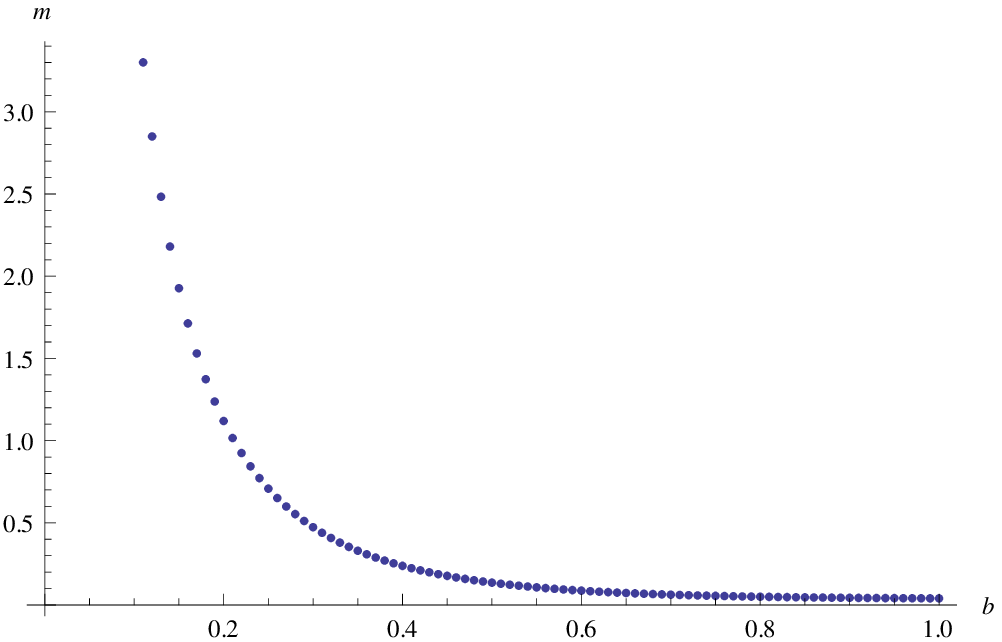}
    \caption{Plot of $m^2_{A_1^3}/q$ vs $\sqrt{q}\beta$}
\label{ma13b}    
\end{minipage}
\end{psfrags}  
\hspace{0.5in}
\begin{psfrags}
\psfrag{m}[][]{\scalebox{0.8}{$m^2_{A_1^3}/q$}}  
\psfrag{t}[][]{\scalebox{0.8}{$\frac{T}{\sqrt{q}}$}}
\begin{minipage}[b]{0.3\textwidth}
    \includegraphics[width=\textwidth]{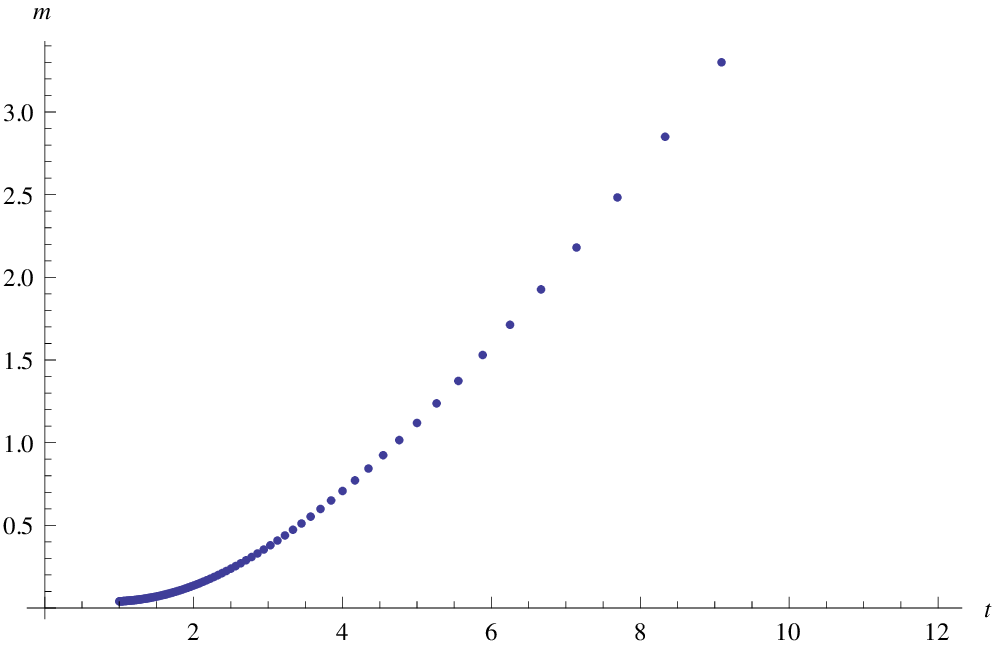}
    \caption{Plot of $m^2_{A_1^3}/q$ vs $\frac{T}{\sqrt{q}}$}
\label{ma13t}   
 \end{minipage}
\end{psfrags}
\end{figure}

\begin{figure}[htbp]
  \centering
\begin{psfrags}
\psfrag{m}[][]{\scalebox{0.8}{$m^2_{A_2^3}/q$}}
\psfrag{b}[][]{\scalebox{0.8}{$\sqrt{q}\beta$}}
  \begin{minipage}[b]{0.3\textwidth}
    \includegraphics[width=\textwidth]{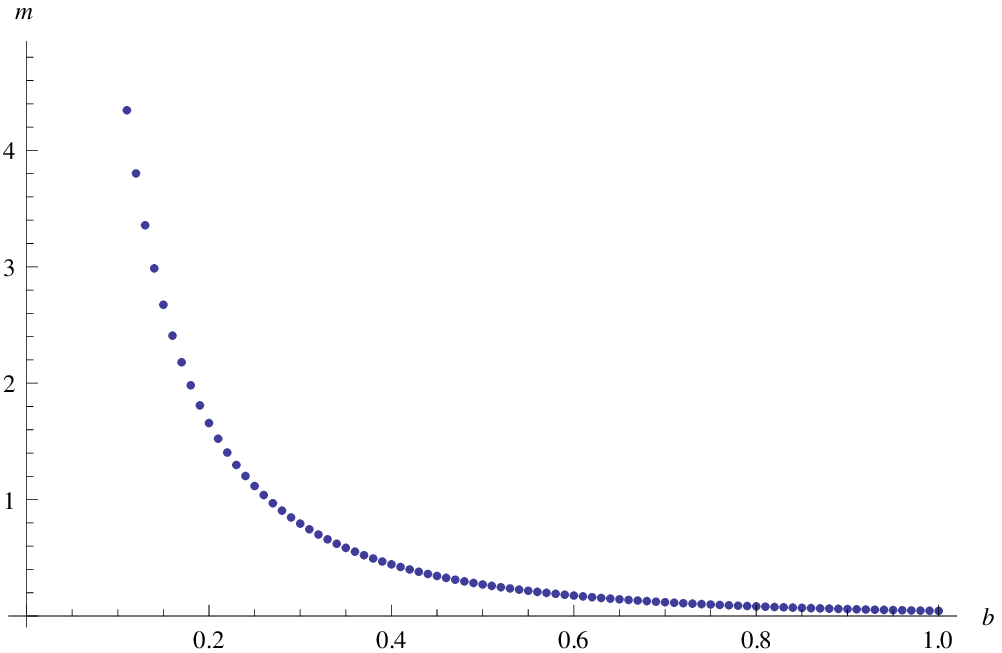}
    \caption{Plot of $m^2_{A_2^3}/q$ vs $\sqrt{q}\beta$}
\label{ma23b}  
\end{minipage}
\end{psfrags}  
\hspace{0.5in}
\begin{psfrags}
\psfrag{m}[][]{\scalebox{0.8}{$m^2_{A_2^3}/q$}}  
\psfrag{t}[][]{\scalebox{0.8}{$\frac{T}{\sqrt{q}}$}}
\begin{minipage}[b]{0.3\textwidth}
    \includegraphics[width=\textwidth]{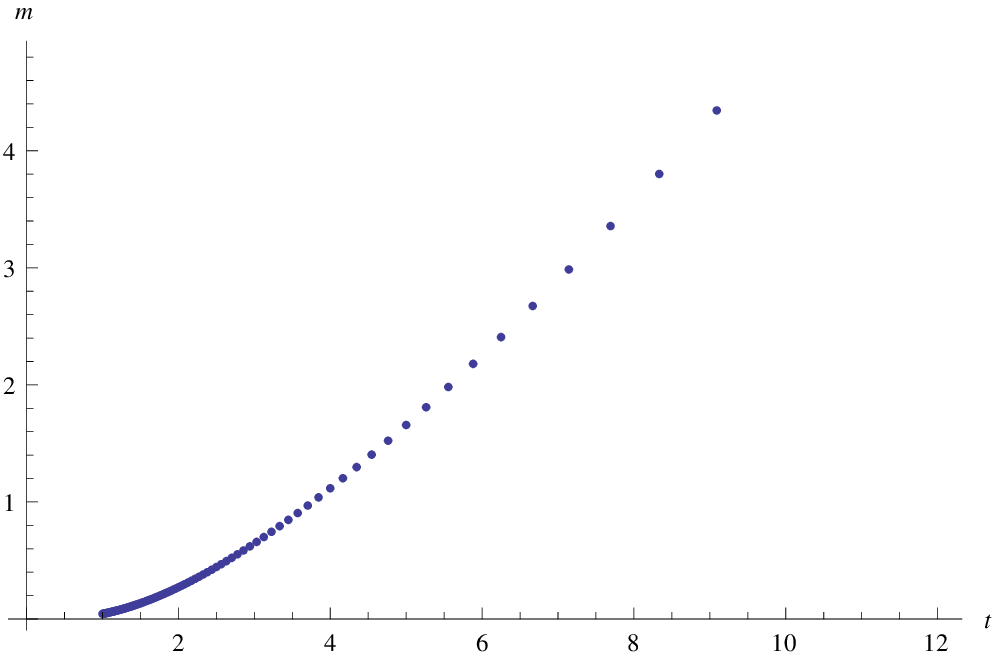}
    \caption{Plot of $m^2_{A_2^3}/q$ vs $\frac{T}{\sqrt{q}}$}
\label{ma23t} 
\end{minipage}
\end{psfrags}
\end{figure}

The numerical plots for the various fields at finite temperature are shown in the Figures \ref{m13b} to \ref{ma23t}. At finite temperature there are two dimensionful parameters $q, T$. The one loop $mass^2$ corrections go as $m^2=q f(T/\sqrt{q})$. For $T/\sqrt{q}>>1$ we expect the $m^2$ to go as $T^2$ in $3+1$ dimensions. To see this explicitly using the standard scaling arguments consider a simple representative of the temperature dependent part of the amplitudes for the tree-level massless fields,

\beqa\label{scaling1}
\Sigma_{massless}^2(\beta)=q\sum_n \int \frac{d^2 {\bf k}}{(2\pi\sqrt{q})^2}\frac{\sqrt{q}}{\sqrt{|{\bf k}|^2+\lambda_n}}\frac{1}{e^{\beta\sqrt{|{\bf k}|^2+\lambda_n}}-1}
\eeqa

Let us now rescale the momenta as ${\bf k}^{'}=\beta{\bf k}$. Further the the term $\beta^2\lambda_n$ in the exponent, in the limit $\beta^2 q \rightarrow 0$ can be replaced by a continuous variable, say $l^{'2}$. Thus replacing

\beqa
\sum_n \rightarrow \frac{1}{\beta}\int \frac{d l^{'}}{(2\pi\sqrt{q})}
\eeqa

we may write (\ref{scaling1}) as 

\beqa\label{scaling2}
\Sigma_{massless}^2(\beta)\sim \frac{1}{\beta^2}\int \frac{d^2 {\bf k}^{'}d l^{'}}{(2\pi)^3}\frac{1}{\sqrt{|{\bf k}|^{'2}+l^{'2}}}\frac{1}{e^{\sqrt{|{\bf k}|^{'2}+l^{'2}}}-1}
\eeqa

The integrals are independent of $\beta$ and hence we expect the $T^2$ behaviour. This assumes that the momentum limits in (\ref{scaling1}) go from zero to infinity. In the numerics however we have to work with finite limits on the momenta. The plots as shown in the figures here are for $|{\bf k}|/{\sqrt{q}}$ integrated over $0$ to $1000$ and $n$ is summed over up to $10$.

\subsection{Numerical Results: Tachyon}

The tachyon two-point one-loop amplitude has various massless fields propagating in the loop. As discussed earlier, 
to regulate the infrared divergences due to the massless fields we incorporate the one-loop corrected masses for the tree-level massless fields. 
The behaviour of these masses as a function of temperature is computed numerically in the previous section. Here however in the study of the behaviour of the tachyon mass as a function of temperature, to simplify the numerical computation, we take fixed non-zero values for masses of the fields. 
This $mass^2/q$ value is chosen to be $3.14$ for all the fields $\Phi_1^3, \Phi_I^3, A_1^3, A_i^3$. The numerical plot for the tachyon mass including the tree-level 
$mass^2/q =-1$ is as follows

\begin{figure}[htbp]
  \centering
\begin{psfrags}
\psfrag{m}[][]{\scalebox{0.8}{$m^2_{\text{tach}}/q$}}
\psfrag{b}[][]{\scalebox{0.8}{$\sqrt{q}\beta$}}
  \begin{minipage}[b]{0.3\textwidth}
    \includegraphics[width=\textwidth]{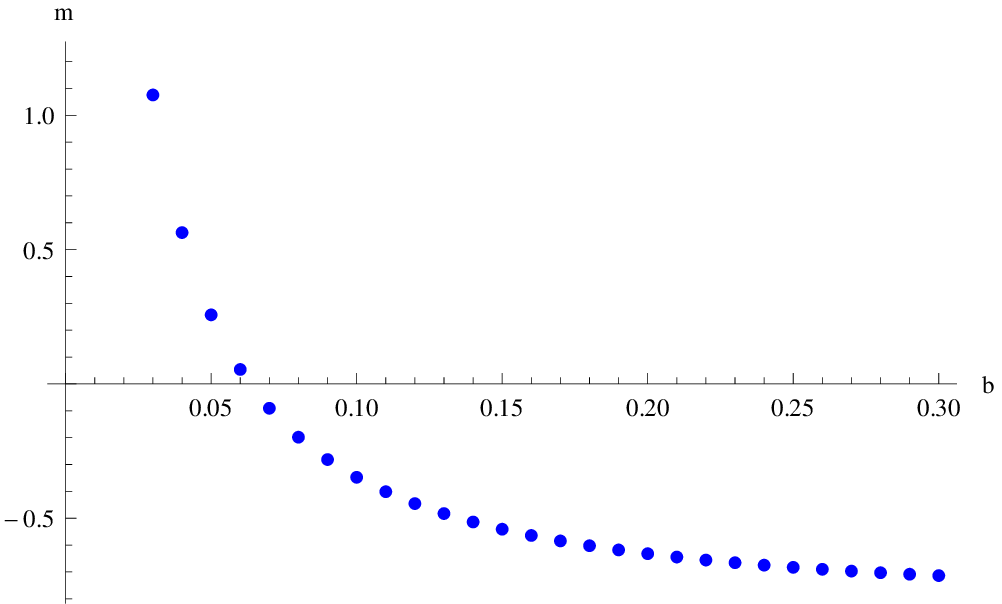}
    \caption{Plot of $m^2_{\text{tach}}/q$ vs $\sqrt{q}\beta$}
\label{mtachb}  
\end{minipage}
\end{psfrags}  
\hspace{0.5in}
\begin{psfrags}
\psfrag{m}[][]{\scalebox{0.8}{$m^2_{\text{tach}}/q$}}  
\psfrag{t}[][]{\scalebox{0.8}{$\frac{T}{\sqrt{q}}$}}
\begin{minipage}[b]{0.3\textwidth}
    \includegraphics[width=\textwidth]{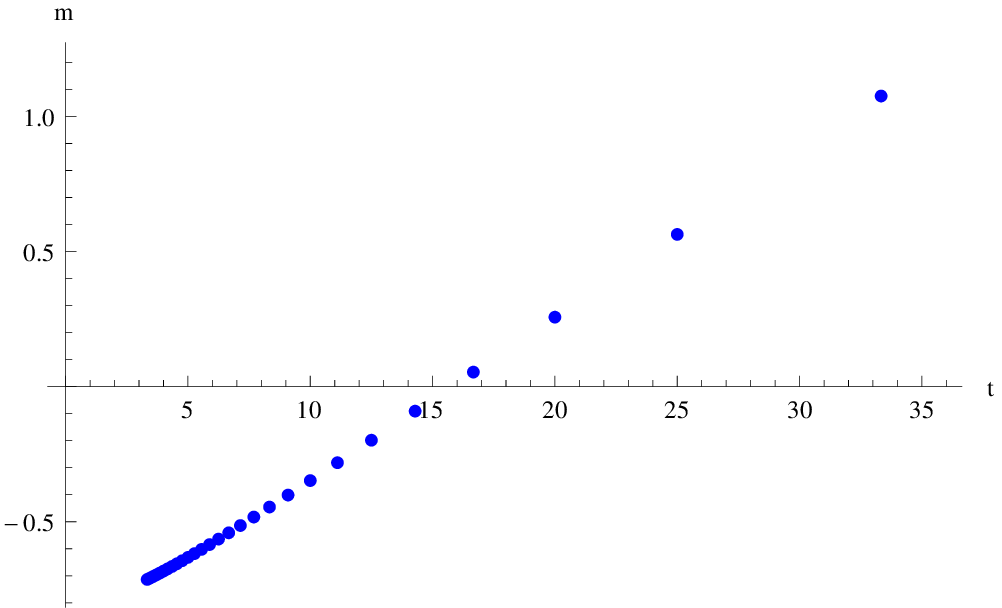}
    \caption{Plot of $m^2_{\text{tach}}/q$ vs $\frac{T}{\sqrt{q}}$}
\label{mtacht} 
\end{minipage}
\end{psfrags}
\end{figure}

There exists a critical temperature $T_c$, where the one-loop effective mass of the tree-level tachyon is zero. 
In the plots (\ref{mtachb}) and (\ref{mtacht}), the critical value of $\sqrt{q} \beta$ namely $\sqrt{q} \beta_c=0.062$ or the critical temperature
$\frac{T_c}{\sqrt{q}} = 15.96$. The zero temperature quantum correction for the tachyon mass-squared is found to be $m^2_{0\text{tach}}/q= 0.259$.
This behaviour bears the hallmark of a phase-transition. In the brane picture the mass squared becoming positive indicates the dissolution of the tachyonic mode, 
by way of which stability is achieved. This is also analogous to the transition from the superconducting BCS phase to the normal phase as studied in \cite{1,2}. 

Following the scaling arguments similar to that for the massless fields we here too we expect a $T^2$ behaviour for the tachyon mass for 
$T>>\sqrt{q}$. However note that the plot in fig (\ref{mtacht}) is almost linear in $\frac{T}{\sqrt{q}}$ which is a deviation from the
$T^2$ behaviour. This is an artifact of the numerical computation. The $T^2$ behaviour and the scalings of the momenta are valid for infinite 
sums and integrals. In order to perform the the numerical calculations, due to technical limitations we had to simplify the computations 
by summing and integrating over relatively small range for the momenta, namely $n, l, k\sim 0-10$. For sufficiently large range of momentum sums and integrations, the $T^2$ behaviour can presumably be reproduced in the numerics.

\section{Discussion and Summary}\label{conclusion}

In this paper we have computed the one-loop correction to the tachyon mass for the intersecting $D3$ branes at finite temperature. The calculation is extension of our previous work \cite{1}, wherein a similar computation was done for the case of intersecting $D1$ branes when the Yang-Mills approximation is valid. The underlying ${\cal N}=4$ Super Yang-Mills for the case of $D3$ branes makes the theory finite in the ultraviolet. Nevertheless we have explicitly shown that all the one-loop amplitudes computed in the paper are finite in the UV. The intersecting brane configuration breaks the spatial $SO(3)$ invariance and the straight intersecting branes implies that in the Yang-Mills approximation the background values of $\Phi_1^3$  is linear in $x$. The momentum modes along $x$ direction for the fields that couple to this background are thus discretized and are labeled as $n$.  
We have shown the finiteness in the UV for large $n$ irrespective of the values of the momentum $|{\bf k}|$ in the other directions for all the one-loop amplitudes computed in the paper. Further for the one-loop amplitudes for the tree-level massless fields we have shown that for fixed $n$ the amplitudes are finite for large values of $|{\bf k}|$.

The infrared divergence appearing in the one-loop tachyon amplitudes are of two kinds. Ones that appear as artifacts of the $A_0^a=0$ gauge can be removed prescriptions as discussed in \cite{kapusta1}-\cite{Leib1}. The genuine infrared divergences appearing from massless modes in the loop are cured by using the one-loop corrected propagator that incorporates the one-loop masses for the tree-level massless fields. Finally the amplitudes are all finite both in the infrared as well as in the ultraviolet. 

It can be seen that the analytic expressions for the amplitudes obtained in the paper reduce to the ones for the $D1$ branes. Similarly one can  easily find the corresponding expressions for the case of intersecting $D2$ branes as well by setting one of the momentum modes transverse to the $x$ direction to zero. In this paper we have only presented here the numerical results for the case of $D3$ branes. It is shown that all the masses including the Tachyon mass grows as $T^2$ as expected. From the numerics we have found the critical temperature $T_c$ when the Tachyon becomes massless. Thus for temperatures above the critical temperature the intersecting configuration is favoured.
Although the configuration of intersecting branes here is in flat space the results obtained in this simplified model is consistent with the 
strong coupling BCS holographic model proposed in \cite{2}. It was shown in \cite{2} that at zero temperature there exists solutions in the Yang-Mills approximation which correspond to breaking of $U(1)$ symmetry thus signaling condensate formation and the brane configuration is smoothed out.

The existence of the critical temperature denotes a phase transition. To analyze the nature of the transition at lower temperatures one needs to analyze the full tachyon potential. This knowledge would in turn shed light on the the BCS transition in the dual theory. In the following we conclude with some observations on the effective potential. Here we restrict ourselves to the case of $D3$ branes. In the Yang-Mills approximation, from the point of view of $(2+1)D$, the modes $C(m,n,{\bf k})$ have masses $(2n-1)q$. Thus there is an infinite tower of states including the Tachyon with masses proportional to $q$. In this scenario it might appear that the notion of Tachyon effective action obtained by integrating out massive modes is ill-defined. There have however been extensive studies of Tachyon effective action in open string theory where the effective actions were proposed from the consistency with world-sheet conformal field theory (see \cite{Sen} for review). Similar techniques should be employed for the construction and further study of effective potential in the present case. For approaches along these directions see \cite{Hashimoto:2003xz}-\cite{J2}. It would further be interesting to generalize such potential to the case of finite temperature and for nontrivial background geometries. 

Alternatively from the point of view of $(3+1)D$ the discrete $n$-modes are momenta for the field $\zeta(z)$ (see eqn (\ref{zeta1})) along the $x$ direction. We can thus construct a $(3+1)D$ effective action for the field $\zeta(z)$. This in the zero external-momentum limit including $n=0$ reduces to the potential in terms of only the $C(0,0,0)$ modes. This potential in principle be computed using perturbative techniques utilized in the paper and would involve computation of $N$-point amplitudes for the $C(0,0,0)$ modes. Compared to the effective potential in the $(2+1)D$ theory this would thus give the effective potential when all the modes with positive mass-squared are set to zero i.e. the massive modes only appear in loops. This is exactly what is done in this paper. Here for simplicity we have ignored $\expect{C(0,n,0)C(0,n^{'},0)}$, amplitudes for $n \ne n^{'} $.  Nevertheless computation of an effective potential only in terms of $C(0,0,0)$ for the $(3+1)D$ theory should be interesting in its own right.

\vspace{0.8cm}
\noindent
{\large {\bf Acknowledgments:}}
\\
We would like to thank Balachandran Sathiapalan for useful discussions and suggestions. V.S. acknowledges CSIR, India, for support through JRF grant. The work of S.P.C. is supported in part  by the DST-Max Planck Partner Group ``Quantum Black Holes" 
between IOP Bhubaneswar and AEI Golm. The work of S.S. is partially supported by the Research and Development Grant, 
University of Delhi.

\newpage
\appendix

\section{Summary of notations}\label{notations}

\begin{table}[h]

\begin{center}
\begin{tabular}{|c|c|}
\hline
\hline
\multicolumn{2}{ |c| }{{\bf Constants}}\\
\hline
$q$&Slope of intersecting brane configuration\\
\hline
$\beta$&Inverse of temperature $T$\\
\hline
$g^2$&Yang-Mills coupling constant\\

\hline
\hline


\multicolumn{2}{ |c| }{{\bf Variables}}\\
\hline
$\o_m=(2 m\pi/\beta);((2m+1)\pi/\beta) $& Matsubara Frequencies, (Bosons);(Fermions)\\
\hline
$\hat{l}=l/\sqrt{q}$&Continuous momentum along $1(x)$ direction\\
\hline
$n$&Discrete momentum along $1(x)$ direction\\
\hline
${\bf k}\equiv (k_2,k_3)$& Momentum along directions $2$ and $3$\\
\hline
$\hat{k_i}=k_i/\sqrt{q}$&Dimensionless momentum\\

\hline
\hline
\multicolumn{2}{ |c| }{{\bf Eigenvalues}}\\
\hline
$\lambda_n$&$(2n-1)q$\\
\hline
$\gamma_n$&$(2n+1)q$\\
\hline
$\lambda^{'}_n$&$2nq$\\
\hline
\hline
\multicolumn{2}{ |c| }{{\bf Normalizations}}\\
\hline
$N$&$\sqrt{q}/\beta$\\
\hline
${\cal N}(n)$&$\f{1}{\sqrt{\sqrt{\pi} 2^n (4n^2-2)(n-2)!}}$\\
\hline
$\tilde{{\cal N}}(n)$&$\f{1}{\sqrt{\sqrt{\pi} 2^n (4n-2)(n-1)!}}$\\
\hline
$\mathcal{N}^{'}(n)$&$\f{1}{\sqrt{\sqrt{\pi}2^n n!}}$\\
\hline
${\cal N}_F(n)$&$\f{1}{\sqrt{\sqrt{\pi} 2^{n+1} (n-1)!}}$\\
\hline
\hline
\end{tabular}
\end{center}
\caption{Constants and normalizations.}
\label{t1}
\end{table}

{\bf Matsubara Sums} \cite{kapusta} (Also see Appendix F of \cite{1})

Sum over bosonic frequencies:
\beqa
f(\gamma_n,\beta)=N\sum_m \f{1}{\o_m^2+\gamma_n+|{\bf k}|^2}=\f{\sqrt{q}}{\sqrt{\gamma_n+|{\bf k}|^2}}\left(\hf+\f{1}{e^{\beta\sqrt{\gamma_n+|{\bf k}|^2}}-1}\right)
\eeqa

Sum over fermionic frequencies:
\beqa
g(\lambda^{'}_n,\beta)=N\sum_m \f{1}{\o_m^2+\lambda^{'}_n+|{\bf k}|^2}=\f{\sqrt{q}}{\sqrt{\lambda^{'}_n+|{\bf k}|^2}}\left(\hf-\f{1}{e^{\beta\sqrt{\lambda^{'}_n+|{\bf k}|^2}}+1}\right)
\eeqa

\section{Dimensional Reduction}\label{reduction}

Following \cite{dimred1} and \cite{dimred2} we give below the details of the dimensional reduction of $D=10$, ${\cal N}=1$ to $D=4$.
The action for $D=10$, ${\cal N}=1$ SYM \footnote{We will use the metric $\mbox{diagonal}(-1, +1, \cdots, +1)$.},

\beqa
S_{10}=\f{1}{g^2}\mbox{tr}\int d^{10}x \left[-\f{1}{2}F_{MN}F^{MN}-i\bar{\Psi}\Gamma^{M}D_{M}\Psi\right]
\eeqa

\beqa
F_{MN}&=&\partial_{M}A_{N}-\partial_{N}A_{M}+i\left[A_{M},A_{N}\right]\\
D_{M}\Psi&=&\partial_{M}\Psi+i\left[A_{M},\Psi\right]
\eeqa

where $M,N=0, \cdot\cdot\cdot 9$ with $A_{M}=\f{\sigma^a}{2}A^a_{M}$, $\Psi=\f{\sigma^a}{2}\Psi^a_{M}$ and,

\beqa
\left[\f{\sigma^a}{2},\f{\sigma^b}{2}\right]=i\epsilon^{abc}\f{\sigma^c}{2} \mbox{~~;~~}
\f{1}{2} \mbox{tr} \left(\sigma^a\sigma^b\right)=\delta^{ab}
\eeqa

To proceed with the reduction, the gamma-matrices are first decomposed as,

\beqa
&&\Gamma^{\mu}=\gamma^{\mu}\otimes \left(\begin{array}{cc}\mathbb{I}_4&0\\0&-\mathbb{I}_4\end{array}\right)
~~~~;~~~~
\Gamma^{3+i}=\gamma^{5}\otimes \left(\begin{array}{cc}\rho_i&0\\0&\rho^{'}_i\end{array}\right) ~~~\mbox{$i=1,2,3$}\non
&&\rho_1=\rho^{'}_1=\gamma^0~~~;~~~
\rho_2=\rho^{'}_2=\gamma^5~~~;~~~
\rho_3=-\rho^{'}_3=\gamma^0\gamma^5.
\eeqa

\beqa
\Gamma^{6+i}&=&\mathbb{I}_4\otimes \left(\begin{array}{cc}0&\zeta_i\\\zeta_i&0\end{array}\right)~~~\mbox{$i=1,2,3$}\\
\zeta_1&=&\gamma^1~~~\zeta_2=\gamma^2~~~\zeta_3=\gamma^3
\eeqa

In this representation, 

\beqa
\Gamma^{11}=\Gamma^0 \cdots \Gamma^9 = \mathbb{I}_4\otimes \left(\begin{array}{cc}0&\rho_3\\-\rho_3&0\end{array}\right)
\eeqa

with $\Gamma^{11}\Psi=\Psi$

and
\beqa
\Psi=\left(\begin{array}{c}\lambda\\-\rho_3\lambda\end{array}\right)
\eeqa

The dimensionally reduced action can now be written as,

\beqa
S_4=S_4^1+S_4^2
\eeqa

where the Bosonic part if the action is,

\beqa\label{actionboson}
S^1_{4}=\f{1}{g^2}\mbox{tr}\int d^{4}x \left[-\f{1}{2}F_{\mu\nu}F^{\mu\nu}-D_{\mu}\Phi_ID^{\mu}\Phi_I-D_{\mu}\tilde{\Phi}_ID^{\mu}\tilde{\Phi}_I+\f{1}{2}\left([\Phi_I,\Phi_J]^2+[\tilde{\Phi}_I,\tilde{\Phi}_J]^2+2[\Phi_I,\tilde{\Phi}_J]^2\right)\right]
\eeqa

and the Fermionic part is,

\beqa\label{actionfermion}
S^2_{4}=\f{1}{g^2}\mbox{tr}\int d^{4}x \left[-i\bar{\lambda}_k\gamma^{\mu}D_{\mu}\lambda_k+\bar{\lambda}_k[(\alpha_{kl}^I\Phi_I+\beta_{kl}^I\gamma^5\tilde{\Phi}),\lambda_l]\right]
\eeqa

The $\alpha$ and $\beta$ matrices satisfy
\beqa
\{\alpha^I,\alpha^J\}=\{\beta^I,\beta^J\}=-2\delta^{IJ} ~~~[\alpha^I,\beta^J]=0
\eeqa

An explicit representation of these matrices is given by the following
\beqa
\alpha^1=\left(\begin{array}{cc}0&\sigma^1\\-\sigma^1&0\end{array}\right)~~~\alpha^2=\left(\begin{array}{cc}0&\sigma^3\\-\sigma^3&0\end{array}\right)~~~\alpha^3=\left(\begin{array}{cc}i\sigma^2&0\\0&i\sigma^2\end{array}\right)
\eeqa

\beqa
\beta^1=\left(\begin{array}{cc}0&i\sigma^2\\i\sigma^2&0\end{array}\right)~~~\beta^2=\left(\begin{array}{cc}0&1\\-1&0\end{array}\right)~~~\beta^3=\left(\begin{array}{cc}-i\sigma^2&0\\0&i\sigma^2\end{array}\right)
\eeqa

We end this section by writing down the explicit forms of the 4D $\gamma$-matrices

\beqa
\gamma^0=\left(\begin{array}{cc}0&i\sigma^2\\i\sigma^2&0\end{array}\right)~~~\gamma^1=\left(\begin{array}{cc}0&-\sigma_3\\-\sigma_3&0\end{array}\right)~~~\gamma^2=\left(\begin{array}{cc}0&-\sigma^1\\-\sigma^1&0\end{array}\right)~~~\gamma^3=\left(\begin{array}{cc}\mathbb{I}_2&0\\0&-\mathbb{I}_2 \end{array}\right)
\eeqa

\newpage

\section{Propagators and Vertices for computation of two point $C(m,n,{\bf k})$ amplitude}\label{tachyonvertices}

\subsection{Bosons}\label{bosons}

Propagator for $C(m,n,{\bf k})$ field:

\beqa
\label{propzeta}
\expect{C(m,n,{\bf k})C(m^{'},n^{'},{\bf k}^{'})}=qg^2\f{\delta_{m,-m^{'}}\delta_{n,n^{'}}(2\pi)^2\delta^2({\bf k}+{\bf k}^{'})}{\o_m^2+\lambda_n+|{\bf k}|^2}
\eeqa

Propagator for the $\Phi_I^{(1,2)}$ ($I=2,3$) and  $\tilde{\Phi}_J^{(1,2)}$ ($J=1,2,3$) fields are identical and are given by:
\beqa
\label{propphi1I}
\expect{\Phi_I^{(1,2)}(m,n,{\bf k})\Phi_I^{(1,2)}(m^{'},n^{'},{\bf k}^{'})}=qg^2\f{\delta_{m,-m^{'}}\delta_{n,n^{'}}(2\pi)^2\delta^2({\bf k}+{\bf k}^{'})}{\o_m^2+\gamma_n+|{\bf k}|^2}
\eeqa

Propagator for the $\Phi_I^3/\tilde{\Phi}_I^3/$ fields for  $I=1,2,3$:
\beqa
\label{propphi3I}
\expect{\Phi_I^3(m,l,{\bf k})\Phi_I^3(m^{'},l^{'},{\bf k}^{'})}=qg^2 \f{\delta_{m,-m^{'}}2\pi\delta(l+l^{'})(2\pi)^2\delta^2({\bf k}+{\bf k}^{'})}{\o_m^2+l^2+|{\bf k}|^2}
\eeqa

Propagator for the $\tilde{A}^2_i$ and  $\tilde{A}^1_i$ fields ($i=1,2,3$):

\beqa\label{proppab}
(qg^2)P^{ab}_{ij}: 
\expect{\tilde{A}_i^{a}(m,n,{\bf k})\tilde{A}_j^{b}(-m,n,-{\bf k})}=\f{qg^2}{k^2}\left(\delta_{ij}-\f{k_ik_j}{k_0^2}\right)\delta^{ab}~~~~~(a,b=1,2)
\eeqa

where, $k^2=\o_m^2+\gamma_n+|{\bf k}|^2$, $k_0=i\o_m$ and $k_x=\sqrt{\gamma_n}$.

Propagator for the $A^3_i$ fields ($i=1,2,3$):

\beqa\label{propp33}
(qg^2) P^{33}_{ij}:\expect{\tilde{A}_i^{3}(m,l,{\bf k})\tilde{A}_j^{3}(-m,-l,-{\bf k})}=\f{qg^2}{k^2}\left(\delta_{ij}-\f{k_ik_j}{k_0^2}\right)
\eeqa

where, $k^2=\o_m^2+l^2+|{\bf k}|^2$, $k_0=i\o_m$.

\newpage

\begin{table}[H]
\begin{center}
\begin{tabular}{lcc}
\begin{psfrags}
\psfrag{a11}[][]{\scalebox{0.8}{$C(m,n,{\bf k})$}}
\psfrag{a12}[][]{\scalebox{0.8}{$C(m^{'},n^{'},{\bf k}^{'})$}}
\psfrag{c1}[][]{\scalebox{0.8}{$C(m^{''},n^{''},{\bf k}^{''})$}}
\psfrag{c2}[][]{\scalebox{0.8}{$C(\tilde{m}^{''},\tilde{n}^{''},\tilde{{\bf k}}^{''})$}}
\psfrag{v2}[][]{$V_1$}
\parbox[c]{3cm}{\includegraphics[width= 2.5cm,angle=0]{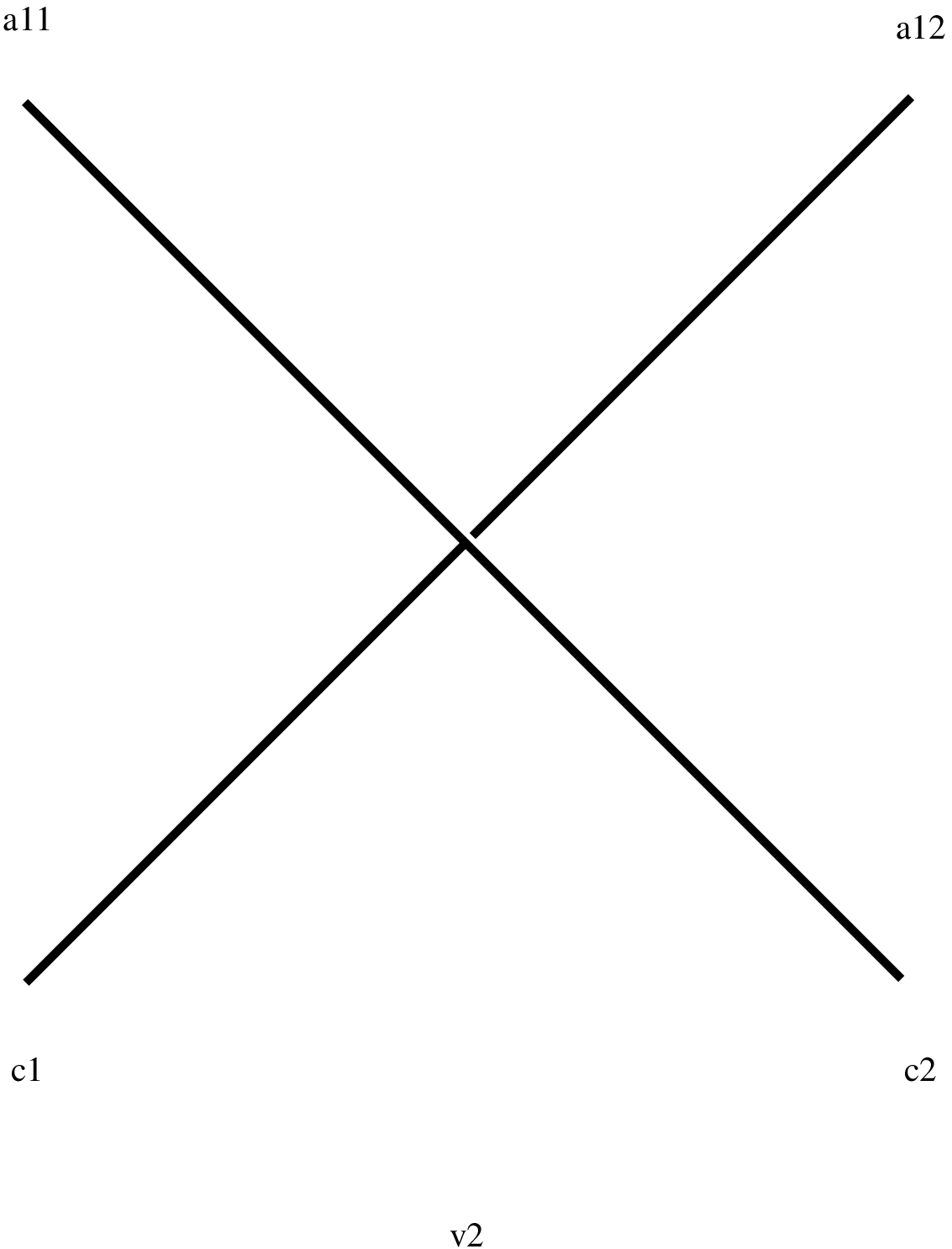}}
\label{v1}
\end{psfrags}
&~~~&
\parbox[c]{13cm}{$\begin{array}{c}V_1=-\f{N}{2qg^2} F_1(n,n^{'},n^{''},\tilde{n}^{''}) (2\pi)^2\delta^2({\bf k}+{\bf k}^{'}+{\bf k}^{''}+\tilde{{\bf k}}^{''})\delta_{m+m^{'}+m^{''}+\tilde{m}^{''}}\\ 
F_1(n,n^{'},n^{''},\tilde{n}^{''})=\sqrt{q}\int dx [\phi_n(x)\phi_{n^{'}}(x)A_{n^{''}}(x) A_{\tilde{n}^{''}}(x) + A_n(x) A_{n^{'}}(x) \phi_{n^{''}}(x) \phi_{\tilde{n}^{''}}(x)] 
\end{array}$}
\end{tabular}
\end{center}
\end{table}

\begin{table}[H]
\begin{center}
\begin{tabular}{lcc}
\begin{psfrags}
\psfrag{a11}[][]{\scalebox{0.8}{$C^{'}(m,n,{\bf k})$}}
\psfrag{a12}[][]{\scalebox{0.8}{$C^{'}(m^{'},n^{'},{\bf k}^{'})$}}
\psfrag{c1}[][]{\scalebox{0.8}{$C(m^{''},n^{''},{\bf k}^{''})$}}
\psfrag{c2}[][]{\scalebox{0.8}{$C(\tilde{m}^{''},\tilde{n}^{''},\tilde{{\bf k}}^{''})$}}
\psfrag{v2}[][]{$V^{'}_1$}
\parbox[c]{3cm}{\includegraphics[width= 2.5cm,angle=0]{vertex3.eps}}
\label{v1}
\end{psfrags}
&~~~&
\parbox[c]{13cm}{$\begin{array}{c}V^{'}_1=-\f{N}{2qg^2} F^{'}_1(n,n^{'},n^{''},\tilde{n}^{''}) (2\pi)^2\delta^2({\bf k}+{\bf k}^{'}+{\bf k}^{''}+\tilde{{\bf k}}^{''})\delta_{m+m^{'}+m^{''}+\tilde{m}^{''}}\\ 
F^{'}_1(n,n^{'},n^{''},\tilde{n}^{''})=2\sqrt{q}\int dx [ A_n(x)\phi_{n^{'}}(x)\phi_{n^{''}}(x) A_{\tilde{n}^{''}}(x)

+   \phi_n(x) A_{n^{'}}(x)\phi_{n^{''}}(x) A_{\tilde{n}^{''}}(x) ] 
\end{array}$}
\end{tabular}
\end{center}
\end{table}

\begin{table}[H]
\begin{center}
\begin{tabular}{lcc}
\begin{psfrags}
\psfrag{a11}[][]{\scalebox{0.8}{$\tilde{A^2_1}(m,n,{\bf k})$}}
\psfrag{a12}[][]{\scalebox{0.8}{$\tilde{A^2_1}(m^{'},n^{'},{\bf k}^{'})$}}
\psfrag{c1}[][]{\scalebox{0.8}{$C(m^{''},n^{''},{\bf k}^{''})$}}
\psfrag{c2}[][]{\scalebox{0.8}{$C(\tilde{m}^{''},\tilde{n}^{''},\tilde{{\bf k}}^{''})$}}
\psfrag{v2}[][]{$\tilde{V}_1$}
\parbox[c]{3cm}{\includegraphics[width= 2.5cm,angle=0]{vertex3.eps}}
\label{v1}
\end{psfrags}
&~~~&
\parbox[c]{11cm}{$\begin{array}{c}\tilde{V}_1=-\f{N}{2qg^2} \tilde{F}_1(n,n^{'},n^{''},\tilde{n}^{''}) (2\pi)^2\delta^2({\bf k}+{\bf k}^{'}+{\bf k}^{''}+\tilde{{\bf k}}^{''})\delta_{m+m^{'}+m^{''}+\tilde{m}^{''}}\\ 
\tilde{F}_1(n,n^{'},n^{''},\tilde{n}^{''})=-\sqrt{q}\int dx [\tilde{\phi}_{n+1}(x)\tilde{\phi}_{n^{'}+1}(x)A_{n^{''}}(x) A_{\tilde{n}^{''}}(x)  \\
+ \tilde{A}_{n+1}(x) \tilde{A}_{n^{'}+1}(x) \phi_{n^{''}}(x) \phi_{\tilde{n}^{''}}(x)] 
\end{array}$}
\end{tabular}
\end{center}
\end{table}

\begin{table}[H]
\begin{center}
\begin{tabular}{lcc}
\begin{psfrags}
\psfrag{a11}[][]{\scalebox{0.8}{$\tilde{A^1_1}(m,n,{\bf k})$}}
\psfrag{a12}[][]{\scalebox{0.8}{$\tilde{A^1_1}(m^{'},n^{'},{\bf k}^{'})$}}
\psfrag{c1}[][]{\scalebox{0.8}{$C(m^{''},n^{''},{\bf k}^{''})$}}
\psfrag{c2}[][]{\scalebox{0.8}{$C(\tilde{m}^{''},\tilde{n}^{''},\tilde{{\bf k}}^{''})$}}
\psfrag{v2}[][]{$\tilde{V}^{'}_1$}
\parbox[c]{3cm}{\includegraphics[width= 2.5cm,angle=0]{vertex3.eps}}
\label{v1}
\end{psfrags}
&~~~&
\parbox[c]{11cm}{$\begin{array}{c}\tilde{V}^{'}_1=-\f{N}{2qg^2} \tilde{F}^{'}_1(n,n^{'},n^{''},\tilde{n}^{''}) (2\pi)^2\delta^2({\bf k}+{\bf k}^{'}+{\bf k}^{''}+\tilde{{\bf k}}^{''})\delta_{m+m^{'}+m^{''}+\tilde{m}^{''}}\\ 
\tilde{F}^{'}_1(n,n^{'},n^{''},\tilde{n}^{''})=-2 \sqrt{q}\int dx [\tilde{A}_{n+1}(x)\tilde{\phi}_{n^{'}+1}(x)\phi_{n^{''}}(x) A_{\tilde{n}^{''}}(x)
\\ + \tilde{A}_{n+1}(x) \tilde{A}_{n^{'}+1}(x) \phi_{n^{''}}(x) \phi_{\tilde{n}^{''}}(x)] 
\end{array}$}
\end{tabular}
\end{center}
\end{table}

\begin{table}[H]
\begin{center}
\begin{tabular}{lcc}
\begin{psfrags}
\psfrag{a11}[][]{\scalebox{0.8}{$\Phi_I^{1}/\tilde{\Phi_I}^{1}$}}
\psfrag{a12}[][]{\scalebox{0.8}{$\Phi_I^{1}/\tilde{\Phi}_I^{1}$}}
\psfrag{c1}[][]{\scalebox{0.8}{$C(m^{''},n^{''},{\bf k}^{''})$}}
\psfrag{c2}[][]{\scalebox{0.8}{$C(\tilde{m}^{''},\tilde{n}^{''},\tilde{{\bf k}}^{''})$}}
\psfrag{v2}[][]{$V^{1}_2$}
\parbox[c]{3cm}{\includegraphics[width= 2.5cm,angle=0]{vertex3.eps}}
\label{v1}
\end{psfrags}
&~~~&
\parbox[c]{13cm}{$\begin{array}{c}V^{1}_2=-\f{N}{2qg^2} F^{1}_2(n,n^{'},n^{''},\tilde{n}^{''}) (2\pi)^2\delta^2({\bf k}+{\bf k}^{'}+{\bf k}^{''}+\tilde{{\bf k}}^{''})\delta_{m+m^{'}+m^{''}+\tilde{m}^{''}}\\ 
F^{1}_2(n,n^{'},n^{''},\tilde{n}^{''})=\sqrt{q}\int dx e^{-qx^2}\left [A_{n^{''}}(x) A_{\tilde{n}^{''}}(x)\right] \left[H_n(x)  H_{n^{'}}(x)\right]\\
\\
\mbox{We also have additional vertices as $V^{1}_2$ for the $A_i^{1}$ $(i\ne1)$ fields in place of} \\ \mbox{$\Phi_I^{1}/\tilde{\Phi_I}^{1}$ fields.}
\end{array}$}

\end{tabular}
\end{center}
\end{table}

\begin{table}[H]
\begin{center}
\begin{tabular}{lcc}
\begin{psfrags}
\psfrag{a11}[][]{\scalebox{0.8}{$\Phi_I^{2}/\tilde{\Phi_I}^{2}$}}
\psfrag{a12}[][]{\scalebox{0.8}{$\Phi_I^{2}/\tilde{\Phi}_I^{2}$}}
\psfrag{c1}[][]{\scalebox{0.8}{$C(m^{''},n^{''},{\bf k}^{''})$}}
\psfrag{c2}[][]{\scalebox{0.8}{$C(\tilde{m}^{''},\tilde{n}^{''},\tilde{{\bf k}}^{''})$}}
\psfrag{v2}[][]{$V^2_2$}
\parbox[c]{3cm}{\includegraphics[width= 2.5cm,angle=0]{vertex3.eps}}
\label{v1}
\end{psfrags}
&~~~&
\parbox[c]{13cm}{$\begin{array}{c}V^2_2=-\f{N}{2qg^2} F^2_2(n,n^{'},n^{''},\tilde{n}^{''}) (2\pi)^2\delta^2({\bf k}+{\bf k}^{'}+{\bf k}^{''}+\tilde{{\bf k}}^{''})\delta_{m+m^{'}+m^{''}+\tilde{m}^{''}}\\ 
F^2_2(n,n^{'},n^{''},\tilde{n}^{''})=\sqrt{q}\int dx e^{-qx^2}\left [\phi_{n^{''}}(x)\phi_{\tilde{n}^{''}}(x)\right] \left[H_n(x)  H_{n^{'}}(x)\right]\\
\\
\mbox{We also have additional vertices as $V^2_2$ for the $A_i^{2}$ $(i\ne1)$ fields in place of} \\ \mbox{$\Phi_I^{2}/\tilde{\Phi_I}^{2}$ fields.}
\end{array}$}

\end{tabular}
\end{center}
\end{table}

\begin{table}[H]
\begin{center}
\begin{tabular}{lcc}
\begin{psfrags}
\psfrag{a11}[][]{\scalebox{0.8}{$\Phi_I^{3}/\tilde{\Phi_I}^{3}$}}
\psfrag{a12}[][]{\scalebox{0.8}{$\Phi_I^{3}/\tilde{\Phi}_I^{3}$}}
\psfrag{c1}[][]{\scalebox{0.8}{$C(m^{''},n^{''},{\bf k}^{''})$}}
\psfrag{c2}[][]{\scalebox{0.8}{$C(\tilde{m}^{''},\tilde{n}^{''},\tilde{{\bf k}}^{''})$}}
\psfrag{v2}[][]{$V^{3}_2$}
\parbox[c]{3cm}{\includegraphics[width= 2.5cm,angle=0]{vertex3.eps}}
\label{v1}
\end{psfrags}
&~~~&
\parbox[c]{13cm}{$\begin{array}{c}V^{3}_2=-\f{N}{2qg^2} F^{3}_2(l,l^{'},n^{''},\tilde{n}^{''}) (2\pi)^2\delta^2({\bf k}+{\bf k}^{'}+{\bf k}^{''}+\tilde{{\bf k}}^{''})\delta_{m+m^{'}+m^{''}+\tilde{m}^{''}}\\ 
F^{3}_2(l,l^{'},n^{''},\tilde{n}^{''})=\sqrt{q}\int dx \left [A_{n^{''}}(x) A_{\tilde{n}^{''}}(x)+\phi_{n^{''}}(x)\phi_{\tilde{n}^{''}}(x)\right] \left[e^{-ilx}e^{-il^{'}x}\right]\\
\\
\mbox{We also have additional vertices as $V^{3}_2$ for the $A_i^{3}$ $(i\ne1)$ fields in place of} \\ \mbox{$\Phi_I^{3}/\tilde{\Phi_I}^{3}$ fields.}
\end{array}$}

\end{tabular}
\end{center}
\end{table}

\begin{table}[H]
\begin{center}
\begin{tabular}{lcc}
\begin{psfrags}
\psfrag{a11}[][]{\scalebox{0.8}{$A_1^{3}(m,l,{\bf k})$}}
\psfrag{a12}[][]{\scalebox{0.8}{$A_1^{3}(m^{'},l^{'},{\bf k}^{'})$}}
\psfrag{c1}[][]{\scalebox{0.8}{$C(m^{''},n^{''},{\bf k}^{''})$}}
\psfrag{c2}[][]{\scalebox{0.8}{$C(\tilde{m}^{''},\tilde{n}^{''},\tilde{{\bf k}}^{''})$}}
\psfrag{v2}[][]{$V_3$}
\parbox[c]{3cm}{\includegraphics[width= 2.5cm,angle=0]{vertex3.eps}}
\label{v1}
\end{psfrags}
&~~~&
\parbox[c]{13cm}{$\begin{array}{c}V_3=-\f{N}{2qg^2} F_3(l,l^{'},n^{''},\tilde{n}^{''}) (2\pi)^2\delta^2({\bf k}+{\bf k}^{'}+{\bf k}^{''}+\tilde{{\bf k}}^{''})\delta_{m+m^{'}+m^{''}+\tilde{m}^{''}}\\ 
F_3(l,l^{'},n^{''},\tilde{n}^{''})=\sqrt{q}\int dx \left [\phi_{n^{''}}(x)\phi_{\tilde{n}^{''}}(x)\right] \left[e^{-ilx}e^{-il^{'}x}\right]
\end{array}$}
\end{tabular}
\end{center}
\end{table}

\begin{table}[H]
\begin{center}
\begin{tabular}{lcc}
\begin{psfrags}
\psfrag{a11}[][]{\scalebox{0.8}{$\Phi_1^{3}(m,l,{\bf k})$}}
\psfrag{a12}[][]{\scalebox{0.8}{$\Phi_1^{3}(m^{'},l^{'},{\bf k}^{'})$}}
\psfrag{c1}[][]{\scalebox{0.8}{$C(m^{''},n^{''},{\bf k}^{''})$}}
\psfrag{c2}[][]{\scalebox{0.8}{$C(\tilde{m}^{''},\tilde{n}^{''},\tilde{{\bf k}}^{''})$}}
\psfrag{v2}[][]{$V^{'}_3$}
\parbox[c]{3cm}{\includegraphics[width= 2.5cm,angle=0]{vertex3.eps}}
\label{v1}
\end{psfrags}
&~~~&
\parbox[c]{13cm}{$\begin{array}{c}V^{'}_3=-\f{N}{2qg^2} F^{'}_3(l,l^{'},n^{''},\tilde{n}^{''}) (2\pi)^2\delta^2({\bf k}+{\bf k}^{'}+{\bf k}^{''}+\tilde{{\bf k}}^{''})\delta_{m+m^{'}+m^{''}+\tilde{m}^{''}}\\ 
F^{'}_3(l,l^{'},n^{''},\tilde{n}^{''})=\sqrt{q}\int dx \left [A_{n^{''}}(x)A_{\tilde{n}^{''}}(x)\right] \left[e^{-ilx}e^{-il^{'}x}\right]
\end{array}$}
\end{tabular}
\end{center}
\end{table}

\begin{table}[H]
\begin{center}
\begin{tabular}{lcc}
\begin{psfrags}
\psfrag{c}[][]{$C(m^{''},n^{''},{\bf k}^{''})$}
\psfrag{a1}[][]{$C^{'}(m,n,{\bf k})$}
\psfrag{a3}[][]{$A_i^3(m^{'},l^{'},{\bf k}^{'})$}
\psfrag{v4}[][]{$V^{i}_4$}
\parbox[c]{6cm}{\includegraphics[width= 5 cm,angle=0]{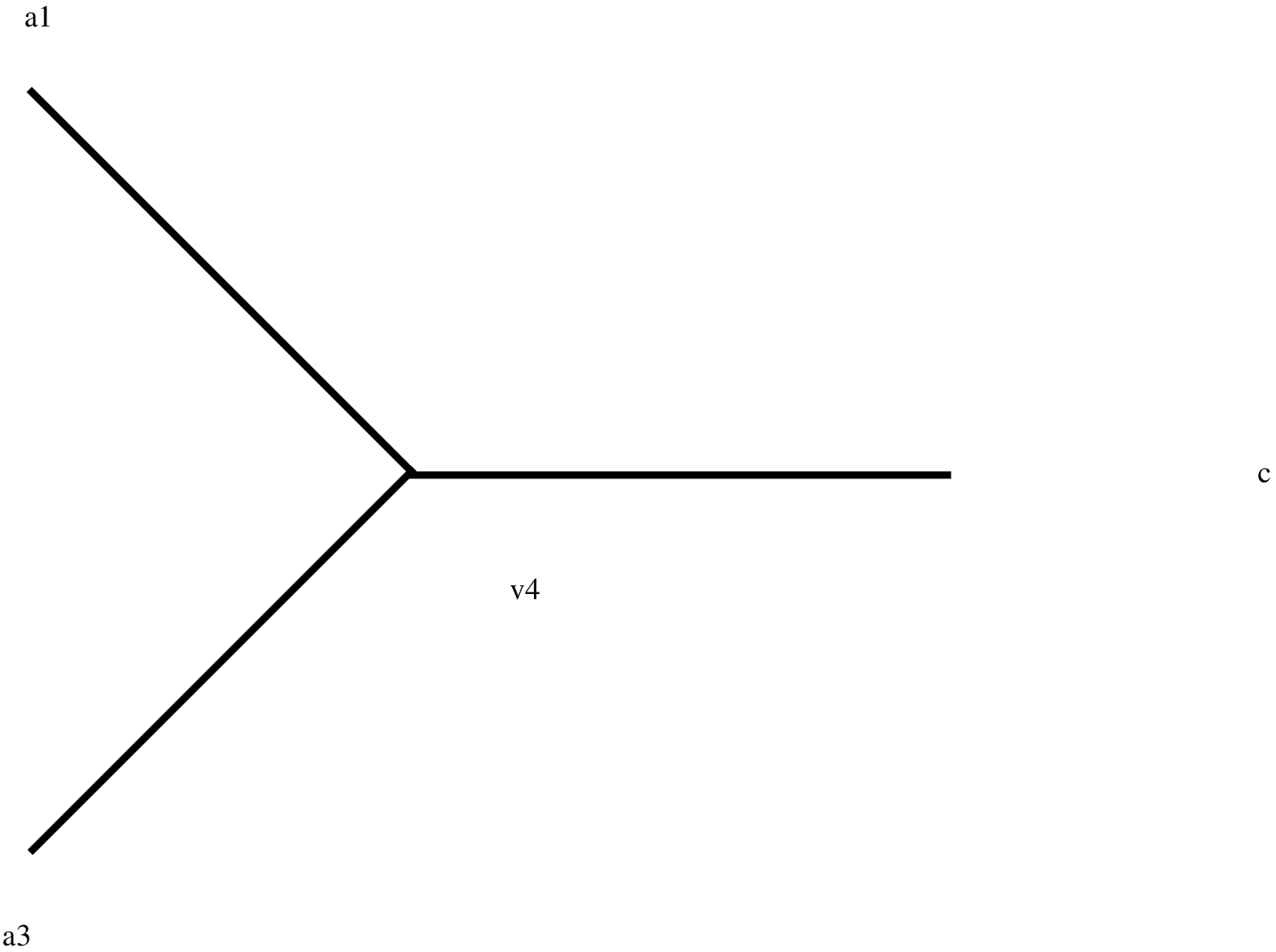}} 
\end{psfrags}
&~~~&
\parbox[c]{9cm}{$\begin{array}{c}V^{i}_4 =-\f{N^{3/2}}{qg^2} \beta F^{i}_4(n,l^{'},n^{''}) (2\pi)^2\delta^2({\bf k}+{\bf k}^{'}+{\bf k}^{''})\delta_{m+m^{'}+m^{''}}\\ ~~~\mbox{for}~(i=1,2,3)\\
F^{1}_4(n,l^{'},n^{''})=\int dx \left[\phi_n(x)\partial_x \phi_{n^{''}}(x)-\phi_{n^{''}}(x)\partial_x
\phi_n(x)\right. \\ \left.+\phi_n(x)A_{n^{''}}(x) (qx)
-  A_n(x)\phi_{n^{''}}(x) (qx) \right]e^{-il^{'}x}\\
F^{i}_4(n,l^{'},n^{''})=\int dx i(k_i-k_i^{''})\left[A_n(x)A_{n^{''}}(x)+\phi_n(x)\phi_{n^{''}}(x)\right]e^{-il^{'}x}\\
~~~\mbox{for}~(i=2,3)
\end{array}$}
\end{tabular}
\end{center}
\end{table}

\begin{table}[H]
\begin{center}
\begin{tabular}{lcc}
\begin{psfrags}
\psfrag{c}[][]{$C(m^{''},n^{''},{\bf k}^{''})$}
\psfrag{a1}[][]{$\Phi_I^{1}(m,n,{\bf k})$}
\psfrag{a3}[][]{$\Phi_I^3(m^{'},l^{'},{\bf k}^{'})$}
\psfrag{v4}[][]{$V_5$}
\parbox[c]{6cm}{\includegraphics[width= 5 cm,angle=0]{vertex1.eps}} 
\end{psfrags}
&~~~&
\parbox[c]{7cm}{$\begin{array}{c}V_5 =-\f{N^{3/2}}{qg^2} \beta F_5(n,l^{'},n^{''}) (2\pi)^2\delta^2({\bf k}+{\bf k}^{'}+{\bf k}^{''})\delta_{m+m^{'}+m^{''}}\\ 
F_5(n,l^{'},n^{''})=\int dx   \left[A_{n^{''}}(x)\partial_x H_n(x)-(qx)A_{n^{''}}(x) H_n(x)\right.\\
+ \left.(il^{'})A_{n^{''}}(x) H_n(x)-
(qx) \phi_{n^{''}}(x) H_n(x)\right]e^{-qx^2/2}e^{-il^{'}x}{\cal N}^{'}(n)\end{array}$}
\end{tabular}
\end{center}
\end{table}

\begin{table}[H]
\begin{center}
\begin{tabular}{lcc}
\begin{psfrags}
\psfrag{c}[][]{$C(m^{''},n^{''},{\bf k}^{''})$}
\psfrag{a1}[][]{$C(m,n,{\bf k})$}
\psfrag{a3}[][]{$\Phi_1^3(m^{'},l^{'},{\bf k}^{'})$}
\psfrag{v4}[][]{$V^{1}_5$}
\parbox[c]{6cm}{\includegraphics[width= 5 cm,angle=0]{vertex1.eps}} 
\end{psfrags}
&~~~&
\parbox[c]{9cm}{$\begin{array}{c}V^{1}_5 =-\f{N^{3/2}}{qg^2} \beta F^{1}_5(n,l^{'},n^{''}) (2\pi)^2\delta^2({\bf k}+{\bf k}^{'}+{\bf k}^{''})\delta_{m+m^{'}+m^{''}}\\ 
F^{1}_5(n,l^{'},n^{''})=\int dx \left[(il^{'})\phi_{n^{''}}(x)A_n(x)+(il^{'})A_{n^{''}}(x)\phi_n(x)\right.\\
+\left.\partial_x\phi_{n^{''}}(x) A_n(x)+\partial_x\phi_n(x) A_{n^{''}}(x)+2(qx)A_n(x)A_{n^{''}}(x)\right]e^{-il^{'}x}\end{array}$}
\end{tabular}
\end{center}
\end{table}

\begin{table}[H]
\begin{center}
\begin{tabular}{lcc}
\begin{psfrags}
\psfrag{c}[][]{$C(m^{''},n^{''},{\bf k}^{''})$}
\psfrag{a1}[][]{$\tilde{A}^2_i(m,n,{\bf k})$}
\psfrag{a3}[][]{$\Phi_1^3(m^{'},l^{'},{\bf k}^{'})$}
\psfrag{v4}[][]{$\tilde{V}^{i}_5$}
\parbox[c]{6cm}{\includegraphics[width= 5 cm,angle=0]{vertex1.eps}} 
\end{psfrags}
&~~~&
\parbox[c]{11cm}{$\begin{array}{c}\tilde{V}^{i}_5 =-\f{N^{3/2}}{qg^2} \beta \tilde{F}^{i}_5(n,l^{'},n^{''}) (2\pi)^2\delta^2({\bf k}+{\bf k}^{'}+{\bf k}^{''})\delta_{m+m^{'}+m^{''}}\\ ~\mbox{for}~(i=1,2,3)\\
\tilde{F}^{1}_5(n,l^{'},n^{''})=i\int dx \left[(il^{'})\phi_{n^{''}}(x)\tilde{A}_{n+1}(x)+(il^{'})A_{n^{''}}(x)\tilde{\phi}_{n+1}(x)\right.\\
\left. +\partial_x\phi_{n^{''}}(x) \tilde{A}_{n+1}(x)+\partial_x\tilde{\phi}_{n+1}(x) A_{n^{''}}(x)+2(qx)A_{n^{''}}(x)\tilde{A}_{n+1}(x)\right]e^{-il^{'}x}\\
\tilde{F}^{i}_5(n,l^{'},n^{''})=i\int dx \left[(k^{''}_i-k^{'}_i)\phi_{n^{''}}(x)H_n(x)\right]e^{-qx^2/2}e^{-il^{'}x}{\cal N}^{'}(n)\\~\mbox{for}~(i=2,3)
\end{array}$}
\end{tabular}
\end{center}
\end{table}

\begin{table}[H]
\begin{center}
\begin{tabular}{lcc}
\begin{psfrags}
\psfrag{c}[][]{$C(m^{''},n^{''},{\bf k}^{''})$}
\psfrag{a1}[][]{$\tilde{A}^1_i(m,n,{\bf k})$}
\psfrag{a3}[][]{$A_j^3(m^{'},l^{'},{\bf k}^{'})$}
\psfrag{v4}[][]{$\tilde{V}^{ij}_5$}
\parbox[c]{6cm}{\includegraphics[width= 5 cm,angle=0]{vertex1.eps}} 
\end{psfrags}
&~~~&
\parbox[c]{11cm}{$\begin{array}{l}\tilde{V}^{ij}_5 =-\f{N^{3/2}}{qg^2} \beta \tilde{F}^{ij}_5(n,l^{'},n^{''}) (2\pi)^2\delta^2({\bf k}+{\bf k}^{'}+{\bf k}^{''})\delta_{m+m^{'}+m^{''}}\\ 
\tilde{F}^{11}_5(n,l^{'},n^{''})=-i\int dx \left[A_{n^{''}}(x)\tilde{\phi}_{n+1}(x)(qx)-\phi_{n^{''}}(x)\tilde{A}_{n+1}(x)(qx)\right.\\
~~~~~~~~\left. +\partial_x\phi_{n^{''}}(x) \tilde{\phi}_{n+1}(x)-\partial_x\tilde{\phi}_{n+1}(x) \phi_{n^{''}}(x)\right]e^{-il^{'}x}\\
\tilde{F}^{i1}_5(n,l^{'},n^{''})=i\int dx \left[(k^{''}_i-k^{'}_i)A_{n^{''}}(x)H_n(x)\right]e^{-qx^2/2}e^{-il^{'}x}{\cal N}^{'}(n)\\~~~~~~~\mbox{for}~(i=2,3)\\
\tilde{F}^{1i}_5(n,l^{'},n^{''})=\int dx (k_i-k^{''}_i)\left[A_{n^{''}}(x)\tilde{A}_{n+1}(x)+\phi_{n^{''}}(x)\tilde{\phi}_{n+1}(x)\right]\times\\~~~~~~~~\times e^{-il^{'}x}~~~\mbox{for}~(i=2,3)\\
\tilde{F}^{ij}_5(n,l^{'},n^{''})=\delta^{ij}\int dx \left[(il^{'})A_{n^{''}}(x)H_n(x)e^{-qx^2/2}\right.\\ ~~~~\left.+\partial_x(e^{-qx^2/2}H_n(x))A_{n^{''}}(x)
 -(qx)\phi_{n^{''}}(x) e^{-qx^2/2} H_n(x)\right]e^{-il^{'}x}{\cal N}^{'}(n)\\
~~~~~~~~~~~~~~~~~~~~~~~~~~~~~~~~~~~~~~~~~~~(i,j=2,3)
\end{array}$}
\end{tabular}
\end{center}
\end{table}

\subsection{Fermions} \label{fermions}

The terms in the action (\ref{actionfermion}) that contain $C(m,n,{\bf k})$ are

\beqa\label{fermionsetstachyon}
&~&\f{1}{g^2}\mbox{tr} \int d^4z \left(\bar{\lambda_I}[\alpha^{1}_{IJ}\Phi_{1},\lambda_J]+\lambda_I\gamma^{1}[A_1,\lambda_I]\right)\\\nonumber
&=& \f{i}{2g^2}\int d^4z \left(\bar{\lambda}^2_4\Phi^{1}_{1}\lambda^3_1+\bar{\lambda}^3_1\Phi^{1}_{1}\lambda^2_4+\bar{\lambda}^1_1\gamma^{1}A^{2}_{1}\lambda^3_1-\bar{\lambda}^3_1\gamma^{1}A^2_1\lambda^1_1\right)+ \cdots
\eeqa 

where $\cdots$ contain other similar terms involving the other three sets $(\lambda^1_2, \lambda^2_3),(\lambda^1_3, \lambda^2_2),(\lambda^1_4, \lambda^2_1)$. Inserting the mode expansions of the fields we can rewrite the action as

\beqa
&~&\sum_{m,m^{'},m^{''}}\sum_{n,n^{'}}\int \f{d^2{\bf k}}{(2\pi\sqrt{q})^2}\f{d^2{\bf k}^{'}}{(2\pi\sqrt{q})^2}\f{d^3{k}^{''}}{(2\pi\sqrt{q})^3}\times\\
&\times&\left[\bar{\lambda}^3_1(m^{''},k^{''})V_f(n,n^{'},k^{''}_x)\theta_1(m^{'},n^{'},{\bf k}^{'})C(m,n,{\bf k}) 
+ \bar{\theta_1}(m^{'},n^{'},{\bf k}^{'})V^{*}_f(n,n^{'},k^{''}_x)\lambda^3_1(m^{''},k^{''})C(m,n,{\bf k})+\cdots\right]\nonumber
\eeqa

The required vertices are now identified as 

\begin{table}[H]
\begin{center}
\begin{tabular}{lcc}
\begin{psfrags}
\psfrag{c}[][]{$C(m,n,{\bf k})$}
\psfrag{a1}[][]{$\theta_1(m^{'},n^{'},{\bf k}^{'})$}
\psfrag{a3}[][]{$\bar{\lambda}^3_1(m^{''},k_x^{''},{\bf k}^{''})$}
\psfrag{v4}[][]{$V_{f}$}
\parbox[c]{5cm}{\includegraphics[width= 4 cm,angle=0]{vertex1.eps}} 
\end{psfrags}
&~~~&
\parbox[c]{8cm}{$\begin{array}{c}V_f =i\f{N}{qg^2}\gamma^{1} F_6(n,n^{'},k^{''}_x)(2\pi)^2\delta^2({\bf k}+{\bf k}^{'}-{\bf k}^{''})\delta_{m+m^{'}-m^{''}}\\ 
F_6(n,n^{'},k^{''}_x)=\sqrt{q}\int dx  \left(L_{n^{'}}(x)A_{n}(x)+R_{n^{'}}(x)\phi_{n}(x)\right)e^{ik^{''}_x x}\end{array}$}

\\
\begin{psfrags}
\psfrag{c}[][]{$C(m,n,{\bf k})$}
\psfrag{a1}[][]{$\bar{\theta}_1(m^{'},n^{'},{\bf k}^{'})$}
\psfrag{a3}[][]{$\lambda^3_1(m^{''},k_x^{''},{\bf k}^{''})$}
\psfrag{v4}[][]{$V^{*}_{f}$}
\parbox[c]{5cm}{\includegraphics[width= 4 cm,angle=0]{vertex1.eps}} 
\end{psfrags}
&~~~&
\parbox[c]{8cm}{$\begin{array}{c}V^{*}_f =-i\f{N}{qg^2}\gamma^{1} F^{*}_6(n,n^{'},k^{''}_x)(2\pi)^2\delta^2({\bf k}-{\bf k}^{'}+{\bf k}^{''})\delta_{m-m^{'}+m^{''}}\\ F^{*}_6(n,n^{'},k^{''}_x)=\sqrt{q}\int dx  \left(L^{*}_{n^{'}}(x)A_{n}(x)+R^{*}_{n^{'}}(x)\phi_{n}(x)\right)e^{-ik^{''}_x x}\end{array}$}

\end{tabular}
\end{center}
\end{table}

and similarly there are three other sets of vertices coming from the other terms in the action.

\section{Modified propagators and Tachyon amplitude}\label{amplitudem}

As mentioned before for the $\Phi_I^3$ fields the mass terms in the propagators in the loop are easy to implement. We thus focus on the 
$A_i^3$ fields. 

We first give some details of derivation of the corrected propagator for the $A_i^3$ fields.
The one-loop effective action is

\beqa\label{Aactionm1}
&&-\f{1}{2qg^2}\int\f{dk_xd^2{\bf k}}{(2\pi\sqrt{q})^3}\sum_{m}\left[\tilde{A}_i^3(m, k){\cal O}^{ij}\tilde{A}_j^3(-m,-k)\right]
\eeqa

where 

\beqa
{\cal O}^{ij}= (k^2+m_2^2)\delta^{ij}-(1-a)k^ik^j++bk_x(u^ik^j+u^jk^i)+(m_1^2-m_2^2+c(k_x)^2)u^iu^j
\eeqa

Here  $(i,j=1,2,3)$, $k^2=(\o_m^2+k_x^2+{\bf k}^2)$ and $u^{i} \equiv (1,0,0)$.  
The inverse of the operator ${\cal O}^{ij}$ can be written in terms of the tensors $\delta^{ij}$, $k^ik^j$, $u^ik^j+u^jk^i$ and  $u^iu^j$. We thus write the inverse as

\beqa
\left({\cal O}^{-1}\right)^{ij}= \mathfrak{A}\delta^{ij}+\mathfrak{B}k^ik^j+\mathfrak{C}(u^ik^j+u^jk^i)+\mathfrak{D}u^iu^j
\eeqa

Demanding that ${\cal O}^{ik}\left({\cal O}^{-1}\right)^{kj}=\delta^{ij}$ and noting that $u.u=1$ and $u.k=k_x$ we get the following equations 

\beqa\label{propeqn}
&&\mathfrak{A}(k^2+m_2^2)=1~~;~~\mathfrak{A}(a-1)+\mathfrak{B}x+\mathfrak{C}y=0~~;~~\mathfrak{A}(m_1^2-m_2^2+c(k_x)^2)+\mathfrak{C}w+ \mathfrak{D}z=0\non
&&\mathfrak{A}bk_x+\mathfrak{B}w+\mathfrak{C}z=0~~;~~\mathfrak{A}bk_x+\mathfrak{C}x+\mathfrak{D}y=0
\eeqa

where 

\beqa
&&x=\o_m^2+(a+b)(k_x)^2+a|{\bf k}|^2+m_2^2~~;~~y=(a+b-1)k_x~~;~~w=k_x((b+c)(k_x)^2+|{\bf k}|^2+m_1^2-m_2^2)\non
&&z=\o_m^2+(b+c+1)(k_x)^2+|{\bf k}|^2+m_1^2).
\eeqa

It can be checked that the equations (\ref{propeqn}) are not all independent. The solution to these equations can be easily worked out which in turn gives the desired propagator. For the case when $a=b=c=0$ the explicit form that is used in the calculations is given by

\beqa
\mathfrak{A}=A~~;~~\mathfrak{B}=AB~~;~~\mathfrak{C}=ABC~~;~~\mathfrak{D}=ABD
\eeqa

where, the functions $A$, $B$, $C$ and $D$ are given by 

\beqa
&& A=\f{1}{k^2+m_2^2}~~~;~~~B=\left(\f{k^2+m_1^2}{k^2+m_1^2+m_2^2}\right)\left(\f{1}{\o_m^2+\f{k_x^2m_1^2+|{\bf k}|^2m_2^2+m_1^2m_2^2}{k^2+m_1^2+m_2^2}}\right)\non
&&C=\f{k_x\left(m_2^2-m_1^2\right)}{k^2+m_1^2}~~;~~D=\left(m_2^2-m_1^2\right)\left(\f{\o_m^2+m_2^2}{k^2+m_1^2}\right)
\eeqa


We now write down parts of the modified tachyon amplitude that essentially contains the $A^3_i$ fields propagating in the loop. 
Writing $k_x=l$

\beqa
(IV))&=&\hf N\sum_{m}\int \f{d l}{(2\pi\sqrt{q})}\int \f{d^2 {\bf k}}{(2\pi\sqrt{q})^2} \left(F_2^3(0,0,l,-l)\right)\left[\f{5}{(\o_m^2+l^2+|{\bf k}|^2+m^2_{\Phi_I})}+\left[2\mathfrak{A}+\mathfrak{B}|{\bf k}|^2\right]\right]\non
\eeqa

\beqa
(V)&=&\hf N\sum_{m}\int \f{d l}{(2\pi\sqrt{q})}\int \f{d^2 {\bf k}}{(2\pi\sqrt{q})^2} \left(F_3^{'}(0,0,l,-l)\right)\left[\mathfrak{A}+\mathfrak{B}l^2+2\mathfrak{C}l+\mathfrak{D}\right]
\eeqa

\beqa
(VII)&=&-\hf N\sum_{m,n}\int \f{d l}{(2\pi\sqrt{q})} \int \f{d^2 {\bf k}}{(2\pi\sqrt{q})^2}\left[\left(q F_4^1(0,n,l)F_4^1(0,n,-l)\right)\left[\f{\mathfrak{A}+\mathfrak{B}l^2+2\mathfrak{C}l+\mathfrak{D}}{(\o_m^2+\lambda_n+|{\bf k}|^2)}\right]\right.\non
&+&\left.\left(- \tilde{{\cal F}}_4(0,n,l)\tilde{{\cal F}}_4(0,n,-l)\right)\left[\f{|{\bf k}|^2(\mathfrak{A}+\mathfrak{B}|{\bf k}|^2)}{(\o_m^2+\lambda_n+|{\bf k}|^2)}\right]\right.\non
&+&\left.\left(2 \sqrt{q} F_4^1(0,n,l)\tilde{{\cal F}}_4(0,n,-l)\right)\left[\f{|{\bf k}|^2(\mathfrak{B}l+\mathfrak{C})}{(\o_m^2+\lambda_n+|{\bf k}|^2)}\right]\right]
\eeqa

\beqa
(XI)&=&-\hf Nq \sum_{m,n}\int \f{d l}{(2\pi\sqrt{q})} \int \f{d^2 {\bf k}}{(2\pi\sqrt{q})^2}\left(\f{1}{(\o_m^2)(\o_m^2+\gamma_n+|{\bf k}|^2)}\right)\times\non
&\times&\left[\left(\tilde{F}^{11}_5(0,n,l)\tilde{F}^{11}_5(0,n,-l)\right)(\o_m^2+\gamma_n)(\mathfrak{A}+\mathfrak{B}l^2+2\mathfrak{C}l+\mathfrak{D})\right.\non
&+&\left.\left(-2\tilde{F}^{11}_5(0,n,l) \tilde{{\cal F}}_5(0,n,-l)\right)\left(\sqrt{\gamma_n} |{\bf k}|^2\right)(\mathfrak{B}l+\mathfrak{C})\right.\non
&+&\left.\left(-(2/\sqrt{q})\tilde{F}^{11}_5(0,n,l)\tilde{{\cal F}}^{1'}_5(0,n,-l)\right)\left( |{\bf k}|^2(\o_m^2+\gamma_n)\right)(\mathfrak{B}l+\mathfrak{C})\right.\non
&+&\left.\left(2\sqrt{(2n+1)}\tilde{F}^{11}_5(0,n,l) \tilde{{\cal F}}^{1}_5(0,n,-l)\right)\left(|{\bf k}|^2(\mathfrak{A}+\mathfrak{B}l^2+2\mathfrak{C}l+\mathfrak{D})\right)\right.\non
&+&\left.\left(-\tilde{{\cal F}}^{1}_5(0,n,l)\tilde{{\cal F}}^{1}_5(0,n,-l)\right)\left( (|{\bf k}|^2/q)(\o_m^2+|{\bf k}|^2)(\mathfrak{A}+\mathfrak{B}l^2+2\mathfrak{C}l+\mathfrak{D})\right)\right.\non
&+&\left.\left(\tilde{{\cal F}}^{1'}_5(0,n,l) \tilde{{\cal F}}^{1'}_5(0,n,-l)\right)\left((|{\bf k}|^2/q)(\o_m^2+\gamma_n)(\mathfrak{A}+\mathfrak{B}|{\bf k}|^2)\right)\right.\non
&+&\left.\left((2/\sqrt{q})\sqrt{(2n+1)}\tilde{{\cal F}}^{1}_5(0,n,l)\tilde{{\cal F}}^{1'}_5(0,n,-l)\right)\left( |{\bf k}|^2\right)^2(\mathfrak{B}l+\mathfrak{C})\right.\non
&+&\left.\left((2/\sqrt{q})\tilde{{\cal F}}_5(0,n,l) \tilde{{\cal F}}^{1}_5(0,n,-l)\right)\left(|{\bf k}|^2(\o_m^2+|{\bf k}|^2)(\mathfrak{B}l+\mathfrak{C})\right)\right.\non
&+&\left.\left(-2\sqrt{(2n+1)}\tilde{{\cal F}}_5(0,n,l) \tilde{{\cal F}}^{1'}_5(0,n,-l)\right)\left(|{\bf k}|^2(\mathfrak{A}+\mathfrak{B}|{\bf k}|^2)\right)\right.\non
&+&\left.\left( {\cal \tilde{F}}_5(0,n,l){\cal \tilde{F}}_5(0,n,-l)\right)\left(\mathfrak{A}(2(\o_m^2)+ |{\bf k}|^2)+\mathfrak{B}|{\bf k}|^2(\o_m^2+|{\bf k}|^2)\right)\right]
\eeqa

\section{Vertices for computation of two point amplitudes for $\Phi_{1,2}^3$ and $A_{\mu}^3$ fields}\label{masslessvertices}

\subsection{$\Phi_1^3$ vertices}\label{verticesphi1}

\begin{table}[H]
\begin{center}
\begin{tabular}{lcc}
\begin{psfrags}
\psfrag{a11}[][]{\scalebox{0.8}{$\Phi_I^{(1,2)/}/\tilde{\Phi}_I^{(1,2)}$}}
\psfrag{a12}[][]{\scalebox{0.8}{$\Phi_I^{(1,2)}/\tilde{\Phi}_I^{(1,2)}$}}
\psfrag{c1}[][]{\scalebox{0.8}{$\Phi_1^3(m^{''},n^{''},{\bf k}^{''})$}}
\psfrag{c2}[][]{\scalebox{0.8}{$\Phi_1^3(\tilde{m}^{''},\tilde{n}^{''},\tilde{{\bf k}}^{''})$}}
\psfrag{v2}[][]{$V^{(1)}_1$}
\parbox[c]{3cm}{\includegraphics[width= 2.5cm,angle=0]{vertex3.eps}}
\label{v1}
\end{psfrags}
&~~~&
\parbox[c]{13cm}{$\begin{array}{c}V^{(1)}_1=-\f{N}{2qg^2} G^{(1)}_1(n,n^{'},k_x^{''},\tilde{k}_x^{''}) (2\pi)^2\delta^2({\bf k}+{\bf k}^{'}+{\bf k}^{''}+\tilde{{\bf k}}^{''})\delta_{m+m^{'}+m^{''}+\tilde{m}^{''}}\\ 
G^{(1)}_1(n,n^{'},k_x^{''},\tilde{k}_x^{''})=\sqrt{q}\int dx [e^{-qx^2}H_n(x)H_{n^{'}}(x)]e^{-(k_x^{''}+\tilde{k}_x^{''})x}
\end{array}$}
\end{tabular}
\end{center}
\end{table}

\begin{table}[H]
\begin{center}
\begin{tabular}{lcc}
\begin{psfrags}
\psfrag{a11}[][]{\scalebox{0.8}{$C/C^{'}(m,n,{\bf k})$}}
\psfrag{a12}[][]{\scalebox{0.8}{$C/C^{'}(m^{'},n^{'},{\bf k}^{'})$}}
\psfrag{c1}[][]{\scalebox{0.8}{$\Phi_1^3(m^{''},n^{''},{\bf k}^{''})$}}
\psfrag{c2}[][]{\scalebox{0.8}{$\Phi_1^3(\tilde{m}^{''},\tilde{n}^{''},\tilde{{\bf k}}^{''})$}}
\psfrag{v2}[][]{$V^{(1)}_2$}
\parbox[c]{3cm}{\includegraphics[width= 2.5cm,angle=0]{vertex3.eps}}
\label{v1}
\end{psfrags}
&~~~&
\parbox[c]{13cm}{$\begin{array}{c}V^{(1)}_2=-\f{N}{2qg^2} G^{(1)}_2(n,n^{'},k_x^{''},\tilde{k}_x^{''}) (2\pi)^2\delta^2({\bf k}+{\bf k}^{'}+{\bf k}^{''}+\tilde{{\bf k}}^{''})\delta_{m+m^{'}+m^{''}+\tilde{m}^{''}}\\ 
G^{(1)}_2(n,n^{'},k_x^{''},\tilde{k}_x^{''})=\sqrt{q}\int dx [A_n(x)A_{n^{'}}(x)]e^{-(k_x^{''}+\tilde{k}_x^{''})x}
\end{array}$}
\end{tabular}
\end{center}
\end{table}

\begin{table}[H]
\begin{center}
\begin{tabular}{lcc}
\begin{psfrags}
\psfrag{a11}[][]{\scalebox{0.8}{$\tilde{A}_{2,,3}^{(1,2)}/\tilde{A}_1^{(1,2)}$}}
\psfrag{a12}[][]{\scalebox{0.8}{$\tilde{A}_{2,,3}^{(1,2)}/\tilde{A}_1^{(1,2)}$}}
\psfrag{c1}[][]{\scalebox{0.8}{$\Phi_1^3(m^{''},n^{''},{\bf k}^{''})$}}
\psfrag{c2}[][]{\scalebox{0.8}{$\Phi_1^3(\tilde{m}^{''},\tilde{n}^{''},\tilde{{\bf k}}^{''})$}}
\psfrag{v2}[][]{$\tilde{V}^{(1)ij}_2$}
\parbox[c]{3cm}{\includegraphics[width= 2.5cm,angle=0]{vertex3.eps}}
\label{v1}
\end{psfrags}
&~~~&
\parbox[c]{13cm}{$\begin{array}{c}\tilde{V}^{(1)ij}_2=-\f{N}{2qg^2} \tilde{G}^{(1)ij}_2(n,n^{'},k_x^{''},\tilde{k}_x^{''}) (2\pi)^2\delta^2({\bf k}+{\bf k}^{'}+{\bf k}^{''}+\tilde{{\bf k}}^{''})\delta_{m+m^{'}+m^{''}+\tilde{m}^{''}}\\ 
\tilde{G}^{(1)11}_2(n,n^{'},k_x^{''},\tilde{k}_x^{''})=-\sqrt{q}\int dx [\tilde{A}_{n+1}(x)\tilde{A}_{n^{'}+1}(x)]e^{-(k_x^{''}+\tilde{k}_x^{''})x}
\\
\tilde{G}^{(1)ij}_2(n,n^{'},k_x^{''},\tilde{k}_x^{''})=\sqrt{q}\int dx [e^{-qx^2}H_n(x)H_{n^{'}}(x)]e^{-(k_x^{''}+\tilde{k}_x^{''})x}\delta_{ij}~~\mbox{for}~~i,j=2,3
\end{array}$}
\end{tabular}
\end{center}
\end{table}

\begin{table}[H]
\begin{center}
\begin{tabular}{lcc}
\begin{psfrags}
\psfrag{c}[][]{$\Phi_1^3(m^{''},k_x^{''},{\bf k}^{''})$}
\psfrag{a1}[][]{$C/C^{'}(m,n,{\bf k})$}
\psfrag{a3}[][]{$C/C^{'}(m^{'},n^{'},{\bf k}^{'})$}
\psfrag{v4}[][]{$V^{(1)}_3$}
\parbox[c]{5.1cm}{\includegraphics[width= 5 cm,angle=0]{vertex1.eps}} 
\end{psfrags}
&~~~&
\parbox[c]{10cm}{$\begin{array}{c}V^{(1)}_3 =-\f{N^{3/2}}{qg^2} \beta G^{(1)}_3(n,n^{'},k_x^{''}) (2\pi)^2\delta^2({\bf k}+{\bf k}^{'}+{\bf k}^{''})\delta_{m+m^{'}+m^{''}}\\ 
G^{(1)}_3(n,n^{'},k_x^{''})=\int dx \left[\partial_x\phi_n(x) A_{n^{'}}(x)+ik_x^{''}\phi_{n}(x)
A_{n^{'}}(x)+\right.\\\left.(qx) A_n(x)A_{n^{'}}(x)  \right]e^{-ik_x^{''}x}
\end{array}$}
\end{tabular}
\end{center}
\end{table}

\begin{table}[H]
\begin{center}
\begin{tabular}{lcc}
\begin{psfrags}
\psfrag{c}[][]{$\Phi_1^3(m^{''},k_x^{''},{\bf k}^{''})$}
\psfrag{a1}[][]{$C/C^{'}(m,n,{\bf k})$}
\psfrag{a3}[][]{$\tilde{A}_{2,3}^{(1,2)}/\tilde{A}_{1}^{(1,2)}$}
\psfrag{v4}[][]{$\tilde{V}^{(1)i}_3$}
\parbox[c]{5.1cm}{\includegraphics[width= 5 cm,angle=0]{vertex1.eps}} 
\end{psfrags}
&~~~&
\parbox[c]{10cm}{$\begin{array}{c}\tilde{V}^{(1)i}_3 =-\f{N^{3/2}}{qg^2} \beta \tilde{G}^{(1)i}_3(n,n^{'},k_x^{''}) (2\pi)^2\delta^2({\bf k}+{\bf k}^{'}+{\bf k}^{''})\delta_{m+m^{'}+m^{''}}\\ 
\tilde{G}^{(1)1}_3(n,n^{'},k_x^{''})=\pm i\int dx \left[\partial_x\phi_n(x) \tilde{A}_{n^{'}+1}(x)+\partial_x\tilde{\phi}_{n^{'}+1}(x) A_{n}(x)\right.\\\left.+ik_x^{''}\phi_{n}(x)\tilde{A}_{n^{'}+1}(x)+ik_x^{''}\tilde{\phi}_{n^{'}+1}(x)A_n(x)+2(qx) A_n(x)\tilde{A}_{n^{'}+1}(x)  \right]e^{-ik_x^{''}x}\\
\tilde{G}^{(1)i}_3(n,n^{'},k_x^{''})=i(k_i-k^{''}_i)\int dx \left[\phi_n(x)H_{n^{'}}(x)e^{-qx^2/2}\right]e^{-ik_x^{''}x}\\~~\mbox{for}~~i=2,3
\end{array}$}
\end{tabular}
\end{center}
\end{table}

\begin{table}[H]
\begin{center}
\begin{tabular}{lcc}
\begin{psfrags}
\psfrag{c}[][]{$\Phi_1^3(m^{''},k_x^{''},{\bf k}^{''})$}
\psfrag{a1}[][]{$\tilde{A}_{2,3}^{(1,2)}/\tilde{A}_{1}^{(1,2)}$}
\psfrag{a3}[][]{$\tilde{A}_{2,3}^{(1,2)}/\tilde{A}_{1}^{(1,2)}$}
\psfrag{v4}[][]{$\tilde{V}^{(1)ij}_3$}
\parbox[c]{5.1cm}{\includegraphics[width= 5 cm,angle=0]{vertex1.eps}} 
\end{psfrags}
&~~~&
\parbox[c]{10cm}{$\begin{array}{c}\tilde{V}^{(1)ij}_3 =-\f{N^{3/2}}{qg^2} \beta \tilde{G}^{(1)ij}_3(n,n^{'},k_x^{''}) (2\pi)^2\delta^2({\bf k}+{\bf k}^{'}+{\bf k}^{''})\delta_{m+m^{'}+m^{''}}\\ 
\tilde{G}^{(1)11}_3(n,n^{'},k_x^{''})=-\int dx \left[\partial_x\tilde{\phi}_{n+1}(x) \tilde{A}_{n^{'}+1}(x)+ik_x^{''}\tilde{\phi}_{n+1}(x)
\tilde{A}_{n^{'}+1}(x)+\right.\\\left.(qx) \tilde{A}_{n+1}(x)\tilde{A}_{n^{'}+1}(x)  \right]e^{-ik_x^{''}x}\\
\tilde{G}^{(1)1i}_3(n,n^{'},k_x^{''})=\pm i(k_i-k^{''}_i)\int dx \left[\tilde{\phi}_{n+1}(x)H_{n^{'}}(x)e^{-qx^2/2}\right]e^{-ik_x^{''}x}\\~~\mbox{for}~~i=2,3\\
\tilde{G}^{(1)ij}_3(n,n^{'},k_x^{''})=\int dx \left[(qx) e^{-qx^2}H_n(x)H_{n^{'}}(x)\right]e^{-ik_x^{''}x}\times\delta_{ij}\\~~\mbox{for}~~i,j=2,3
\\
\mbox{We also have additional vertices  $V^{(1)}_4$ with functions $G^{(1)}_4$ same as }\\ \mbox{$\tilde{G}^{(1)ij}$ for the $\Phi_I^{(1,2)}/\tilde{\Phi}_I^{(1,2)}$  fields in place of $\tilde{A}_{2,3}^{(1,2)}$ fields.}
\end{array}$}
\end{tabular}
\end{center}
\end{table}

We now compute the vertices involving the fermions. The relevant terms in the action that contains this $\Phi_1^3$ field are
\beqa\label{fermionsetsphi13}
&~&\f{1}{g^2}\mbox{tr} \int d^4z \left(\bar{\lambda_I}[\alpha^{1}_{IJ}\Phi_{1},\lambda_J]\right)\\\nonumber
&=& -\f{i}{2g^2}\int d^4z \left(\bar{\lambda}^1_1\Phi^{3}_{1}\lambda^2_4+\bar{\lambda}^2_4\Phi^{3}_{1}\lambda^1_1\right)+ \cdots
\eeqa 

where the omitted terms correspond to the other three pairs $(\lambda^1_2, \lambda^2_3),(\lambda^1_3, \lambda^2_2),(\lambda^1_4, \lambda^2_1)$.
The action can be written in terms of momentum modes as

\beqa
&~&\sum_{m,m^{'},m^{''}}\sum_{n,n^{'}}\int \f{d^2{\bf k}}{(2\pi\sqrt{q})^2}\f{d^2{\bf k}^{'}}{(2\pi\sqrt{q})^2}\f{d^3{k}^{''}}{(2\pi\sqrt{q})^3}\times\non
&\times&\left[\bar{\theta}_1(m,n, {\bf k})V^1_f(n,n^{'},k^{''}_x)\theta_1(m^{'},n^{'},{\bf k}^{'})\Phi^3_1(m^{''}, k^{''}) 
+ \theta_1^{T}\gamma^0(m,n,{\bf k})V^{1'}_f(n,n^{'},k^{''}_x)\theta_1(m^{'},n^{'},{\bf k}^{'})\Phi_1^3(m^{''}, k^{''})\right.\non
&~&\left.+ \bar{\theta}_1(m^{'},n^{'},{\bf k}^{'})V^{1'*}_f(n,n^{'},k^{''}_x)\theta^{*}_1(m,n,{\bf k})\Phi_1^3(m^{''}, k^{''})+\cdots\right]
\eeqa

The required vertices are now identified as 

\begin{table}[H]
\begin{center}
\begin{tabular}{lcc}
\begin{psfrags}
\psfrag{c}[][]{$\Phi_1^3(m^{''}, k^{''})$}
\psfrag{a1}[][]{$\bar{\theta_1}(m,n,{\bf k})$}
\psfrag{a3}[][]{$\theta_1(m^{'},n^{'},{\bf k}^{'})$}
\psfrag{v4}[][]{$V_f^1$}
\parbox[c]{5cm}{\includegraphics[width= 4 cm,angle=0]{vertex1.eps}} 
\end{psfrags}
&~~~&
\parbox[c]{8cm}{$\begin{array}{c}V^{1}_f =i\f{N}{qg^2}\gamma^{1}G^1_f(n,n^{'},k^{''}_x)(2\pi)^2\delta^2({\bf k}-{\bf k}^{'}-{\bf k}^{''})\delta_{m-m^{'}-m^{''}}\\ 
G^1_f(n,n^{'},k^{''}_x)=\sqrt{q}\int dx  \left(L_{n^{'}}(x)R^{*}_{n}(x)-R_{n^{'}}(x)L^{*}_{n}(x)\right)e^{-ik^{''}_x x}\end{array}$}

\\

\begin{psfrags}
\psfrag{c}[][]{$\Phi^3_1(m^{''}, k^{''})$}
\psfrag{a1}[][]{$\theta^{T}_1\gamma^0(m,n,{\bf k})$}
\psfrag{a3}[][]{$\theta_1(m^{'},n^{'},{\bf k}^{'})$}
\psfrag{v4}[][]{$V^{1'}_{f}$}
\parbox[c]{5cm}{\includegraphics[width= 4 cm,angle=0]{vertex1.eps}}
\end{psfrags}
&~~~&
\parbox[c]{8cm}{$\begin{array}{c}V^{1'}_f =i\f{N}{qg^2}\gamma^{1} G^{1'}_f(n,n^{'},k^{''}_x)(2\pi)^2\delta^2({\bf k}+{\bf k}^{'}+{\bf k}^{''})\delta_{m+m^{'}+m^{''}}\\ G^{1'}_f(n,n^{'},k_x^{''})=\sqrt{q}\int dx  R_n(x)L_{n^{'}}(x) e^{-ik^{''}_x x}\end{array}$}
\\

\begin{psfrags}
\psfrag{c}[][]{$\Phi^3_1(m^{''}, k^{''})$}
\psfrag{a1}[][]{$\bar{\theta}_1(m^{'},n^{'},{\bf k}^{'})$}
\psfrag{a3}[][]{$\theta^{*}_1(m,n,{\bf k})$}
\psfrag{v4}[][]{$V^{1'*}_{f}$}
\parbox[c]{5cm}{\includegraphics[width= 4 cm,angle=0]{vertex1.eps}} 
\end{psfrags}
&~~~&
\parbox[c]{8cm}{$\begin{array}{c}V^{1'*}_f =i\f{N}{qg^2}\gamma^{1} G^{1'*}_f(n,n^{'},k_x^{''})(2\pi)^2\delta^2({\bf k}+{\bf k}^{'}-{\bf k}^{''})\delta_{m+m^{'}-m^{''}}\\ G^{1'*}_f(n,n^{'},k_x^{''})=\sqrt{q}\int dx  R^{*}_n(x)L^{*}_{n^{'}}(x) e^{-ik^{''}_x x}\end{array}$}

\end{tabular}
\end{center}
\end{table}

\subsection{$\Phi_2^3$ vertices}{\label{verticesphi32}}

(There are five massless scalar fields corresponding to $\Phi_{2,3}^3$ and $\tilde{\Phi}^3_I, ~(I=1,2,3)$. Due to an unbroken $SO(5)$ invariance 
in the theory the two point amplitudes for all these fields is same. We thus only consider the vertices here for $\Phi_{2}^3$.) We first write down the vertices containing only bosonic fields.

\begin{table}[H]
\begin{center}
\begin{tabular}{lcc}
\begin{psfrags}
\psfrag{a11}[][]{\scalebox{0.8}{$\Phi_I^{(1,2)/}/\tilde{\Phi}_I^{(1,2)}$}}
\psfrag{a12}[][]{\scalebox{0.8}{$\Phi_I^{(1,2)}/\tilde{\Phi}_I^{(1,2)}$}}
\psfrag{c1}[][]{\scalebox{0.8}{$\Phi_2^3(m^{''},n^{''},{\bf k}^{''})$}}
\psfrag{c2}[][]{\scalebox{0.8}{$\Phi_2^3(\tilde{m}^{''},\tilde{n}^{''},\tilde{{\bf k}}^{''})$}}
\psfrag{v2}[][]{$V^{(2)}_1$}
\parbox[c]{3cm}{\includegraphics[width= 2.5cm,angle=0]{vertex3.eps}}
\label{v1}
\end{psfrags}
&~~~&
\parbox[c]{13cm}{$\begin{array}{c}V^{(2)}_1=-\f{N}{2qg^2} G^{(2)}_1(n,n^{'},k_x^{''},\tilde{k}_x^{''}) (2\pi)^2\delta^2({\bf k}+{\bf k}^{'}+{\bf k}^{''}+\tilde{{\bf k}}^{''})\delta_{m+m^{'}+m^{''}+\tilde{m}^{''}}\\ 
G^{(2)}_1(n,n^{'},k_x^{''},\tilde{k}_x^{''})=\sqrt{q}\int dx [e^{-qx^2}H_n(x)H_{n^{'}}(x)]e^{-(k_x^{''}+\tilde{k}_x^{''})x}
\end{array}$}
\end{tabular}
\end{center}
\end{table}

\begin{table}[H]
\begin{center}
\begin{tabular}{lcc}
\begin{psfrags}
\psfrag{a11}[][]{\scalebox{0.8}{$C/C^{'}(m,n,{\bf k})$}}
\psfrag{a12}[][]{\scalebox{0.8}{$C/C^{'}(m^{'},n^{'},{\bf k}^{'})$}}
\psfrag{c1}[][]{\scalebox{0.8}{$\Phi_2^3(m^{''},n^{''},{\bf k}^{''})$}}
\psfrag{c2}[][]{\scalebox{0.8}{$\Phi_2^3(\tilde{m}^{''},\tilde{n}^{''},\tilde{{\bf k}}^{''})$}}
\psfrag{v2}[][]{$V^{(2)}_2$}
\parbox[c]{3cm}{\includegraphics[width= 2.5cm,angle=0]{vertex3.eps}}
\label{v1}
\end{psfrags}
&~~~&
\parbox[c]{13cm}{$\begin{array}{c}V^{(2)}_2=-\f{N}{2qg^2} G^{(2)}_2(n,n^{'},k_x^{''},\tilde{k}_x^{''}) (2\pi)^2\delta^2({\bf k}+{\bf k}^{'}+{\bf k}^{''}+\tilde{{\bf k}}^{''})\delta_{m+m^{'}+m^{''}+\tilde{m}^{''}}\\ 
G^{(2)}_2(n,n^{'},k_x^{''},\tilde{k}_x^{''})=\sqrt{q}\int dx [A_n(x)A_{n^{'}}(x)+\phi_n(x)\phi_{n^{'}}(x)]e^{-(k_x^{''}+\tilde{k}_x^{''})x}
\end{array}$}
\end{tabular}
\end{center}
\end{table}

\begin{table}[H]
\begin{center}
\begin{tabular}{lcc}
\begin{psfrags}
\psfrag{a11}[][]{\scalebox{0.8}{$\tilde{A}_{2,,3}^{(1,2)}/\tilde{A}_1^{(1,2)}$}}
\psfrag{a12}[][]{\scalebox{0.8}{$\tilde{A}_{2,,3}^{(1,2)}/\tilde{A}_1^{(1,2)}$}}
\psfrag{c1}[][]{\scalebox{0.8}{$\Phi_2^3(m^{''},n^{''},{\bf k}^{''})$}}
\psfrag{c2}[][]{\scalebox{0.8}{$\Phi_2^3(\tilde{m}^{''},\tilde{n}^{''},\tilde{{\bf k}}^{''})$}}
\psfrag{v2}[][]{$\tilde{V}^{(2)ij}_2$}
\parbox[c]{3cm}{\includegraphics[width= 2.5cm,angle=0]{vertex3.eps}}
\label{v1}
\end{psfrags}
&~~~&
\parbox[c]{13cm}{$\begin{array}{c}\tilde{V}^{(2)ij}_2=-\f{N}{2qg^2} \tilde{G}^{(2)ij}_2(n,n^{'},k_x^{''},\tilde{k}_x^{''}) (2\pi)^2\delta^2({\bf k}+{\bf k}^{'}+{\bf k}^{''}+\tilde{{\bf k}}^{''})\delta_{m+m^{'}+m^{''}+\tilde{m}^{''}}\\ 
\tilde{G}^{(2)11}_2(n,n^{'},k_x^{''},\tilde{k}_x^{''})=-\sqrt{q}\int dx [\tilde{A}_{n+1}(x)\tilde{A}_{n^{'}+1}(x)+\tilde{\phi}_{n+1}(x)\tilde{\phi}_{n^{'}+1}(x)]e^{-(k_x^{''}+\tilde{k}_x^{''})x}
\\
\tilde{G}^{(2)ij}_2(n,n^{'},k_x^{''},\tilde{k}_x^{''})=\sqrt{q}\int dx [e^{-qx^2}H_n(x)H_{n^{'}}(x)]e^{-(k_x^{''}+\tilde{k}_x^{''})x}\delta_{ij}~~\mbox{for}~~i,j=2,3
\end{array}$}
\end{tabular}
\end{center}
\end{table}

\begin{table}[H]
\begin{center}
\begin{tabular}{lcc}
\begin{psfrags}
\psfrag{c}[][]{$\Phi_2^3(m^{''},k_x^{''},{\bf k}^{''})$}
\psfrag{a1}[][]{$\Phi_2^{(1,2)}(m,n,{\bf k})$}
\psfrag{a3}[][]{$C/C^{'}(m^{'},n^{'},{\bf k}^{'})$}
\psfrag{v4}[][]{$V^{(2)}_3$}
\parbox[c]{5.1cm}{\includegraphics[width= 5 cm,angle=0]{vertex1.eps}} 
\end{psfrags}
&~~~&
\parbox[c]{11cm}{$\begin{array}{c}V^{(2)}_3 =-\f{N^{3/2}}{qg^2} \beta G^{(2)}_3(n,n^{'},k_x^{''}) (2\pi)^2\delta^2({\bf k}+{\bf k}^{'}+{\bf k}^{''})\delta_{m+m^{'}+m^{''}}\\ 
G^{(2)}_3(n,n^{'},k_x^{''})=\int dx \left[\partial_x(e^{-qx^2/2}H_n(x)) A_{n^{'}}(x)+ik_x^{''}H_{n}(x)e^{-qx^2/2}
A_{n^{'}}(x)\right.\\\left.-(qx) e^{-qx^2/2}H_n(x)\phi_{n^{'}}(x)  \right]e^{-ik_x^{''}x}
\end{array}$}
\end{tabular}
\end{center}
\end{table}

\begin{table}[H]
\begin{center}
\begin{tabular}{lcc}
\begin{psfrags}
\psfrag{c}[][]{$\Phi_2^3(m^{''},k_x^{''},{\bf k}^{''})$}
\psfrag{a1}[][]{$\Phi_2^{(1,2)}(m,n,{\bf k})$}
\psfrag{a3}[][]{$\tilde{A}_{2,3}^{(1,2)}/\tilde{A}_{1}^{(1,2)}$}
\psfrag{v4}[][]{$\tilde{V}^{(2)i}_3$}
\parbox[c]{5.1cm}{\includegraphics[width= 5 cm,angle=0]{vertex1.eps}} 
\end{psfrags}
&~~~&
\parbox[c]{10cm}{$\begin{array}{c}\tilde{V}^{(2)i}_3 =-\f{N^{3/2}}{qg^2} \beta \tilde{G}^{(2)i}_3(n,n^{'},k_x^{''}) (2\pi)^2\delta^2({\bf k}+{\bf k}^{'}+{\bf k}^{''})\delta_{m+m^{'}+m^{''}}\\ 
\tilde{G}^{(2)1}_3(n,n^{'},k_x^{''})=\pm i\int dx \left[\partial_x(e^{-qx^2/2}H_n(x)) \tilde{A}_{n^{'}+1}(x)\right.\\\left.+ik_x^{''}e^{-qx^2/2}H_{n}(x)\tilde{A}_{n^{'}+1}(x)-(qx) e^{-qx^2/2}H_n(x)\tilde{\phi}_{n^{'}+1}(x)  \right]e^{-ik_x^{''}x}\\
\tilde{G}^{(2)i}_3(n,n^{'},k_x^{''})=i(k_i-k^{''}_i)\int dx \left[H_n(x)H_{n^{'}}(x)e^{-qx^2}\right]e^{-ik_x^{''}x}\\~~\mbox{for}~~i=2,3
\end{array}$}
\end{tabular}
\end{center}
\end{table}

We now compute the vertices consisting of the fermion fields.
The relevant terms in the action that contain $\Phi_2^3$ field are
\beqa\label{fermionsetsphi23}
&~&\f{1}{g^2}\mbox{tr} \int d^4z \left(\bar{\lambda_I}[\alpha^{2}_{IJ}\Phi_{2},\lambda_J]\right)\\\nonumber
&=& \f{i}{2g^2}\int d^4z \left[\left(\bar{\lambda}^1_1\Phi^{3}_{2}\lambda^2_3+\bar{\lambda}^2_3\Phi^{3}_{2}\lambda^1_1\right)-
\left(\bar{\lambda}^1_2\Phi^{3}_{2}\lambda^2_4+\bar{\lambda}^2_4\Phi^{3}_{2}\lambda^1_2\right)
\right]+ \cdots
\eeqa 

In the above expression the two terms correspond to the coupling between two sets of fermions $(\lambda^1_1, \lambda^2_4)$ and $(\lambda^1_2, \lambda^2_3)$ the omitted terms similarly correspond to couplings between $(\lambda^1_3, \lambda^2_2)$ and $(\lambda^1_4, \lambda^2_1)$.

Inserting the corresponding mode expansions we get

\beqa
&~&\sum_{m,m^{'},m^{''}}\sum_{n,n^{'}}\int \f{d^2{\bf k}}{(2\pi\sqrt{q})^2}\f{d^2{\bf k}^{'}}{(2\pi\sqrt{q})^2}\f{d^3{k}^{''}}{(2\pi\sqrt{q})^3}\times\\
&\times&\left[\bar{\theta}_1(m,n, {\bf k})V^2_f(n,n^{'},k^{''}_x)\theta_2(m^{'},n^{'},{\bf k}^{'})\Phi^3_2(m^{''}, k^{''}) 
+\bar{\theta}_2(m^{'},n^{'}, {\bf k}^{'})V^{2*}_f(n,n^{'},k^{''}_x)\theta_1(m,n,{\bf k})\Phi^3_2(m^{''}, k^{''}) \right.\non
&~&\left.+ \theta_1^{T}\gamma^0(m,n,{\bf k})V^{2^{'}}_f(n,n^{'},k^{''}_x)\theta_2(m^{'},n^{'},{\bf k}^{'})\Phi_2^3(m^{''}, k^{''})
+ \bar{\theta}_1(m,n,{\bf k})V^{2'*}_f(n,n^{'},k^{''}_x)\theta^{*}_2(m^{'},n^{'},{\bf k}^{'})\Phi_2^3(m^{''}, k^{''})+\cdots\right]\nonumber
\eeqa

The required vertices are now identified as 

\begin{table}[H]
\begin{center}
\begin{tabular}{lcc}
\begin{psfrags}
\psfrag{c}[][]{$\Phi_2^3(m^{''}, k^{''})$}
\psfrag{a1}[][]{$\bar{\theta_1}(m,n,{\bf k})$}
\psfrag{a3}[][]{$\theta_2(m^{'},n^{'},{\bf k}^{'})$}
\psfrag{v4}[][]{$V_f^2$}
\parbox[c]{5cm}{\includegraphics[width= 4 cm,angle=0]{vertex1.eps}} 
\end{psfrags}
&~~~&
\parbox[c]{8cm}{$\begin{array}{c}V^{2}_f =i\f{N}{qg^2}\gamma^{1}G^2_f(n,n^{'},k^{''}_x)(2\pi)^2\delta^2({\bf k}-{\bf k}^{'}-{\bf k}^{''})\delta_{m-m^{'}-m^{''}}\\ 
G^2_f(n,n^{'},k^{''}_x)=\sqrt{q}\int dx  \left(L^{*}_n(x)R_{n^{'}}(x)+R^{*}_{n}(x)L_{n^{'}}(x)\right)e^{-ik^{''}_x x}\end{array}$}
\\
\begin{psfrags}
\psfrag{c}[][]{$\Phi_2^3(m^{''}, k^{''})$}
\psfrag{a1}[][]{$\theta_1(m,n,{\bf k})$}
\psfrag{a3}[][]{$\bar{\theta}_2(m^{'},n^{'},{\bf k}^{'})$}
\psfrag{v4}[][]{$V_f^{2*}$}
\parbox[c]{5cm}{\includegraphics[width= 4 cm,angle=0]{vertex1.eps}} 
\end{psfrags}
&~~~&
\parbox[c]{8cm}{$\begin{array}{c}V^{2*}_f =-i\f{N}{qg^2}\gamma^{1}G^{2*}_f(n,n^{'},k^{''}_x)(2\pi)^2\delta^2({\bf k}-{\bf k}^{'}+{\bf k}^{''})\delta_{m-m^{'}+m^{''}}\\ 
G^{2*}_f(n,n^{'},k^{''}_x)=\sqrt{q}\int dx  \left(L_n(x)R^{*}_{n^{'}}(x)+R_{n}(x)L^{*}_{n^{'}}(x)\right)e^{-ik^{''}_x x}\end{array}$}
\end{tabular}
\end{center}
\end{table}

\begin{table}[H]
\begin{center}
\begin{tabular}{lcc}
\begin{psfrags}
\psfrag{c}[][]{$\Phi^3_2(m^{''}, k^{''})$}
\psfrag{a1}[][]{$\theta_{1}(m,n,{\bf k})$}
\psfrag{a3}[][]{$\theta_2^{T}\gamma^0(m^{'},n^{'},{\bf k}^{'})$}
\psfrag{v4}[][]{$V^{2'}_{f}$}
\parbox[c]{5cm}{\includegraphics[width= 4 cm,angle=0]{vertex1.eps}} 
\end{psfrags}
&~~~&
\parbox[c]{8cm}{$\begin{array}{c}V^{2'}_f =-i\f{N}{qg^2}\gamma^{1} G^{2'}_f(n,n^{'},k_x^{''})(2\pi)^2\delta^2({\bf k}+{\bf k}^{'}+{\bf k}^{''})\delta_{m+m^{'}+m^{''}}\\ G^{2'}_f(n,n^{'},k_x^{''})=\sqrt{q}\int dx  \left(L_n(x)R_{n^{'}}(x)+R_n(x)L_{n^{'}}(x) \right)e^{-ik^{''}_x x}\end{array}$}

\\

\begin{psfrags}
\psfrag{c}[][]{$\Phi^3_2(m^{''}, k^{''})$}
\psfrag{a1}[][]{$\bar{\theta}_1(m,n,{\bf k})$}
\psfrag{a3}[][]{$\theta^{*}_2(m^{'},n^{'},{\bf k}^{'})$}
\psfrag{v4}[][]{$V^{2'*}_{f}$}
\parbox[c]{5cm}{\includegraphics[width= 4 cm,angle=0]{vertex1.eps}} 
\end{psfrags}
&~~~&
\parbox[c]{8cm}{$\begin{array}{c}V^{2'*}_f =i\f{N}{qg^2}\gamma^{1} G^{2'*}_f(n,n^{'},k_x^{''})(2\pi)^2\delta^2({\bf k}+{\bf k}^{'}-{\bf k}^{''})\delta_{m+m^{'}-m^{''}}\\ G^{2'*}_f(n,n^{'},k_x^{''})=\sqrt{q}\int dx  \left(L^{*}_n(x)R^{*}_{n^{'}}(x)+R^{*}_n(x)L^{*}_{n^{'}}(x)\right) e^{-ik^{''}_x x}\end{array}$}
\end{tabular}
\end{center}
\end{table}

\subsection{$A_{\mu}^3$ vertices}\label{verticesA3mu}

{\bf Vertices for $A_{1}^3$ consisting of only bosons:}\\

\begin{table}[H]
\begin{center}
\begin{tabular}{lcc}
\begin{psfrags}
\psfrag{a11}[][]{\scalebox{0.8}{$\Phi_I^{(1,2)/}/\tilde{\Phi}_I^{(1,2)}$}}
\psfrag{a12}[][]{\scalebox{0.8}{$\Phi_I^{(1,2)}/\tilde{\Phi}_I^{(1,2)}$}}
\psfrag{c1}[][]{\scalebox{0.8}{$A_1^3(m^{''},n^{''},{\bf k}^{''})$}}
\psfrag{c2}[][]{\scalebox{0.8}{$A_1^3(\tilde{m}^{''},\tilde{n}^{''},\tilde{{\bf k}}^{''})$}}
\psfrag{v2}[][]{$V^{(A_1^3)}_1$}
\parbox[c]{3cm}{\includegraphics[width= 2.5cm,angle=0]{vertex3.eps}}
\label{v1}
\end{psfrags}
&~~~&
\parbox[c]{13cm}{$\begin{array}{c}V^{(A_1^3)}_1=-\f{N}{2qg^2} G^{(A_1^3)}_1(n,n^{'},k_x^{''},\tilde{k}_x^{''}) (2\pi)^2\delta^2({\bf k}+{\bf k}^{'}+{\bf k}^{''}+\tilde{{\bf k}}^{''})\delta_{m+m^{'}+m^{''}+\tilde{m}^{''}}\\ 
G^{(A_1^3)}_1(n,n^{'},k_x^{''},\tilde{k}_x^{''})=\sqrt{q}\int dx [e^{-qx^2}H_n(x)H_{n^{'}}(x)]e^{-(k_x^{''}+\tilde{k}_x^{''})x}
\end{array}$}
\end{tabular}
\end{center}
\end{table}

\begin{table}[H]
\begin{center}
\begin{tabular}{lcc}
\begin{psfrags}
\psfrag{a11}[][]{\scalebox{0.8}{$C/C^{'}(m,n,{\bf k})$}}
\psfrag{a12}[][]{\scalebox{0.8}{$C/C^{'}(m^{'},n^{'},{\bf k}^{'})$}}
\psfrag{c1}[][]{\scalebox{0.8}{$A_1^3(m^{''},n^{''},{\bf k}^{''})$}}
\psfrag{c2}[][]{\scalebox{0.8}{$A_1^3(\tilde{m}^{''},\tilde{n}^{''},\tilde{{\bf k}}^{''})$}}
\psfrag{v2}[][]{$V^{(A_1^3)}_2$}
\parbox[c]{3cm}{\includegraphics[width= 2.5cm,angle=0]{vertex3.eps}}
\label{v1}
\end{psfrags}
&~~~&
\parbox[c]{13cm}{$\begin{array}{c}V^{(A_1^3)}_2=-\f{N}{2qg^2} G^{(A_1^3)}_2(n,n^{'},k_x^{''},\tilde{k}_x^{''}) (2\pi)^2\delta^2({\bf k}+{\bf k}^{'}+{\bf k}^{''}+\tilde{{\bf k}}^{''})\delta_{m+m^{'}+m^{''}+\tilde{m}^{''}}\\ 
G^{(A_1^3)}_2(n,n^{'},k_x^{''},\tilde{k}_x^{''})=\sqrt{q}\int dx [\phi_n(x)\phi_{n^{'}}(x)]e^{-(k_x^{''}+\tilde{k}_x^{''})x}
\end{array}$}
\end{tabular}
\end{center}
\end{table}

\begin{table}[H]
\begin{center}
\begin{tabular}{lcc}
\begin{psfrags}
\psfrag{a11}[][]{\scalebox{0.8}{$\tilde{A}_{2,,3}^{(1,2)}/\tilde{A}_1^{(1,2)}$}}
\psfrag{a12}[][]{\scalebox{0.8}{$\tilde{A}_{2,,3}^{(1,2)}/\tilde{A}_1^{(1,2)}$}}
\psfrag{c1}[][]{\scalebox{0.8}{$A_1^3(m^{''},n^{''},{\bf k}^{''})$}}
\psfrag{c2}[][]{\scalebox{0.8}{$A_1^3(\tilde{m}^{''},\tilde{n}^{''},\tilde{{\bf k}}^{''})$}}
\psfrag{v2}[][]{$\tilde{V}^{(A_1^3)ij}_2$}
\parbox[c]{3cm}{\includegraphics[width= 2.5cm,angle=0]{vertex3.eps}}
\label{v1}
\end{psfrags}
&~~~&
\parbox[c]{13cm}{$\begin{array}{c}\tilde{V}^{(A_1^3)ij}_2=-\f{N}{2qg^2} \tilde{G}^{(A_1^3)ij}_2(n,n^{'},k_x^{''},\tilde{k}_x^{''}) (2\pi)^2\delta^2({\bf k}+{\bf k}^{'}+{\bf k}^{''}+\tilde{{\bf k}}^{''})\delta_{m+m^{'}+m^{''}+\tilde{m}^{''}}\\ 
\tilde{G}^{(A_1^3)11}_2(n,n^{'},k_x^{''},\tilde{k}_x^{''})=-\sqrt{q}\int dx [\tilde{\phi}_{n+1}(x)\tilde{\phi}_{n^{'}+1}(x)]e^{-(k_x^{''}+\tilde{k}_x^{''})x}
\\
\tilde{G}^{(A_1^3)ij}_2(n,n^{'},k_x^{''},\tilde{k}_x^{''})=\sqrt{q}\int dx [e^{-qx^2}H_n(x)H_{n^{'}}(x)]e^{-(k_x^{''}+\tilde{k}_x^{''})x}\delta_{ij}~~\mbox{for}~~i,j=2,3
\end{array}$}
\end{tabular}
\end{center}
\end{table}

\begin{table}[H]
\begin{center}
\begin{tabular}{lcc}
\begin{psfrags}
\psfrag{c}[][]{$A_1^3(m^{''},k_x^{''},{\bf k}^{''})$}
\psfrag{a1}[][]{$C(m,n,{\bf k})$}
\psfrag{a3}[][]{$C^{'}(m^{'},n^{'},{\bf k}^{'})$}
\psfrag{v4}[][]{$V^{(A_1^3)}_3$}
\parbox[c]{5.1cm}{\includegraphics[width= 5 cm,angle=0]{vertex1.eps}} 
\end{psfrags}
&~~~&
\parbox[c]{10cm}{$\begin{array}{c}V^{(A_1^3)}_3 =-\f{N^{3/2}}{qg^2} \beta G^{(A_1^3)}_3(n,n^{'},k_x^{''}) (2\pi)^2\delta^2({\bf k}+{\bf k}^{'}+{\bf k}^{''})\delta_{m+m^{'}+m^{''}}\\ 
G^{(A_1^3)}_3(n,n^{'},k_x^{''})=\int dx \left[\partial_x\phi_{n^{'}}(x) \phi_{n}(x)-\partial_x\phi_{n}(x) \phi_{n^{n}}(x)
\right.\\-\left.(qx) A_n(x)\phi_{n^{'}}(x)+ (qx) \phi_n(x)A_{n^{'}}(x) \right]e^{-ik_x^{''}x}
\end{array}$}
\end{tabular}
\end{center}
\end{table}

\begin{table}[H]
\begin{center}
\begin{tabular}{lcc}
\begin{psfrags}
\psfrag{c}[][]{$A_1^3(m^{''},k_x^{''},{\bf k}^{''})$}
\psfrag{a1}[][]{$C^{'}(m,n,{\bf k})$}
\psfrag{a3}[][]{$\tilde{A}_{i}^{2}(m^{'},n^{'},{\bf k}^{'})$}
\psfrag{v4}[][]{$\tilde{V}^{(A_1^3)i}_3$}
\parbox[c]{5.1cm}{\includegraphics[width= 5 cm,angle=0]{vertex1.eps}} 
\end{psfrags}
&~~~&
\parbox[c]{10cm}{$\begin{array}{c}\tilde{V}^{(A_1^3)i}_3 =-\f{N^{3/2}}{qg^2} \beta \tilde{G}^{(A_1^3)i}_3(n,n^{'},k_x^{''}) (2\pi)^2\delta^2({\bf k}+{\bf k}^{'}+{\bf k}^{''})\delta_{m+m^{'}+m^{''}}\\ 
\tilde{G}^{(A_1^3)1}_3(n,n^{'},k_x^{''})=\pm i\int dx \left[\partial_x\phi_n(x) \tilde{\phi}_{n^{'}+1}(x)-\partial_x\tilde{\phi}_{n^{'}+1}(x) \phi_{n}(x)\right.\\\left.+(qx) A_n(x)\tilde{\phi}_{n^{'}+1}(x)- (qx) \phi_n(x)\tilde{A}_{n^{'}+1}(x) \right]e^{-ik_x^{''}x}\\
\tilde{G}^{(A_1^3)i}_3(n,n^{'},k_x^{''})=i(k^{''}_i-k_i)\int dx \left[A_n(x)H_{n^{'}}(x)e^{-qx^2/2}\right]e^{-ik_x^{''}x}\\~~\mbox{for}~~i=2,3
\end{array}$}
\end{tabular}
\end{center}
\end{table}

\begin{table}[H]
\begin{center}
\begin{tabular}{lcc}
\begin{psfrags}
\psfrag{c}[][]{$A_1^3(m^{''},k_x^{''},{\bf k}^{''})$}
\psfrag{a1}[][]{$\tilde{A}_{i}^{1}(m,n,{\bf k})$}
\psfrag{a3}[][]{$\tilde{A}_{i}^{2}(m^{'},n^{'},{\bf k}^{'})$}
\psfrag{v4}[][]{$\tilde{V}^{(A_1^3)ij}_3$}
\parbox[c]{5.1cm}{\includegraphics[width= 5 cm,angle=0]{vertex1.eps}} 
\end{psfrags}
&~~~&
\parbox[c]{10cm}{$\begin{array}{c}\tilde{V}^{(A_1^3)ij}_3 =-\f{N^{3/2}}{qg^2} \beta \tilde{G}^{(A_1^3)ij}_3(n,n^{'},k_x^{''}) (2\pi)^2\delta^2({\bf k}+{\bf k}^{'}+{\bf k}^{''})\delta_{m+m^{'}+m^{''}}\\ 
\tilde{G}^{(A_1^3)11}_3(n,n^{'},k_x^{''})=\int dx \left[\partial_x\tilde{\phi}_{n+1}(x) \tilde{\phi}_{n^{'}+1}(x)-\partial_x\tilde{\phi}_{n^{'}+1}(x) \tilde{\phi}_{n+1}(x)\right.\\\left.+(qx) \tilde{A}_{n+1}(x)\tilde{\phi}_{n^{'}+1}(x)- (qx) \tilde{\phi}_{n+1}(x)\tilde{A}_{n^{'}+1}(x) \right]e^{-ik_x^{''}x}\\
\tilde{G}^{(A_1^3)1i}_3(n,n^{'},k_x^{''})=(k^{''}_i-k_i)\int dx \left[\tilde{A}_{n+1}(x)H_{n^{'}}(x)e^{-qx^2/2}\right]e^{-ik_x^{''}x}\\~~\mbox{for}~~i=2,3\\
\tilde{G}^{(A_1^3)ij}_3(n,n^{'},k_x^{''})=\int dx \left[\partial_x(e^{-qx^2/2}H_{n^{'}}(x)H_{n}(x)e^{-qx^2/2}-\right.\\\left.\partial_x(e^{-qx^2/2}H_{n}(x)H_{n^{'}}(x)e^{-qx^2/2}\right]e^{-ik_x^{''}x}\times\delta_{ij}\\~~\mbox{for}~~i,j=2,3
\\
\mbox{We also have additional vertices  $V^{(A_1^3)}_4$ with functions $G^{(A_1^3)}_4$ same as }\\ \mbox{$\tilde{G}^{(A_1^3)ij}$ for the $\Phi_I^{(1,2)}/\tilde{\Phi}_I^{(1,2)}$  fields in place of $\tilde{A}_{2,3}^{(1,2)}$ fields.}
\end{array}$}
\end{tabular}
\end{center}
\end{table}

{\bf Vertices for $A_{2}^3$ consisting of only bosons:} \\

\begin{table}[H]
\begin{center}
\begin{tabular}{lcc}
\begin{psfrags}
\psfrag{a11}[][]{\scalebox{0.8}{$\Phi_I^{(1,2)/}/\tilde{\Phi}_I^{(1,2)}$}}
\psfrag{a12}[][]{\scalebox{0.8}{$\Phi_I^{(1,2)}/\tilde{\Phi}_I^{(1,2)}$}}
\psfrag{c1}[][]{\scalebox{0.8}{$A_2^3(m^{''},n^{''},{\bf k}^{''})$}}
\psfrag{c2}[][]{\scalebox{0.8}{$A_2^3(\tilde{m}^{''},\tilde{n}^{''},\tilde{{\bf k}}^{''})$}}
\psfrag{v2}[][]{$V^{(A_2^3)}_1$}
\parbox[c]{3cm}{\includegraphics[width= 2.5cm,angle=0]{vertex3.eps}}
\label{v1}
\end{psfrags}
&~~~&
\parbox[c]{13cm}{$\begin{array}{c}V^{(A_2^3)}_1=-\f{N}{2qg^2} G^{(A_2^3)}_1(n,n^{'},k_x^{''},\tilde{k}_x^{''}) (2\pi)^2\delta^2({\bf k}+{\bf k}^{'}+{\bf k}^{''}+\tilde{{\bf k}}^{''})\delta_{m+m^{'}+m^{''}+\tilde{m}^{''}}\\ 
G^{(A_2^3)}_1(n,n^{'},k_x^{''},\tilde{k}_x^{''})=\sqrt{q}\int dx [e^{-qx^2}H_n(x)H_{n^{'}}(x)]e^{-(k_x^{''}+\tilde{k}_x^{''})x}
\end{array}$}
\end{tabular}
\end{center}
\end{table}

\begin{table}[H]
\begin{center}
\begin{tabular}{lcc}
\begin{psfrags}
\psfrag{a11}[][]{\scalebox{0.8}{$C/C^{'}(m,n,{\bf k})$}}
\psfrag{a12}[][]{\scalebox{0.8}{$C/C^{'}(m^{'},n^{'},{\bf k}^{'})$}}
\psfrag{c1}[][]{\scalebox{0.8}{$A_2^3(m^{''},n^{''},{\bf k}^{''})$}}
\psfrag{c2}[][]{\scalebox{0.8}{$A_2^3(\tilde{m}^{''},\tilde{n}^{''},\tilde{{\bf k}}^{''})$}}
\psfrag{v2}[][]{$V^{(A_2^3)}_2$}
\parbox[c]{3cm}{\includegraphics[width= 2.5cm,angle=0]{vertex3.eps}}
\label{v1}
\end{psfrags}
&~~~&
\parbox[c]{13cm}{$\begin{array}{c}V^{(A_2^3)}_2=-\f{N}{2qg^2} G^{(A_2^3)}_2(n,n^{'},k_x^{''},\tilde{k}_x^{''}) (2\pi)^2\delta^2({\bf k}+{\bf k}^{'}+{\bf k}^{''}+\tilde{{\bf k}}^{''})\delta_{m+m^{'}+m^{''}+\tilde{m}^{''}}\\ 
G^{(A_2^3)}_2(n,n^{'},k_x^{''},\tilde{k}_x^{''})=\sqrt{q}\int dx [A_n(x)A_{n^{'}}(x)+\phi_n(x)\phi_{n^{'}}(x)]e^{-(k_x^{''}+\tilde{k}_x^{''})x}
\end{array}$}
\end{tabular}
\end{center}
\end{table}

\begin{table}[H]
\begin{center}
\begin{tabular}{lcc}
\begin{psfrags}
\psfrag{a11}[][]{\scalebox{0.8}{$\tilde{A}_{2,,3}^{(1,2)}/\tilde{A}_1^{(1,2)}$}}
\psfrag{a12}[][]{\scalebox{0.8}{$\tilde{A}_{2,,3}^{(1,2)}/\tilde{A}_1^{(1,2)}$}}
\psfrag{c1}[][]{\scalebox{0.8}{$A_2^3(m^{''},n^{''},{\bf k}^{''})$}}
\psfrag{c2}[][]{\scalebox{0.8}{$A_2^3(\tilde{m}^{''},\tilde{n}^{''},\tilde{{\bf k}}^{''})$}}
\psfrag{v2}[][]{$\tilde{V}^{(A_2^3)ij}_2$}
\parbox[c]{3cm}{\includegraphics[width= 2.5cm,angle=0]{vertex3.eps}}
\label{v1}
\end{psfrags}
&~~~&
\parbox[c]{13cm}{$\begin{array}{c}\tilde{V}^{(A_2^3)ij}_2=-\f{N}{2qg^2} \tilde{G}^{(A_2^3)ij}_2(n,n^{'},k_x^{''},\tilde{k}_x^{''}) (2\pi)^2\delta^2({\bf k}+{\bf k}^{'}+{\bf k}^{''}+\tilde{{\bf k}}^{''})\delta_{m+m^{'}+m^{''}+\tilde{m}^{''}}\\ 
\tilde{G}^{(A_2^3)11}_2(n,n^{'},k_x^{''},\tilde{k}_x^{''})=-\sqrt{q}\int dx [\tilde{A}_{n+1}(x)\tilde{A}_{n^{'}+1}(x)+\tilde{\phi}_{n+1}(x)\tilde{\phi}_{n^{'}+1}(x)]e^{-(k_x^{''}+\tilde{k}_x^{''})x}
\\
\tilde{G}^{(A_1^3)ij}_2(n,n^{'},k_x^{''},\tilde{k}_x^{''})=\sqrt{q}\int dx [e^{-qx^2}H_n(x)H_{n^{'}}(x)]e^{-(k_x^{''}+\tilde{k}_x^{''})x}\delta_{ij}~~\mbox{for}~~i,j=2,3
\end{array}$}
\end{tabular}
\end{center}
\end{table}

\begin{table}[H]
\begin{center}
\begin{tabular}{lcc}
\begin{psfrags}
\psfrag{c}[][]{$A_2^3(m^{''},k_x^{''},{\bf k}^{''})$}
\psfrag{a1}[][]{$C(m,n,{\bf k})$}
\psfrag{a3}[][]{$C^{'}(m^{'},n^{'},{\bf k}^{'})$}
\psfrag{v4}[][]{$V^{(A_2^3)}_3$}
\parbox[c]{5.1cm}{\includegraphics[width= 5 cm,angle=0]{vertex1.eps}} 
\end{psfrags}
&~~~&
\parbox[c]{10cm}{$\begin{array}{c}V^{(A_2^3)}_3 =-\f{N^{3/2}}{qg^2} \beta G^{(A_2^3)}_3(n,n^{'},k_x^{''}) (2\pi)^2\delta^2({\bf k}+{\bf k}^{'}+{\bf k}^{''})\delta_{m+m^{'}+m^{''}}\\ 
G^{(A_2^3)}_3(n,n^{'},k_x^{''})=i(k_2^{'}-k_2)\int dx \left[A_n(x)A_{n^{'}}(x)+\phi_n(x)\phi_{n^{'}}(x) \right]e^{-ik_x^{''}x}
\end{array}$}
\end{tabular}
\end{center}
\end{table}

\begin{table}[H]
\begin{center}
\begin{tabular}{lcc}
\begin{psfrags}
\psfrag{c}[][]{$A_2^3(m^{''},k_x^{''},{\bf k}^{''})$}
\psfrag{a1}[][]{$\tilde{A}_{1,2}^{2}{(m,n,{\bf k})}$}
\psfrag{a3}[][]{$C^{'}(m^{'},n^{'},{\bf k}^{'})$}
\psfrag{v4}[][]{$\tilde{V}^{(A_2^3)i}_3$}
\parbox[c]{5.1cm}{\includegraphics[width= 5 cm,angle=0]{vertex1.eps}} 
\end{psfrags}
&~~~&
\parbox[c]{10cm}{$\begin{array}{c}\tilde{V}^{(A_2^3)i}_3 =-\f{N^{3/2}}{qg^2} \beta \tilde{G}^{(A_2^3)i}_3(n,n^{'},k_x^{''}) (2\pi)^2\delta^2({\bf k}+{\bf k}^{'}+{\bf k}^{''})\delta_{m+m^{'}+m^{''}}\\ 
\tilde{G}^{(A_2^3)1}_3(n,n^{'},k_x^{''})=(k_2-k_2^{'})\int dx \left[\tilde{A}_{n+1}(x)A_{n^{'}}(x)+\tilde{\phi}_{n+1}(x)\phi_{n^{'}}(x) \right]\\
\times e^{-ik_x^{''}x} \\
\tilde{G}^{(A_2^3)2}_3(n,n^{'},k_x^{''}=\int dx \left[\partial_x(e^{-qx^2/2}H_{n}(x))A_{n^{'}}(x)\right.\\
\left.+ik^{''}_xA_{n^{'}}(x)e^{-qx^2/2}H_{n}(x)-(qx)e^{-qx^2/2}H_{n}(x)\phi_{n^{'}}(x) \right]e^{-ik_x^{''}x}\\
\end{array}$}
\end{tabular}
\end{center}
\end{table}

\begin{table}[H]
\begin{center}
\begin{tabular}{lcc}
\begin{psfrags}
\psfrag{c}[][]{$A_2^3(m^{''},k_x^{''},{\bf k}^{''})$}
\psfrag{a1}[][]{$\tilde{A}_{i}^{1}(m,n,{\bf k})$}
\psfrag{a3}[][]{$\tilde{A}_{j}^{2}(m^{'},n^{'},{\bf k}^{'})$}
\psfrag{v4}[][]{$\tilde{V}^{(A_1^3)ij}_3$}
\parbox[c]{5.1cm}{\includegraphics[width= 5 cm,angle=0]{vertex1.eps}} 
\end{psfrags}
&~~~&
\parbox[c]{11cm}{$\begin{array}{c}\tilde{V}^{(A_2^3)ij}_3 =-\f{N^{3/2}}{qg^2} \beta \tilde{G}^{(A_2^3)ij}_3(n,n^{'},k_x^{''}) (2\pi)^2\delta^2({\bf k}+{\bf k}^{'}+{\bf k}^{''})\delta_{m+m^{'}+m^{''}}\\ 
\tilde{G}^{(A_2^3)11}_3(n,n^{'},k_x^{''})=i(k_2^{'}-k_2)\int dx \left[\tilde{A}_{n+1}(x)A_{n^{'}+1}(x)+\tilde{\phi}_{n+1}(x)\phi_{n^{'}+1}(x) \right]\\
\times e^{-ik_x^{''}x} \\
\tilde{G}^{(A_2^3)12}_3(n,n^{'},k_x^{''})=i\int dx \left[\partial_x(e^{-qx^2/2}H_{n}(x))\tilde{A}_{n^{'}+1}(x)\right.\\
\left.+ik^{''}_x\tilde{A}_{n^{'}+1}(x)e^{-qx^2/2}H_{n}(x)-(qx)e^{-qx^2/2}H_{n}(x)\tilde{\phi}_{n^{'}+1}(x) \right]e^{-ik_x^{''}x}\\
\tilde{G}^{(A_2^3)32}_3(n,n^{'},k_x^{''})=i(k_3^{'}-k_3^{''})\int dx \left[e^{-qx^2}H_n(x)H_{n^{'}}(x)\right]e^{-ik_x^{''}x}\\
\tilde{G}^{(A_2^3)23}_3(n,n^{'},k_x^{''})=i(k_3^{''}-k_3^{'})\int dx \left[e^{-qx^2}H_n(x)H_{n^{'}}(x)\right]e^{-ik_x^{''}x}\\
\tilde{G}^{(A_2^3)33}_3(n,n^{'},k_x^{''})=i(k_2^{'}-k_2)\int dx \left[e^{-qx^2}H_n(x)H_{n^{'}}(x)\right]e^{-ik_x^{''}x}\\
\mbox{For other values of $(i,j)$, $\tilde{G}^{(A_2^3)ij}_3=0$}\\ \mbox{We also have additional vertices  $V^{(A_2^3)}_4$ with functions $G^{(A_2^3)}_4$ same as }\\ \mbox{$\tilde{G}^{(A_2^3)33}_3$ for the $\Phi_I^{(1,2)}/\tilde{\Phi}_I^{(1,2)}$  fields in place of $\tilde{A}_{3}^{1,2}$ fields.}
\end{array}$}
\end{tabular}
\end{center}
\end{table}

{\bf Vertices for $A_{\mu}^3$ involving fermions:}\\

The relevant terms in the action that contain $A_{\mu}^3$ field are
\beqa\label{fermionsetsA3}
&~&\f{1}{g^2}\mbox{tr} \int d^4z \left(\bar{\lambda_I}\gamma^{\mu}[A_{\mu},\lambda_I]\right)\\\nonumber
&=& \f{i}{2g^2}\int d^4z \left[\bar{\lambda}^2_I\gamma^{\mu}A^{3}_{\mu}\lambda^1_I-\bar{\lambda}^1_I\gamma^{\mu}A^{3}_{\mu}\lambda^2_I
\right]+ \cdots
\eeqa

For $I=1$ and $I=4$, the coupled sets of fermions are $(\lambda^1_1, \lambda^2_4)$ and 
$(\lambda^1_4, \lambda^2_1)$. Similarly for $I=2$ and $I=3$, the other two sets namely  $(\lambda^1_2, \lambda^2_3)$ and $(\lambda^1_3, \lambda^2_2)$ are coupled to each other.

In the following we evaluate the vertices involving $I=1$ and $I=4$. The corresponding vertices for $I=2$ and $I=3$ are similar to these. Putting in the mode expansions we get

\beqa
&~&\sum_{m,m^{'},m^{''}}\sum_{n,n^{'}}\int \f{d^2{\bf k}}{(2\pi\sqrt{q})^2}\f{d^2{\bf k}^{'}}{(2\pi\sqrt{q})^2}\f{d^3{k}^{''}}{(2\pi\sqrt{q})^3}\times\\
&\times&\left[\bar{\theta}_1(m,n, {\bf k})V^{\mu}_f(n,n^{'},k^{''}_x)\theta_4(m^{'},n^{'},{\bf k}^{'})A^3_{\mu}(m^{''}, k^{''}) 
+\bar{\theta}_4(m^{'},n^{'}, {\bf k}^{'})V^{\mu*}_f(n,n^{'},k^{''}_x)\theta_1(m,n,{\bf k})A^3_{\mu}(m^{''}, k^{''}) \right.\non
&~&\left.+ \theta_1^{T}\gamma^0(m,n,{\bf k})V^{'\mu}_f(n,n^{'},k^{''}_x)\theta_4(m^{'},n^{'},{\bf k}^{'})A_{\mu}^3(m^{''}, k^{''})
+ \bar{\theta}_1(m,n,{\bf k})V^{'\mu*}_f(n,n^{'},k^{''}_x)\theta^{*}_4(m^{'},n^{'},{\bf k}^{'})A_{\mu}^3(m^{''}, k^{''})+\cdots\right]\nonumber
\eeqa

Defining $(-1)^{\mu}=+1$ for $\mu=2,3$ and $(-1)^{\mu}=-1$ for $\mu=1$, the vertices are written down below.

\begin{table}[H]
\begin{center}
\begin{tabular}{lcc}
\begin{psfrags}
\psfrag{c}[][]{$A_{\mu}^3(m^{''}, k^{''})$}
\psfrag{a1}[][]{$\bar{\theta_1}(m,n,{\bf k})$}
\psfrag{a3}[][]{$\theta_4(m^{'},n^{'},{\bf k}^{'})$}
\psfrag{v4}[][]{$V_f^{\mu}$}
\parbox[c]{5cm}{\includegraphics[width= 4 cm,angle=0]{vertex1.eps}} 
\end{psfrags}
&~~~&
\parbox[c]{8cm}{$\begin{array}{c}V^{\mu}_f =i\f{N}{qg^2}\gamma^{\mu}\gamma^{1}G_f(n,n^{'},k^{''}_x)(2\pi)^2\delta^2({\bf k}-{\bf k}^{'}-{\bf k}^{''})\delta_{m-m^{'}-m^{''}}\\ 
G_f(n,n^{'},k^{''}_x)=\sqrt{q}\int dx  \left(L^{*}_n(x)R_{n^{'}}(x) \pm R^{*}_{n}(x)L_{n^{'}}(x)\right)e^{-ik^{''}_x x}\end{array}$}
\\
\begin{psfrags}
\psfrag{c}[][]{$A_{\mu}^3(m^{''}, k^{''})$}
\psfrag{a1}[][]{$\theta_1(m,n,{\bf k})$}
\psfrag{a3}[][]{$\bar{\theta}_4(m^{'},n^{'},{\bf k}^{'})$}
\psfrag{v4}[][]{$V_f^{\mu*}$}
\parbox[c]{5cm}{\includegraphics[width= 4 cm,angle=0]{vertex1.eps}} 
\end{psfrags}
&~~~&
\parbox[c]{8cm}{$\begin{array}{c}V^{\mu*}_f =-(-1)^{\mu}i\f{N}{qg^2}\gamma^{\mu}\gamma^{1}G^{*}_f(n,n^{'},k^{''}_x)(2\pi)^2\delta^2({\bf k}-{\bf k}^{'}+{\bf k}^{''})\delta_{m-m^{'}+m^{''}}\\ 
G^{*}_f(n,n^{'},k^{''}_x)=\sqrt{q}\int dx  \left(L_n(x)R^{*}_{n^{'}}(x) +(-1)^{\mu} R_{n}(x)L^{*}_{n^{'}}(x)\right)e^{-ik^{''}_x x}\end{array}$}
\end{tabular}
\end{center}
\end{table}

\begin{table}[H]
\begin{center}
\begin{tabular}{lcc}

\begin{psfrags}
\psfrag{c}[][]{$A^3_{\mu}(m^{''}, k^{''})$}
\psfrag{a1}[][]{$\theta_1(m,n,{\bf k})$}
\psfrag{a3}[][]{$\theta_4^{T}\gamma^0(m^{'},n^{'},{\bf k}^{'})$}
\psfrag{v4}[][]{$V^{'\mu}_{f}$}
\parbox[c]{5cm}{\includegraphics[width= 4 cm,angle=0]{vertex1.eps}} 
\end{psfrags}
&~~~&
\parbox[c]{8cm}{$\begin{array}{c}V^{'\mu}_f =-i\f{N}{qg^2}\gamma^{\mu}\gamma^{1} G^{'}_f(n,n^{'},k_x^{''})(2\pi)^2\delta^2({\bf k}+{\bf k}^{'}+{\bf k}^{''})\delta_{m+m^{'}+m^{''}}\\ G^{'}_f(n,n^{'},k_x^{''})=\sqrt{q}\int dx  \left(L_n(x)R_{n^{'}}(x) +(-1)^{\mu} R_n(x)L_{n^{'}}(x) \right)e^{-ik^{''}_x x}\end{array}$}

\\

\begin{psfrags}
\psfrag{c}[][]{$A^3_{\mu}(m^{''}, k^{''})$}
\psfrag{a1}[][]{$\bar{\theta}_1(m,n,{\bf k})$}
\psfrag{a3}[][]{$\theta^{*}_4(m^{'},n^{'},{\bf k}^{'})$}
\psfrag{v4}[][]{$V^{'\mu*}_{f}$}
\parbox[c]{5cm}{\includegraphics[width= 4 cm,angle=0]{vertex1.eps}} 
\end{psfrags}
&~~~&
\parbox[c]{8cm}{$\begin{array}{c}V^{'\mu *}_f =(-1)^{\mu}i\f{N}{qg^2}\gamma^{\mu}\gamma^{1} G^{'*}_f(n,n^{'},k_x^{''})(2\pi)^2\delta^2({\bf k}+{\bf k}^{'}-{\bf k}^{''})\delta_{m+m^{'}-m^{''}}\\ G^{'*}_f(n,n^{'},k_x^{''})=\sqrt{q}\int dx  \left(L^{*}_n(x)R^{*}_{n^{'}}(x)+(-1)^{\mu} R^{*}_n(x)L^{*}_{n^{'}}(x)\right) e^{-ik^{''}_x x}\end{array}$}

\end{tabular}
\end{center}
\end{table}



\begin{thebibliography}{99}




\bibitem{1}
  S.~P.~Chowdhury, S.~Sarkar and B.~Sathiapalan,
  ``BCS Instability and Finite Temperature Corrections to Tachyon Mass in Intersecting D1-Branes,''
  JHEP {\bf 1409} (2014) 063
  [arXiv:1403.0389 [hep-th]].

\bibitem{2}
  S.~Kalyana Rama, S.~Sarkar, B.~Sathiapalan and N.~Sircar,
  ``Strong Coupling BCS Superconductivity and Holography,''
  Nucl.\ Phys.\ B {\bf 852} (2011) 634
  [arXiv:1104.2843 [hep-th]].



\bibitem{LMV}
H.~Liu, J.~McGreevy and D.~Vegh,
``Non-Fermi liquids from holography,''
arXiv:0903.2477 [hep-th].

\bibitem{FLMV}
T.~Faulkner, H.~Liu, J.~McGreevy and D.~Vegh,
``Emergent quantum criticality, Fermi surfaces, and AdS2,''
arXiv:0907.2694 [hep-th].

\bibitem{FP}
T.~Faulkner and J.~Polchinski,
``Semi-Holographic Fermi Liquids,''
arXiv:1001.5049 [hep-th].

\bibitem{NS}
D.~Nickel and D.~T.~Son,
``Deconstructing holographic liquids,''
arXiv:1009.3094 [hep-th].

\bibitem{FLR}
T.~Faulkner, H.~Liu and M.~Rangamani,
``Integrating out geometry: Holographic Wilsonian RG and the
membrane paradigm,''
arXiv:1010.4036 [hep-th].

\bibitem{HP}
I.~Heemskerk and J.~Polchinski,
``Holographic and Wilsonian Renormalization Groups,''
arXiv:1010.1264 [hep-th].

\bibitem{HKSS}
C.~P.~Herzog, P.~Kovtun, S.~Sachdev and D.~T.~Son,
``Quantum critical transport, duality, and M-theory,''
Phys.\ Rev.\  D {\bf 75}, 085020 (2007) 
[arXiv:hep-th/0701036].

\bibitem{DHS}
F.~Denef, S.~A.~Hartnoll and S.~Sachdev,
``Quantum oscillations and black hole ringing,''
Phys.\ Rev.\  D {\bf 80}, 126016 (2009)
[arXiv:0908.1788 [hep-th]].

\bibitem{G}
S.~S.~Gubser,
``TASI lectures: Collisions in anti-de Sitter space, conformal
symmetry, and holographic superconductors,''
arXiv:1012.5312 [hep-th].

\bibitem{GPR}
S.~S.~Gubser, S.~S.~Pufu and F.~D.~Rocha,
``Quantum critical superconductors in string theory and
M-theory,''
Phys.\ Lett.\  B {\bf 683}, 201 (2010)
[arXiv:0908.0011 [hep-th]].

\bibitem{HHH}
S.~A.~Hartnoll, C.~P.~Herzog and G.~T.~Horowitz,
``Building a Holographic Superconductor,''
Phys.\ Rev.\ Lett.\  {\bf 101}, 031601 (2008)
[arXiv:0803.3295 [hep-th]]; 
\\
S.~A.~Hartnoll, C.~P.~Herzog and G.~T.~Horowitz,
``Holographic Superconductors,''
JHEP {\bf 12}, 015 (2008)
[arXiv:0810.1563 [hep-th]].



\bibitem{SS1}
T.~Sakai and S.~Sugimoto,
``Low energy hadron physics in holographic QCD,''
Prog.\ Theor.\ Phys.\  {\bf 113}, 843 (2005)
[arXiv:hep-th/0412141]; 



\bibitem{EW1}
E.~Witten,
``Anti-de Sitter space, thermal phase transition, and
confinement in gauge theories,''
Adv.\ Theor.\ Math.\ Phys.\  {\bf 2}, 505 (1998)
[arXiv:hep-th/9803131].


\bibitem{KK}
A.~Karch and E.~Katz,
``Adding flavor to AdS/CFT,''
JHEP {\bf 06}, 043 (2002)
[arXiv:hep-th/0205236].

\bibitem{KMMW}
M.~Kruczenski, D.~Mateos, R.~C.~Myers and D.~J.~Winters,
``Towards a holographic dual of large-N(c) QCD,''
JHEP {\bf 05}, 041 (2004)
[arXiv:hep-th/0311270].

\bibitem{SS}
T.~Sakai and J.~Sonnenschein,
``Probing flavored mesons of confining gauge theories by
supergravity,''
JHEP {\bf 09}, 047 (2003)
[arXiv:hep-th/0305049].

\bibitem{EKSS}
J.~Erlich, E.~Katz, D.~T.~Son and M.~A.~Stephanov,
``QCD and a Holographic Model of Hadrons,''
Phys.\ Rev.\ Lett.\  {\bf 95}, 261602 (2005)
[arXiv:hep-ph/0501128].

\bibitem{KKSS}
A.~Karch, E.~Katz, D.~T.~Son and M.~A.~Stephanov,
``Linear Confinement and AdS/QCD,''
Phys.\ Rev.\  D {\bf 74}, 015005 (2006)
[arXiv:hep-ph/0602229].

\bibitem{AHJK}
E.~Antonyan, J.~A.~Harvey, S.~Jensen and D.~Kutasov,
``NJL and QCD from string theory,''
arXiv:hep-th/0604017; 


E.~Antonyan, J.~A.~Harvey and D.~Kutasov,
``Chiral symmetry breaking from intersecting D-branes,''
Nucl.\ Phys.\  B {\bf 784}, 1 (2007)
[arXiv:hep-th/0608177].


\bibitem{Hata}
H.~Hata, T.~Sakai, S.~Sugimoto and S.~Yamato,
``Baryons from instantons in holographic QCD,''
Prog.\ Theor.\ Phys.\  {\bf 117}, 1157 (2007)
[arXiv:hep-th/0701280].


\bibitem{BLL}
O.~Bergman, G.~Lifschytz and M.~Lippert,
``Holographic Nuclear Physics,''
JHEP {\bf 11}, 056 (2007)
[arXiv:0708.0326 [hep-th]].



\bibitem{KMMMT}
S.~Kobayashi, D.~Mateos, S.~Matsuura, R.~C.~Myers and
R.~M.~Thomson,
``Holographic phase transitions at finite baryon density,''
JHEP {\bf 02}, 016 (2007)
[arXiv:hep-th/0611099].


\bibitem{ASY}
O.~Aharony, J.~Sonnenschein and S.~Yankielowicz,
``A holographic model of deconfinement and chiral symmetry
restoration,''
Annals Phys.\  {\bf 322}, 1420 (2007)
[arXiv:hep-th/0604161].

\bibitem{BSS}
O.~Bergman, S.~Seki and J.~Sonnenschein,
``Quark mass and condensate in HQCD,''
JHEP {\bf 12}, 037 (2007)
[arXiv:0708.2839 [hep-th]].

\bibitem{Dhar}
A.~Dhar and P.~Nag,
``Sakai-Sugimoto model, Tachyon Condensation and Chiral
symmetry Breaking,''
JHEP {\bf 01}, 055 (2008)
[arXiv:0708.3233 [hep-th]]; 
\\
A.~Dhar and P.~Nag,
``Tachyon condensation and quark mass in modified
Sakai-Sugimoto model,''
Phys.\ Rev.\  D {\bf 78}, 066021 (2008)
[arXiv:0804.4807 [hep-th]].

\bibitem{J1}
  N.~Jokela, M.~Jarvinen and S.~Nowling,
  ``Winding effects on brane/anti-brane pairs,''
  JHEP {\bf 0907} (2009) 085
  [arXiv:0901.0281 [hep-th]].


\bibitem{BDL}
M.~Berkooz, M.~R.~Douglas and R.~G.~Leigh,
``Branes intersecting at angles,''
Nucl.\ Phys.\  B {\bf 480}, 265 (1996)
[arXiv:hep-th/9606139].


\bibitem{SJ1}
  H.~Arfaei and M.~M.~Sheikh Jabbari,
  ``Different d-brane interactions,''
  Phys.\ Lett.\ B {\bf 394} (1997) 288
  doi:10.1016/S0370-2693(97)00022-1
  [hep-th/9608167].

\bibitem{SJ2}
  M.~M.~Sheikh Jabbari,
  ``Classification of different branes at angles,''
  Phys.\ Lett.\ B {\bf 420} (1998) 279
  doi:10.1016/S0370-2693(97)01550-5
  [hep-th/9710121].


\bibitem{HT}
A.~Hashimoto and W.~Taylor,
``Fluctuation spectra of tilted and intersecting D-branes from
the Born-Infeld action,''
Nucl.\ Phys.\  B {\bf 503}, 193 (1997)
[arXiv:hep-th/9703217].

\bibitem{Hashimoto:2003pu} 
  K.~Hashimoto and W.~Taylor,
  ``Strings between branes,''
  JHEP {\bf 0310}, 040 (2003)
  [hep-th/0307297].


\bibitem{Hashimoto:2003xz}
  K.~Hashimoto and S.~Nagaoka,
  ``Recombination of intersecting D-branes by local tachyon condensation,''
  JHEP {\bf 0306} (2003) 034
  [hep-th/0303204].

\bibitem{Nagaoka:2003zn}
  S.~Nagaoka,
  ``Higher dimensional recombination of intersecting D-branes,''
  JHEP {\bf 0402} (2004) 063
  [hep-th/0312010].




\bibitem{Epple:2003xt}
  F.~T.~J.~Epple and D.~Lust,
  ``Tachyon condensation for intersecting branes at small and large angles,''
  Fortsch.\ Phys.\  {\bf 52} (2004) 367
  [hep-th/0311182].

\bibitem{Huang}
  W.~H.~Huang,
  ``Recombination of intersecting D-branes in tachyon field theory,''
  Phys.\ Lett.\ B {\bf 564} (2003) 155
  doi:10.1016/S0370-2693(03)00706-8
  [hep-th/0304171].

\bibitem{Jones}
  N.~T.~Jones and S.~H.~H.~Tye,
  ``Spectral flow and boundary string field theory for angled d-branes,''
  JHEP {\bf 0308} (2003) 037
  doi:10.1088/1126-6708/2003/08/037
  [hep-th/0307092].

\bibitem{J2}
  N.~Jokela and M.~Lippert,
  ``Inhomogeneous tachyon dynamics and the zipper,''
  JHEP {\bf 0908} (2009) 024
  [arXiv:0906.0317 [hep-th]].

\bibitem{kapusta1}
  K.~Kajantie and J.~I.~Kapusta,
  ``Infrared Limit of the Axial Gauge Gluon Propagator at High Temperature,''
  Phys.\ Lett.\ B {\bf 110} (1982) 299.

\bibitem{Landshoff:1985fv}
  P.~V.~Landshoff,
  ``The Propagator in Axial Gauge,''
  Phys.\ Lett.\ B {\bf 169} (1986) 69.

\bibitem{kapusta2}
K.~Kajantie and J.~Kapusta 
``Behaviour of gluons at high temperature,''
Ann.\ Phys.\ {bf 160} (1985) 477

\bibitem{Leib1}
  G.~Leibbrandt,
  ``Introduction to Noncovariant Gauges,''
  Rev.\ Mod.\ Phys.\  {\bf 59} (1987) 1067.
  
\bibitem{Sen}
  A.~Sen,
  ``Tachyon dynamics in open string theory,''
  Int.\ J.\ Mod.\ Phys.\ A {\bf 20} (2005) 5513
  doi:10.1142/S0217751X0502519X
  [hep-th/0410103].

\bibitem{kapusta}
J. Kapusta,
``Finite Temperature Field Theory",
Publisher: Cambridge Monographs on Mathematical Physics.

\bibitem{dimred1} 
  L.~Brink, J.~H.~Schwarz and J.~Scherk,
  ``Supersymmetric Yang-Mills Theories,''
  Nucl.\ Phys.\ B {\bf 121}, 77 (1977).

\bibitem{dimred2} 
  F.~Gliozzi, J.~Scherk and D.~I.~Olive,
  ``Supersymmetry, Supergravity Theories and the Dual Spinor Model,''
  Nucl.\ Phys.\ B {\bf 122}, 253 (1977).





\end{thebibliography}
\end{document}